\documentclass[journal]{IEEEtran}
\usepackage[english]{babel}
\usepackage{changes}
\usepackage{flushend}
\usepackage{amssymb}
\usepackage[cmex10]{amsmath}
\usepackage{overpic}
\usepackage{algorithm}
\usepackage{algorithmic}
\usepackage{multirow}
\usepackage{array}
\usepackage{graphicx}
\usepackage{subfigure}
\usepackage{bm}
\usepackage[caption=false,font=normalsize,labelfont=sf,textfont=sf]{subfig}
\usepackage{color}
\usepackage{tikz}
\usepackage{epstopdf}
\usepackage{stfloats}
\usepackage{cite}
\usepackage{tikz}
\usepackage{amsmath}
\usepackage{psfrag}
\usepackage{acronym}
\usepackage{enumerate}
\usetikzlibrary{arrows}
\usetikzlibrary{decorations}

\definecolor{BLUE}{rgb}{0,0,1}

\newtheorem{corollary}{Corollary}
\newtheorem{proposition}{Proposition}
\newtheorem{remark}{Remark}
\newtheorem{lemma}{Lemma}

\newtheorem{theorem}{Theorem}

\newcommand{\tr}[1]{{\rm tr}\left\{#1\right\}}

\newcommand{\diag}[1]{{\rm diag}\left\{#1\right\}}

\acrodef{aoa}[AOA]{angle-of-arrival}
\acrodef{bcrb}[BCRB]{Bayesian Cram\'{e}r-Rao bound}
\acrodef{bfim}[BFIM]{Bayesian Fisher information matrix}
\acrodef{bp}[BP]{belief propagation}
\acrodef{cdi}[CDI]{cooperative dilution intensity}
\acrodef{cir}[CIR]{channel impulse response}
\acrodef{cl}[CL]{cooperative localization}
\acrodef{cp}[CP]{cyclic prefix}
\acrodef{crb}[CRB]{Cram\'{e}r-Rao bound}
\acrodef{crlb}[CRLB]{Cram\'{e}r-Rao lower bound}
\acrodef{dft}[DFT]{discrete Fourier transform}
\acrodef{dof}[DoF]{degree of freedom}
\acrodef{dpeb}[DPEB]{directional position error bound}
\acrodef{fim}[FIM]{Fisher information matrix}
\acrodef{efim}[EFIM]{equivalent Fisher information matrix}
\acrodef{hogs}[HOGS]{hybrid orthogonal-Gaussian signalling}
\acrodef{hsug}[HSUG]{hybrid semi-unitary--Gaussian}
\acrodef{ici}[ICI]{information coupling intensity}
\acrodef{icrb}[ICRB]{inverse CRB}
\acrodef{iid}[i.i.d.]{independently and identically distributed}
\acrodef{im}[IM]{index modulation}
\acrodef{isac}[ISAC]{Integrated Sensing and Communication}
\acrodef{los}[LoS]{line-of-sight}
\acrodef{mse}[MSE]{mean-squared error}
\acrodef{ofdm}[OFDM]{orthogonal frequency-division multiplexing}
\acrodef{pdf}[PDF]{probability density function}
\acrodef{peb}[PEB]{position error bound}
\acrodef{speb}[SPEB]{squared position error bound}
\acrodef{pll}[PLL]{phase-locked loop}
\acrodef{psk}[PSK]{phase shift keying}
\acrodef{p2p}[P2P]{point-to-point}
\acrodef{qam}[QAM]{quadrature amplitude modulation}
\acrodef{rbs}[RBS]{reference broadcast synchronization}
\acrodef{rhs}[RHS]{right hand side}
\acrodef{rii}[RII]{ranging information intensity}
\acrodef{rss}[RSS]{received signal strength}
\acrodef{rc}[RC]{ranging coefficient}
\acrodef{speb}[SPEB]{squared position error bound}
\acrodef{toa}[TOA]{time-of-arrival}
\acrodef{tdoa}[TDOA]{time-difference-of-arrival}
\acrodef{tpsn}[TPSN]{time synchronization protocol for sensor network}
\acrodef{vmp}[VMP]{variational message passing}
\acrodef{wsn}[WSN]{wireless sensor network}
\acrodef{efim}[EFIM]{equivalent Fisher information matrix}
\acrodef{dio}[DIO]{distance-information-only}
\acrodef{aio}[AIO]{angle-information-only}
\acrodef{saaf}[SAAF]{squared array aperture function}
\acrodef{snc}[S\&C]{sensing and communications}
\acrodef{uoa}[UOA]{uniformly oriented array}
\acrodef{rgg}[RGG]{random geometric graph}
\acrodef{snr}[SNR]{signal-to-noise ratio}
\acrodef{eoc}[EoC]{efficiency of cooperation}
\acrodef{npi}[NPI]{nominal position information}
\acrodef{gnss}[GNSS]{global navigation satellite system}
\acrodef{mimo}[MIMO]{multiple-input multiple-output}
\acrodef{mcs}[MCS]{minimally constrained system}
\acrodef{zzb}[ZZB]{Ziv-Zakai lower bound}
\acrodef{wwb}[WWB]{Weiss-Weinstein lower bound}
\acrodef{nlos}[NLOS]{non-light-of-sight}
\acrodef{mmse}[MMSE]{minimum mean squared error}
\acrodef{uav}[UAV]{unmanned aerial vehicle}
\acrodef{ppp}[PPP]{Poisson point process}
\acrodef{bpp}[BPP]{binomial point process}
\acrodef{cln}[CLN]{cooperative location-aware network}
\acrodef{pdr}[PDR]{pedestrian dead reckoning}
\acrodef{ml}[ML]{maximum likelihood}
\acrodef{map}[MAP]{maximum \textit{a posteriori}}
\acrodef{kkt}[KKT]{Karush-Kuhn-Tucker}
\acrodef{st}[ST]{subspace tradeoff}
\acrodef{drt}[DRT]{deterministic-random tradeoff}
\acrodef{ustm}[USTM]{unitary space-time modulation}

\acrodefplural{dof}[DoFs]{degrees of freedom}
\acrodefplural{drt}[DRTs]{deterministic-random tradeoffs}

\usepackage{winsnotation}

\title{On the Fundamental Tradeoff of Integrated Sensing and Communications Under Gaussian Channels}
\author{Yifeng Xiong, \IEEEmembership{Member, IEEE}, Fan Liu, \IEEEmembership{Member, IEEE}, Yuanhao Cui, \IEEEmembership{Member, IEEE}, \\ Weijie Yuan, \IEEEmembership{Member, IEEE}, Tony Xiao Han, \IEEEmembership{Member, IEEE}, and Giuseppe Caire, \IEEEmembership{Fellow, IEEE}
\thanks{Y. Xiong is with the School of Information and Electronic Engineering, Beijing University of Posts and Telecommunications, Beijing, 100876, China, and is also with the Department of Electronic and Electrical Engineering, Southern University of Science and Technology, Shenzhen 518055, China (e-mail: yifengxiong@bupt.edu.cn).}
\thanks{F. Liu, Y. Cui and W. Yuan are with the Department of Electronic and Electrical Engineering, Southern University of Science and Technology, Shenzhen 518055, China (e-mail: liuf6@sustech.edu.cn, cuiyuanhao@bupt.edu.cn, yuanwj@sustech.edu.cn).}
\thanks{T. X.-Han is with Huawei Technologies Co., Ltd (e-mail: tony.hanxiao@\\huawei.com)}
\thanks{G. Caire is with the Chair of Communications and Information Theory, Technical University of Berlin, 10623 Berlin, Germany (e-mail: caire@tu-berlin.de).}
}

\begin{document}
\maketitle

\begin{abstract}
\ac{isac} is recognized as a promising technology for the next-generation wireless networks, which provides significant performance gains over individual \ac{snc} systems via the shared use of wireless resources. The characterization of the \ac{snc} performance tradeoff is at the core of the theoretical foundation of \ac{isac}. In this paper, we consider a \ac{p2p} \ac{isac} model under vector Gaussian channels, and propose to use the \ac{crb}-rate region as a basic tool for depicting the fundamental \ac{snc} tradeoff. In particular, we consider the scenario where a unified \ac{isac} waveform is emitted from a dual-functional \ac{isac} transmitter (Tx), which simultaneously performs \ac{snc} tasks with a communication receiver (Rx) and a sensing Rx. In order to perform both \ac{snc} tasks, the \ac{isac} waveform is required to be random to convey communication information, with realizations being perfectly known at both the \ac{isac} Tx and the sensing Rx as a reference sensing signal as in typical radar systems. In this context, we treat the \ac{isac} waveform as a random but known nuisance parameter in the sensing signal model, and define a Miller-Chang type \ac{crb} for the analysis of the sensing performance.

As the main contribution of this paper, we characterize the \ac{snc} performance at the two corner points of the \ac{crb}-rate region, namely, $P_{\rm SC}$ indicating the maximum achievable communication rate constrained by the minimum \ac{crb}, and $P_{\rm CS}$ indicating the minimum achievable \ac{crb} constrained by the maximum communication rate. In particular, we derive the high-SNR communication capacity at $P_{\rm SC}$, and provide lower and upper bounds for the sensing \ac{crb} at $P_{\rm CS}$. We show that these two points can be achieved by the conventional Gaussian signalling and a novel strategy relying on the uniform distribution over the set of semi-unitary matrices, i.e., the Stiefel manifold, respectively. Based on the above-mentioned analysis, we provide an outer bound and various inner bounds for the achievable \ac{crb}-rate regions.

Our main results reveal a two-fold tradeoff in \ac{isac} systems, consisting of the \ac{st} and the \ac{drt} that depend on the resource allocation and data modulation schemes employed for \ac{snc}, respectively. Within this framework, we examine the state-of-the-art \ac{isac} signalling strategies and study a number of illustrative examples, which are validated through numerical simulations.
\end{abstract}

\begin{IEEEkeywords}
Integrated sensing and communication, Gaussian channels, CRB-rate region, deterministic-random tradeoff, subspace tradeoff.
\end{IEEEkeywords}

\section{Introduction}
\subsection{Background and Related Works}
\IEEEPARstart{I}{n} an \ac{isac} system, wireless communications and sensing functionalities are performed by using a single hardware platform and a common radio waveform over the same frequency band, which considerably improves the energy-, spectral-, and hardware-efficiencies \cite{dual_functional,isac_network,anliu_fundamantal,cooploc,coopsync}. Due to the substantial performance gain attainable for both \ac{snc} capabilities, \ac{isac} is well-recognized as a key enabler for a variety of emerging applications including vehicular networks, smart home, and smart cities. Despite the fact that \ac{snc} have long been considered as two isolated fields, e.g., wireless networks and radar systems, they are indeed intertwined with each other as an ``odd couple" in an information-theoretic sense \cite{odd_couple}.

For decades, \ac{snc} researchers are mostly working on a generic linear Gaussian model, expressed as
\begin{equation}
\RM{Y} = \RM{H}\RM{X} + \RM{Z},
\end{equation}
where $\RM{Y}\in\mathbb{C}^{N\times T}$, $\RM{H}\in\mathbb{C}^{N\times M}$, $\RM{X}\in\mathbb{C}^{M\times T}$, and $\RM{Z}\in\mathbb{C}^{N\times T}$ denote the received signal, target response/communication channel, transmitted signal, and Gaussian noise, respectively, which can be in scalar, vector, or matrix forms. From the communication perspective, the basic problem is given channel $\RM{H}$ to design an optimal transmission strategy $\RM{X}$ that maximizes the channel capacity. From the sensing perspective, the basic problem is to estimate $\RM{H}$, or, more relevant to radar sensing, to estimate the target parameters (e.g., amplitude, delay, angle, and Doppler) contained in $\RM{H}$ as accurate as possible, based on the knowledge of $\RM{X}$. 
Indeed, \ac{snc} are connected with each other given the duality between signals and linear systems, which are mathematically interchangeable. Nonetheless, the distinct physical roles of $\RM{X}$ and $\RM{H}$ in \ac{snc} and the resulting performance measures lead to a number of unique performance tradeoffs and design criteria \cite{crb_radcom,overview_radcom_sp,radcom_vehicular,bayesian_radcom} to the \ac{isac} signal processing, particularly for simultaneously realizing \ac{snc} functionalities via a unified waveform. Despite that capacity characterization and parameter estimation limits have been independently investigated for \ac{snc} for years, the depiction of the inherent information-estimation tradeoff and the corresponding performance bounds in an \ac{isac} system remains long-standing open.

The pioneering effort to reveal the fundamental connection between information and estimation theory can be traced back to the early results on the I-MMSE equation \cite{immse}, which states that for the linear Gaussian model $\RM{Y} = \sqrt{\text{snr}}\RM{H}\RM{X} + \RM{Z}$, the derivative of the mutual information $I\left(\RM{Y};\RM{X}\right)$ with respect to the \ac{snr} is equal to the minimum mean squared error (MMSE) for estimating $\RM{H}\RM{X}$ by observing $\RM{Y}$.
While the I-MMSE equation implies maximizing the MMSE of $\RM{H}\RM{X}$ also maximizes the scaling ratio of the mutual information $I\left(\RM{Y};\RM{X}\right)$, which bridges the information theory and estimation theory, the equation itself mainly concentrates on the estimation performance of unknown data symbol $\RM{X}$ rather than on the target parameters in $\RM{H}$, since $\RM{H}$ is assumed to be deterministic and known \cite{immse}. Consequently, the connection between $I\left(\RM{Y};\RM{X}\right)$ and the sensing performance limits for $\RM{H}$ was not fully investigated. More relevant to this work, one may model the \ac{isac} channel as a state-dependent channel \cite{1412040}, which characterizes the information-conveyed signal reflected by state-varying targets \cite{cd}. 
The goal for such an \ac{isac} system is to transmit information through the channel while estimating its state under the minimum distortion. On top of that, a capacity-distortion (C-D) tradeoff is acquired to depict the \ac{snc} performance bounds \cite{cdmac,cdbroadcast,cddetection,cdit}. While these studies could capture the \ac{snc} tradeoff from the information-theoretic perspective, the connections among communication capacity, target channel states, and to-be-estimated parameters in sensing tasks are still not clearly indicated. 
    \subsection{Contribution of This Paper}
To gain a better understanding of the interweaved \ac{snc} functionalities, in this paper, we shed light on the fundamental limits of \ac{isac} by analyzing the tradeoff between the communication capacity and sensing \ac{crb}, two key performance metrics at the cores of information theory and estimation theory. We consider a general point-to-point (P2P) system setting, where a random signal $\RM{X}$ is emitted from an \ac{isac} Tx and received both at a communication Rx and a sensing Rx, thus to realize simultaneous \ac{snc} functionalities. Since $\RM{X}$ encodes useful information intended for the communication Rx, it is unknown to the communication Rx, yet is perfectly known to both the ISAC Tx and sensing Rx due to the fact that they are colocated or collaborative in typical monostatic or bistatic radar settings. While the P2P communication capacity is known to be the maximum mutual information over all possible distributions of $\RM{X}$, the optimal achievable sensing performance is less understood, especially when the signal $\RM{X}$ is random. To cope with this issue, we propose to treat $\RM{X}$ as a random known nuisance parameter in the sensing model, and resort to a Miller-Chang type \ac{crb} to measure the average sensing performance \cite{mcb}.

Under this framework, our main contributions are summarized as follows:

\begin{itemize}
    \item We define a Miller-Chang type \ac{crb} for measuring the sensing performance limit in an \ac{isac} system, and show that its optimum is achieved when the sample covariance matrix $\RM{R}_{\RM{X}} = T^{-1} \RM{X}\RM{X}^{\rm H}$ has a deterministic trace, and the support of the distribution $p\left(\RM{R}_{\RM{X}}\right)$ (and hence $p(\RM{X})$) is restricted to the optimal solution set of a deterministic \ac{crb} minimization problem. In particular, if the solution is unique, the sensing-optimal sample covariance matrix $\RM{R}_{\RM{X}}$ itself should be deterministic;
    \item We define the \ac{crb}-rate region as the set of all achievable pairs of the communication rate and the sensing \ac{crb}, and propose a pentagon inner bound of the \ac{crb}-rate region that can be achieved through simple time-sharing strategy. Within this framework, we study the \ac{isac} performance at the two corner points of the \ac{crb}-rate region, namely, $P_{\rm CS}$ indicating the minimum achievable \ac{crb} constrained by the maximum communication rate, and $P_{\rm SC}$ indicating the maximum achievable communication rate constrained by the minimum \ac{crb}.
    \item We derive the high-\ac{snr} communication capacity for the sensing-optimal point $P_{\rm SC}$, and prove that it can be asymptotically achieved by a strategy based on uniform sampling over the set of semi-unitary matrices, i.e., the Stiefel manifold \cite{riemannian_geometry}. As a further step, we provide lower and upper bounds for the sensing \ac{crb} at the communication-optimal point $P_{\rm CS}$;
    \item We reveal that the \ac{snc} tradeoff in \ac{isac} systems is a two-fold tradeoff, namely, the \acf{st} balancing the resource allocation between the subspaces spanned by \ac{snc} channels, and the \acf{drt} depicting the exploitable \ac{snc} \acp{dof} in \ac{isac} signals. Based on these two tradeoffs, we further propose an outer bound and a variety of inner bounds for the \ac{crb}-rate region.
    \item We unveil the connection between the above fundamental tradeoffs and existing \ac{isac} waveform designs, and provide illustrative examples to demonstrate the behaviour of \ac{st} and \ac{drt} in typical \ac{isac} application scenarios.
\end{itemize}

\begin{figure}[t]
\centering
\begin{minipage}{.2\textwidth}
 \centering
\includegraphics[width=.99\textwidth]{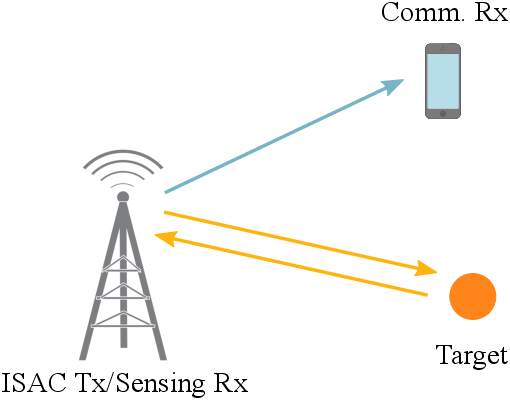}
\vspace{-1mm}
\footnotesize  (a) Monostatic sensing
\end{minipage}
\hspace{5mm}
\begin{minipage}{.21\textwidth}
\centering
\includegraphics[width=.99\textwidth]{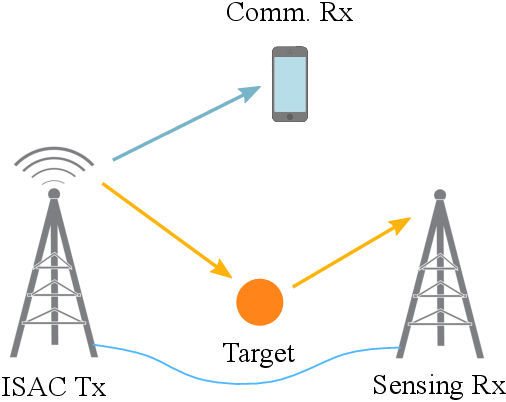}
\footnotesize  (b) Bistatic sensing
\end{minipage}
\caption{The \ac{isac} scenarios considered in this paper, where the dual-functional waveform $\RM{X}$ is known to the sensing receiver.}
\label{fig:scenarios}
\end{figure}

To the best of our knowledge, this is the first work that addresses the fundamental tradeoff in \ac{isac} from both information- and estimation-theoretic perspectives. Our hope is that this paper can serve as a stepping stone towards the fully characterization of the \ac{crb}-rate performance region, as well as towards the design of bound-achieving \ac{isac} transmission strategies.

The rest of this paper is organized as follows. The system model and the performance metrics for both sensing and communication are introduced in Section~\ref{sec:model}. Then, in Section~\ref{sec:results}, we present our main results characterizing the corner points $P_{\rm SC}$ and $P_{\rm CS}$ of the \ac{crb}-rate region, while the corresponding proofs are deferred to the Appendices. Based on these results, we discuss the \ac{st} and \ac{drt} in Section~\ref{ssec:two_fold}, provide some strategies achieving inner bounds of the \ac{crb}-rate region in Section~\ref{ssec:inner_bounds}, and unveil the connections between the \ac{snc} tradeoff and existing \ac{isac} system designs in Section~\ref{ssec:connections}. This is followed by a case study in Section~\ref{sec:case_study}, where we discuss the \ac{st} and \ac{drt} in practical \ac{isac} application scenarios. We illustrate and validate the analytical results using numerical examples in Section~\ref{sec:numerical}, and finally conclude the paper in Section~\ref{sec:numerical}.

\subsection*{Notations}
Throughout this paper, $\rv{a}$, $\RV{a}$, $\RM{A}$, and $\RS{A}$ represent random variables (scalars), random vectors, random matrices and random sets, respectively; The corresponding deterministic quantities, are denoted by $a$, $\V{a}$, $\M{A}$, and $\Set{A}$, respectively. The $m$-by-$n$ matrix of zeros (resp. ones) is denoted by $\M{0}_{m\times n}$ (resp. $\M{1}_{m\times n}$). The $m$-dimensional vector of zeros (resp. ones) is denoted by $\V{0}_{m}$ (resp. $\M{1}_{m}$). The $m$-by-$m$ identity matrix is denoted by $\M{I}_{m}$. The Kronecker product between matrices $\M{A}$ and $\M{B}$ is denoted by $\M{A}\otimes\M{B}$. $\|\V{x}\|_p$ denotes the $l_p$ norm, which represents the $l_2$ norm by default when the subscript is omitted. The notation $\mathbb{E}_{\RV{x}}\{\cdot\}$ denotes the expectation with respect to $\RV{x}$. $[\cdot]^\dagger$ denotes the Moore-Penrose pseudo inverse of its argument, $[\cdot]^\ast$ denotes the complex conjugate of its argument, while $[\cdot]^{\rm H}$ denotes the Hermitian transpose of its argument. $\M{A}_{:,i}$ denotes the $i$-th column of $\M{A}$, while $\M{A}_{i,:}$ denotes the $i$-th row. The notation ${\rm diag}(\cdot)$ denotes the matrix obtained by placing its arguments on the main diagonal of a sqaure matrix, ${\rm mdiag}(\cdot)$ denotes the vector consisting of the main diagonal entries of its argument, while ${\rm blkdiag}(\cdot)$ denotes the matrix obtained similarly, but with matrix arguments. $\M{A}\succeq \M{B}$ implies that $\M{A}-\M{B}$ is positive semidefinite. $\mathrm{tr}\{\cdot\}$ stands for the trace of a square matrix. ${\rm vec}\{\M{A}\}$ denote the column-stacked vector of the matrix $\M{A}$, while ${\rm mat}_{M\times N}\{\V{v}\}$ denotes the $M\times N$ matrix satisfying ${\rm vec}\{{\rm mat}_{M\times N}(\V{v})\}=\V{v}$. The subscripts in the aforementioned notations may be omitted when they are clear from the context.

\section{System Model and Performance Metrics}\label{sec:model}
\subsection{System Model}
Let us consider the general \ac{isac} system model of
\begin{subequations}
\begin{align}
\RM{Y}_{\rm c} &= \RM{H}_{\rm c} \RM{X} + \RM{Z}_{\rm c}, \label{comm_model} \\
\RM{Y}_{\rm s} &= \RM{H}_{\rm s} \RM{X} + \RM{Z}_{\rm s}, \label{sensing_model}
\end{align}
\end{subequations}
where  $\RM{H}_{\rm c}\in\mathbb{C}^{N_{\rm c}\times M}$ and $\RM{H}_{\rm s}\in\mathbb{C}^{N_{\rm s}\times M}$ denote the communication channel and the target response matrix, respectively, $\RM{Y}_{\rm c}\in\mathbb{C}^{N_{\rm c}\times T}$ and $\RM{Y}_{\rm s}\in\mathbb{C}^{N_{\rm s}\times T}$ denote the received communication and sensing signals, respectively. $\RM{X}\in\mathbb{C}^{M\times T}$ denotes the transmitted dual-functional waveform emitted from an ISAC TX for performing both \ac{snc} tasks. In this paper, we assume that $\RM{X}$ is known to both the ISAC Tx and sensing Rx, but unknown to the communication Rx. This is an abstraction of the practical scenarios portrayed in Fig.~\ref{fig:scenarios}. A more general scenario where the sensing task can be performed by the communication receiver itself will be addressed in our future works. The random matrix $\RM{Z}_{\rm c}\in\mathbb{C}^{N_{\rm c}\times T}$ denotes the received communication noise, modelled as \ac{iid} circularly symmetric complex Gaussian random variables with zero mean, namely ${\rm vec}(\RM{Z}_{\rm c})\sim \mathcal{CN}(\V{0},\sigma_{\rm c}^2\M{I}_{N_{\rm c}\times T})$. Similarly, we also model the sensing noise $\RM{Z}_{\rm c}$ as ${\rm vec}(\RM{Z}_{\rm s})\sim \mathcal{CN}(\V{0},\sigma_{\rm s}^2\M{I}_{N_{\rm s}\times T})$. The communication subsystem aims for transmitting as much information as possible (reliably) to the receiver, while the sensing subsystem aims for estimating the sensing parameters $\RV{\eta}\in\mathbb{R}^{K}$ contained in $\RM{H}_{\rm s}$ satisfying
$$
\RM{H}_{\rm s}=\V{g}(\RV{\eta})
$$
to the highest possible accuracy, where $\V{g}(\cdot)$ is an injective mapping from $\mathbb{R}^K$ to $\mathbb{C}^{N_{\rm s}\times M}$.\footnote{This would impose upper bounds for the maximum number $K_{\max}$ of identifiable parameters. A trivial upper bound would be $K_{\max}\leq 2MN_{\rm s}$. Tighter bounds may be obtained by incorporating specific knowledge of the observation model. We refer interested readers to \cite{identifiability}.} We consider a block fading model for both the target response matrix and the communication channel. Specifically, we assume that parameters $\RV{\eta}$ varies every $T$ symbols in an \ac{iid} manner, following a known distribution $p_{\RV{\eta}}(\V{\eta})$ which has a finite variance, and that the communication channel $\RM{H}_{\rm c}$ also varies every $T$ symbols in an \ac{iid} manner.\footnote{These assumptions correspond to the memoryless channel model widely used in information-theoretical \ac{isac} studies (cf. \cite{cdit,inf_isac_correlated}). This model mainly aims for portraying practical scenarios in which the sensing target is relatively close to the sensing link, and hence the sensing parameters vary synchronously with the communication channel \cite{centitrack}. We note that our analysis also applies to the case where the communication channel varies every $kT$ symbols where $k\in\mathbb{Z}_{+}$.} We will refer to $T$ as the \emph{coherent sensing period} in the rest of the paper. We limit the average power of each transmitted symbol to be $P_{\rm T}$, hence we have
\begin{equation}\label{power_constraint}
\tr{\widetilde{\M{R}}_{\RM{X}}}=\mathbb{E}\{\tr{\RM{R}_{\RM{X}}}\} = P_{\rm T}M,
\end{equation}
where $\widetilde{\M{R}}_{\RM{X}}:=\mathbb{E}\{\RM{R}_{\RM{X}}\}$ denotes the covariance matrix of $\RM{X}$, with $\RM{R}_{\RM{X}}:=T^{-1}\RM{X}\RM{X}^{\rm H}$ representing the sample covariance matrix.
\subsection{\ac{snc} Performance Metrics}
The performance of the communication subsystem is conventionally characterized by the ergodic achievable rate (under certain design constraints), which can be expressed as
\begin{equation}
R_{\Set{F}}=\max_{p_{\RM{X}}(\M{X})} ~T^{-1} I(\RM{Y}_{\rm c};\RM{X}|\RM{H}_{\rm c}),~{\rm s.t.}~p_{\RM{X}}(\M{X})\in\Set{F},
\end{equation}
where $I(\RM{Y}_{\rm c};\RM{X}|\RM{H}_{\rm c})$ denotes the mutual information between $\RM{Y}_{\rm c}$ and $\RM{X}$ conditioned on $\RM{H}_{\rm c}$, and $\Set{F}$ denotes the feasibility region of $p_{\RM{X}}(\M{X})$ determined by the design constraints. If not stated otherwise, in the rest of this paper, the feasibility region $\Set{F}$ is the set of all $p_{\RM{X}}(\RM{X})$'s that satisfy the power constraint \eqref{power_constraint}, and the subscript $\Set{F}$ in $R_{\Set{F}}$ is omitted whenever there is no confusion.

The performance of the sensing subsystem is typically characterized by the estimation \ac{mse} of the parameters $\RV{\eta}$, given by
\begin{equation}
{\rm MSE}_{\RV{\eta}}(\hat{\RV{\eta}}):=\mathbb{E}\{\|\RV{\eta}-\hat{\RV{\eta}}\|^2\}.
\end{equation}
However, this metric depends on the specific choice of the estimator $\hat{\RV{\eta}}$, which hinders the essential relationship between the sensing performance and the design of $p_{\RM{X}}(\M{X})$. To this end, we consider the \ac{bcrb} of $\RV{\eta}$, which constitutes a lower bound for the \ac{mse} of weakly unbiased estimators \cite{van_trees}, taking the following form:
\begin{equation}
{\rm MSE}_{\RV{\eta}}(\hat{\RV{\eta}})\geq \mathbb{E}\left(\tr{\RM{J}_{\RV{\eta}|\RM{X}}^{-1}}\right),
\end{equation}
where the expectation is taken with respect to $\RM{X}$, and $\RM{J}_{\RV{\eta}|\RM{X}}$ denotes the \ac{bfim} of $\RV{\eta}$ given by
$$
\begin{aligned}
\RM{J}_{\RV{\eta}|\RM{X}}&\!:= \!\mathbb{E} \!\left\{\!\frac{\partial\ln p_{\RM{Y}_{\rm s}|\RM{X},\RV{\eta}}(\M{Y}_{\rm s}|\M{X},\V{\eta})}{\partial \V{\eta}}\frac{\partial\ln p_{\RM{Y}_{\rm s}|\RM{X},\RV{\eta}}(\M{Y}_{\rm s}|\M{X},\V{\eta})}{\partial \V{\eta}^{\rm T}} \bigg|\RM{X}\right\}\\
&\hspace{5mm}+\mathbb{E}\left\{\frac{\partial \ln p_{\RV{\eta}}(\V{\eta})}{\partial \V{\eta}}\frac{\partial \ln p_{\RV{\eta}}(\V{\eta})}{\partial \V{\eta}^{\rm T}}\right\}.
\end{aligned}
$$
It is well-known that certain practical estimators (e.g. the \ac{map} estimator) are capable of achieving the \ac{bcrb} in the asymptotic limit of high \ac{snr} \cite{van_trees}.\footnote{When the \ac{snr} is lower, \ac{bcrb} is less tight due to the fact that it mainly exploits the local information of the posterior distribution around the mode. Therefore, in this paper, we will focus on the analysis of high-\ac{snr} scenarios.} In light of this, in this paper, we choose the \ac{bcrb} of $\RV{\eta}$, namely
\begin{equation}\label{bcrb_form}
\epsilon:=\mathbb{E}\left(\tr{\RM{J}_{\RV{\eta}|\RM{X}}^{-1}}\right),
\end{equation}
as the metric of sensing accuracy.

At this point, it is worthwhile to clarify our motivation of choosing the specific form of \ac{bcrb} in \eqref{bcrb_form}, as well as to justify this choice. Indeed, there exist multiple versions of the \ac{bcrb} when the parameters (or a part of the parameters) are random \cite{some_classes}.  What we are using is in fact the Miller-Chang type bound \cite{mcb}, which treats $\RM{X}$ as a nuisance parameter, and takes the expectation with respect to the conditional CRB $\tr{\RM{J}_{\RV{\eta}|\RM{X}}^{-1}}$. This specific form is particularly relevant to the scenario where $\RM{X}$ is known to the sensing receiver. To elaborate, let us consider a specific realization of $\RM{X}$, denoted as $\M{X}_i$. Since the sensing receiver always have the full knowledge about $\RM{X}$, the \ac{bcrb} for this realization is given by $\tr{\RM{J}_{\RV{\eta}|\RM{X}=\M{X}_i}^{-1}}$. Taking the expectation over all realizations of $\RM{X}$, we obtain \eqref{bcrb_form}.

\begin{figure}[t]
\centering
\includegraphics[width=.43\textwidth]{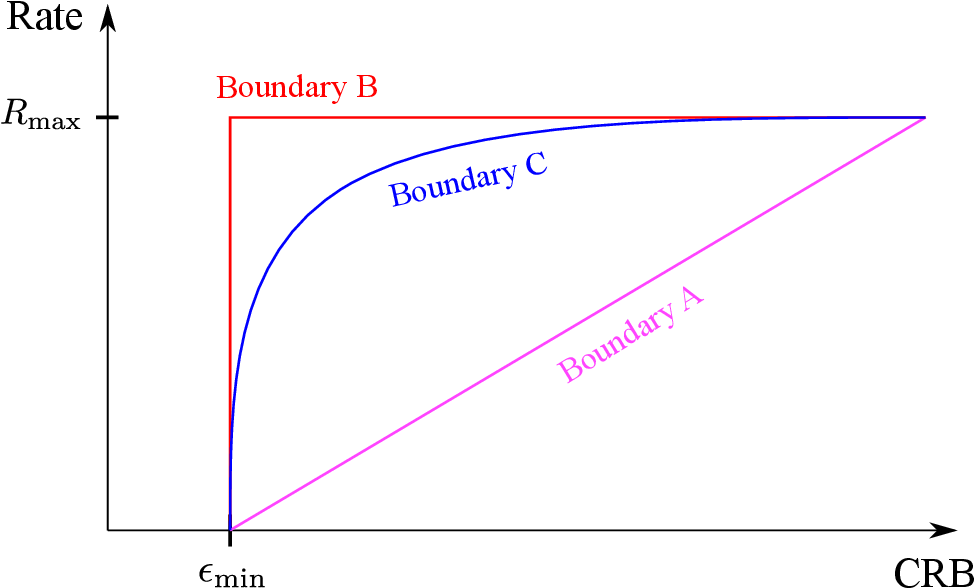}
\caption{Graphical illustration of various possible \ac{crb}-rate regions.}
\label{fig:region_illustration}
\end{figure}

To reveal the fundamental \ac{snc} performance tradeoff, we define the \ac{crb}-rate region, which is the set of all feasible ordered pairs $(\epsilon,R)$, where $R$ and $\epsilon$ are the communication rate and the sensing \ac{crb}, respectively. Typically, we are interested in the boundary of the \ac{crb}-rate region, which may be viewed as the Pareto front constituted by all optimal performance tradeoffs. To provide further intuitions about the boundary, we portray some conceptual \ac{crb}-rate regions in Fig.~\ref{fig:region_illustration}. Specifically, the point $(\epsilon_{\min},0)$ represents the minimum achievable sensing \ac{mse} regardless of the communication performance, while the point $(+\infty,R_{\max})$ represents the maximum achievable rate regardless of the sensing performance. The boundary A may be seen in scenarios where the integration of sensing and communication does not provide additional performance gain (but we have some \textit{a priori} knowledge about the sensing parameters, hence the maximum \ac{crb} is not infinity), while the boundary B may be seen in idealistic scenarios where both sensing and communication performance can achieve their optimum without eroding the other. In most practical scenarios, the boundary of \ac{crb}-rate regions may have a similar shape as that of the boundary C.

\section{Main Results}\label{sec:results}
In this section, we summarize the main analytical results of this paper. Before delving into details, we first give a general description of an inner bound of the \ac{crb}-rate region.
\begin{proposition}[Pentagon Inner Bound]\label{prop:pentagon}
Any point $(\epsilon,R)$ satisfying the inequalities
\begin{subequations}
\begin{align}
\epsilon &\geq \epsilon_{\min},\\
R&\leq R_{\max},\\
\epsilon&\geq \epsilon_{\min}+\frac{\epsilon_{\rm CS}-\epsilon_{\min}}{R_{\max}-R_{\rm SC}}(R-R_{\rm SC}) \label{line_connection}
\end{align}
\end{subequations}
is achievable, where we define
\begin{subequations}\label{crb_rate_region}
\begin{align}
\epsilon_{\min}&:=\min_{p_{\RM{X}}(\M{X})\in\Set{F}} ~\epsilon,\\
R_{\max} &:= \max_{p_{\RM{X}}(\M{X})\in\Set{F}}~T^{-1} I(\RM{Y}_{\rm c};\RM{X}|\RM{H}_{\rm c}),\\
\epsilon_{\rm CS}&:=\!\min_{p_{\RM{X}}(\M{X})\in\Set{F}} \epsilon,~\mathrm{s.t.}~T^{-1}I(\RM{Y}_{\rm c};\RM{X}|\RM{H}_{\rm c})\!=\!R_{\max},\\
R_{\rm SC}&:= \max_{p_{\RM{X}}(\M{X})\in\Set{F}}~T^{-1} I(\RM{Y}_{\rm c};\RM{X}|\RM{H}_{\rm c}),~\mathrm{s.t.}~\epsilon\!=\!\epsilon_{\min},
\end{align}
\end{subequations}
and
\begin{equation}\label{cool_points}
P_{\rm SC}:=(\epsilon_{\min},R_{\rm SC}),~P_{\rm CS}:=(\epsilon_{\rm CS},R_{\max}).
\end{equation}
\begin{IEEEproof}
The points in \eqref{cool_points} are obviously achievable (but note that in certain scenarios we may have $R_{\rm SC}=0$ or $\epsilon_{\rm CS}=\infty$). The line segment connecting $P_{\rm SC}$ and $P_{\rm CS}$ (characterized by \eqref{line_connection}) can be achieved by using the celebrated time-sharing strategy \cite[Chap.~4]{nit}, namely applying the strategy corresponding to $P_{\rm SC}$ with probability $p_1$, while applying the strategy corresponding to $P_{\rm CS}$ with probability $1-p_1$. This completes the proof.
\end{IEEEproof}
\end{proposition}

\subsection{The Structure of Sensing-optimal Signals}
\begin{figure}[t]
\centering
\begin{overpic}[width=.45\textwidth]{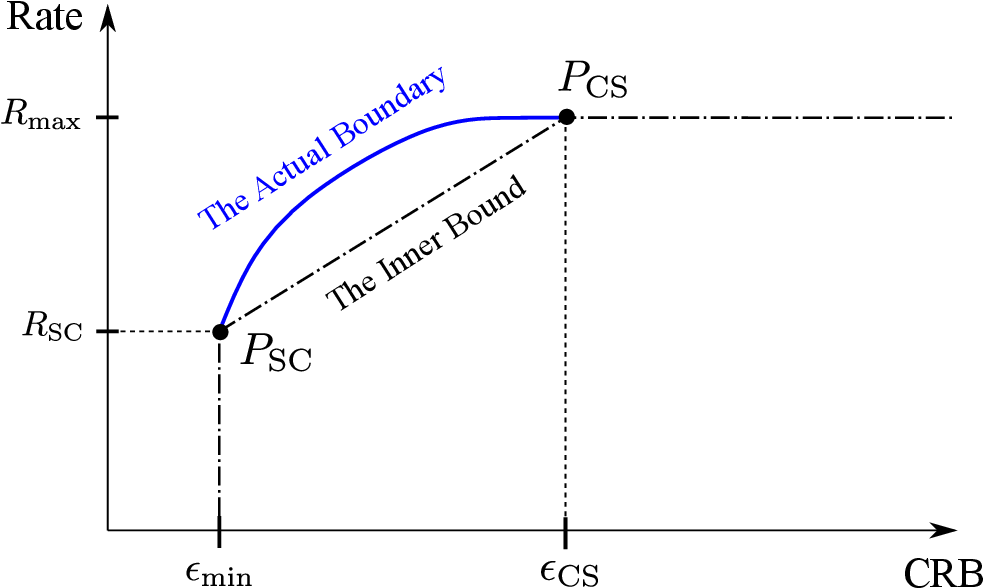}
\put(24,17.5){\footnotesize (Sec. \ref{ssec:psc})}
\put(65.5,50){\footnotesize (Sec. \ref{ssec:pcs})}
\end{overpic}
\caption{The pentagon inner bound of \ac{crb}-rate regions.}
\label{fig:pentagon_inner_bound}
\end{figure}

As it can be observed from Fig.~\ref{fig:pentagon_inner_bound}, the points $P_{\rm SC}$ and $P_{\rm CS}$ correspond to the sensing-optimal and the communication-optimal strategies, respectively, which serve as important control points of the \ac{crb}-rate region. Naturally, a detailed characterization of these points is desired. In particular, to obtain $P_{\rm SC}$--achieving strategies, we should first find the conditions that a sensing-optimal signal $\RM{X}$ must satisfy. Especially, the \ac{isac} scenario imposes a unique challenge that the signal $\RM{X}$ has to be random for carrying information, while existing analysis of pure sensing (e.g. radar sensing) scenarios typically assume deterministic signal or neglect the randomness. To this end, we derive the general form of the \ac{bfim} as follows.

\begin{proposition}[\ac{bfim} Structure]\label{prop:fim}
The \ac{bfim} of $\RV{\eta}$ conditioned on $\RM{X}$ takes the following form
\begin{equation}
\RM{J}_{\RV{\eta}|\RM{X}} =  \frac{T}{\sigma_{\rm s}^2}\M{\Phi}(\RM{R}_{\RM{X}}),
\end{equation}
where $\M{\Phi}(\cdot)$ is an affine map characterized by
\begin{equation}\label{phi_def}
\M{\Phi}(\M{A})=\sum_{i=1}^{r_1}\widetilde{\M{F}}_i\M{A}^{\rm T} \widetilde{\M{F}}_i^{\rm H}+\sum_{j=1}^{r_2}\widetilde{\M{G}}_j\M{A}\widetilde{\M{G}}_j^{\rm H} + \widetilde{\M{J}}_{\rm P},
\end{equation}
where $\widetilde{\M{J}}_{\rm P} = \sigma_{\rm s}^2T^{-1}\M{J}_{\rm P}$, and the term $\M{J}_{\rm P}$ is contributed by the prior distribution $p_{\RV{\eta}}(\V{\eta})$, given by\footnote{In this paper, we consider the scenario where $\RM{H}_{\rm s}$ and $\RM{H}_{\rm c}$ are not statistically correlated, although they may be physically correlated in terms of subspace overlap (as will be detailed in Section \ref{ssec:two_fold}). When they do have a statistical correlation, the knowledge of $\RM{H}_{\rm c}$ may be modelled as an observation of $\RV{\eta}$ via $p(\RM{H}_{\rm c}|\RM{H}_{\rm s})$ and $\RM{H}_{\rm s}=\V{g}(\RV{\eta})$. Let us denote the contribution of $p(\RM{H}_{\rm c}|\RM{H}_{\rm s})$ to the \ac{bfim} as $\RM{J}_{\RV{\eta}|\RM{H}_{\rm c}}$. Since $\RM{J}_{\RV{\eta}|\RM{H}_{\rm c}}$ does not depend on the transmitted signal $\RM{X}$, it can be absorbed into the term $\M{J}_{\rm P}$, and hence the proposed framework also applies to this case after proper modifications.}
$$
\M{J}_{\rm P} = \mathbb{E}\left\{\frac{\partial \ln p_{\RV{\eta}}(\V{\eta})}{\partial \V{\eta}}\frac{\partial \ln p_{\RV{\eta}}(\V{\eta})}{\partial \V{\eta}^{\rm T}}\right\}.
$$
The matrices $\widetilde{\M{F}}_i$ and $\widetilde{\M{G}}_i$ are given by \eqref{kraus_operators_F} and \eqref{kraus_operators_G}. Furthermore, we have $r_1, r_2\leq KM$.
\begin{IEEEproof}
Please refer to Appendix \ref{sec:proof_fim}.
\end{IEEEproof}
\end{proposition}

When the trace of $\RM{R}_{\RM{X}}$ is fixed, we have the following alternative representation of $\M{\Phi}(\cdot)$.
\begin{corollary}\label{coro:supermap}
When the trace of $\RM{R}_{\RM{X}}$ is fixed, i.e., ${\rm tr}\{\RM{R}_{\RM{X}}\}=\gamma$, we may express $\M{\Phi}(\cdot)$ as follows
\begin{equation}\label{asymmetric_representation}
\M{\Phi}(\M{A})=\M{\Phi}_{\gamma} (\M{A}):=\sum_{i=1}^{r_3} \bar{\M{F}}_i \M{A}\bar{\M{G}}_i^{\rm H},
\end{equation}
where $r_3\leq KM$, and the matrices $\bar{\M{F}}_i$ and $\bar{\M{G}}_i$ are given by \eqref{bar_f_g}.
\begin{IEEEproof}
Please refer to Appendix \ref{sec:proof_supermap}.
\end{IEEEproof}
\end{corollary}

Using Proposition \ref{prop:fim} and Corollary \ref{coro:supermap}, we may now characterize the minimum achievable \ac{crb} as follows.
\begin{proposition}[Sensing-optimal $\RM{R}_{\RM{X}}$]\label{prop:max_icrb}
The minimum achievable \ac{crb} $\epsilon_{\min}$ is achieved when the trace of $\RM{R}_{\RM{X}}$ is deterministic, namely when $\tr{\RM{R}_{\RM{X}}}=\tr{\widetilde{\M{R}}_{\RM{X}}}$. Moreover, the support of $p(\RM{R}_{\RM{X}})$ should be restricted to the optimal solution set of the following deterministic convex optimization problem
\begin{subequations}\label{opt_problem}
\begin{align}
\min_{\widetilde{\M{R}}_{\RM{X}}} &~~\tr{\Big(\M{\Phi}_{P_{\rm T}M}(\widetilde{\M{R}}_{\RM{X}})\Big)^{-1}}, \label{obj_function_CRB}\\
{\rm s.t.}&~~\tr{\widetilde{\M{R}}_{\RM{X}}}=P_{\rm T}M,~\widetilde{\M{R}}_{\RM{X}}\succeq \V{0},~\widetilde{\M{R}}_{\RM{X}}=\widetilde{\M{R}}_{\RM{X}}^{\rm H}. \label{opt_constraints}
\end{align}
\end{subequations}
\begin{IEEEproof}
(sketch) We first show that the function
$$
f(\rv{\gamma}):=\min_{\M{R}\succeq \M{0}} \tr{(\M{\Phi}_{\gamma}(\M{R}))^{-1}},~{\rm s.t.}~\tr{\M{R}}=\rv{\gamma},~\M{R}=\M{R}^{\rm H}
$$
is convex. The optimal objective function value of the problem \eqref{opt_problem} can then be expressed as $f(P_{\rm T}M)$. Thus for positive semidefinite Hermitian $\RM{R}_{\RM{X}}$, we obtain
$$
\begin{aligned}
&\min_{p(\RM{R}_{\RM{X}})}\mathbb{E}\left\{\tr{[\M{\Phi}(\RM{R}_{\RM{X}})]^{-1}}\right\},~{\rm s.t.}~\tr{\RM{R}_{\RM{X}}}=P_{\rm T}M\\
&\hspace{3mm}\geq \min_{p(\rv{\gamma})} \mathbb{E}\left\{\min_{\tr{\RM{R}_{\RM{X}}}=\rv{\gamma}}\tr{[\M{\Phi}_{\rv{\gamma}}(\RM{R}_{\RM{X}})]^{-1}}\right\},~{\rm s.t.}~\mathbb{E}\{\rv{\gamma}\}\!=\!P_{\rm T}M\\
&\hspace{3mm}=\min_{p(\rv{\gamma})} \mathbb{E}\left\{f(\rv{\gamma})\right\},~{\rm s.t.}~\mathbb{E}\{\rv{\gamma}\}=P_{\rm T}M\\
&\hspace{3mm}\geq f(P_{\rm T}M),
\end{aligned}
$$
where the last line follows from Jensen's inequality. For a detailed proof, please refer to Appendix \ref{sec:proof_max_icrb}.
\end{IEEEproof}
\end{proposition}

Proposition \ref{prop:max_icrb} tells us that the sensing-optimal $\RM{R}_{\RM{X}}$ has a deterministic trace. Moreover, when the problem \eqref{opt_problem} has a unique optimal solution, it holds that the sensing-optimal $\RM{R}_{\RM{X}}$ is itself deterministic, namely $\RM{R}_{\RM{X}}=\widetilde{\M{R}}_{\RM{X}}$. Next, we present a sufficient and necessary condition for the uniqueness of the optimal solution of \eqref{opt_problem}.

\begin{proposition}[Uniqueness of the Sensing-optimal $\RM{R}_{\RM{X}}$]\label{prop:uniqueness}
Consider a maximum-rank optimal solution of \eqref{opt_problem}, denoted as $\M{R}_{\rm opt}$. It is also the unique optimal solution of \eqref{opt_problem}, if and only if the matrix $\M{\Xi}(\M{U}_{\rm opt}^*\otimes \M{U}_{\rm opt})\in\mathbb{C}^{K^2\times r^2}$ has full column rank, where $\M{\Xi}\in\mathbb{C}^{K^2\times M^2}$ is given by
\begin{equation}
\M{\Xi}=\sum_{i=1}^{r_1} (\widetilde{\M{F}}_i^{\ast} \otimes \widetilde{\M{F}}_i)\M{K} + \sum_{j=1}^{r_2}\widetilde{\M{G}}_j^*\otimes \widetilde{\M{G}}_j,
\end{equation}
$r$ is the rank of $\M{R}_{\rm opt}$, $\M{K}$ is an elementary matrix satisfying $\M{K}{\rm vec}(\M{A})={\rm vec}(\M{A}^{\rm T})$, $\M{U}_{\rm opt}$ is obtained by the eigendecomposition of $\M{R}_{\rm opt}$
\begin{equation}
\M{R}_{\rm opt} = \M{U}_{\rm opt}\M{\Lambda}_{\rm opt}\M{U}_{\rm opt}^{\rm H},
\end{equation}
and $\M{\Lambda}_{\rm opt}\in\mathbb{C}^{r\times r}$ is a diagonal matrix containing all non-zero eigenvalues of $\M{R}_{\rm opt}$.
\begin{IEEEproof}
(sketch) Observe that the objective function \eqref{obj_function_CRB} is convex with respective to $\widetilde{\M{R}}_{\RM{X}}$. Therefore, it suffices to show that it is strictly convex in a set constituted by $\widetilde{\M{R}}_{\RM{X}}$ satisfying the aforementioned conditions. This would in turn require $\M{\Phi}(\cdot)$ be an injective map. For a detailed proof, please refer to Appendix \ref{sec:proof_uniqueness}.
\end{IEEEproof}
\end{proposition}

Proposition \ref{prop:uniqueness} is both sufficient and necessary for the uniqueness of the optimal solution, but may not be convenient for verifying the solution-uniqueness of specific problems, since it requires the information about the maximum-rank optimal solutions. In light of this, next we provide some sufficient conditions.
\begin{proposition}[Sufficient Conditions for Uniqueness]\label{prop:sufficient}
The problem \eqref{opt_problem} has a unique optimal solution, if any of the following conditions holds:
\begin{enumerate}
  \item (Generic) $\M{\Xi}(\M{U}_{\rm R}^*\otimes \M{U}_{\rm R})$ has full column rank, where $\M{U}_{\rm R}$ contains the eigenvectors corresponding to the maximum eigenvalue of $\M{\Phi}_{P_{\rm T}M}^{\rm a}(\M{Z}_{\rm A})$, namely
\begin{equation}
\M{U}_{\rm R}^{\rm H}\M{\Phi}_{P_{\rm T}M}^{\rm a}(\M{Z}_{\rm A})\M{U}_{\rm R} = \M{I}\cdot \max_{\|\V{v}\|=1} \V{v}^{\rm H}\M{\Phi}_{P_{\rm T}M}^{\rm a}(\M{Z}_{\rm A})\V{v},
\end{equation}
with $\M{Z}_{\rm A}$ being the optimal solution of the following problem
        \begin{align}\label{dual_problem}
        \max_{\M{Z}_{\rm A}} &~~\Big(\tr{\M{Z}_{\rm A}^{\frac{1}{2}}}\Big)^2,\nonumber \\
        {\rm s.t.}&~~P_{\rm T}M\M{\Phi}_{P_{\rm T}M}^{\rm a}(\M{Z}_{\rm A})\!\preceq\! \M{I},~\M{Z}_{\rm A}\!=\!\M{Z}_{\rm A}^{\rm H},~\M{Z}_{\rm A}\!\succeq\! \M{0},
        \end{align}
        where $\M{\Phi}_{P_{\rm T}M}^{\rm a}(\cdot)$ is the adjoint operator of $\M{\Phi}_{P_{\rm T}M}(\cdot)$, characterized by
        \begin{equation}\label{adjoint_phi}
        \M{\Phi}_{P_{\rm T}M}^{\rm a}(\M{A}) = \sum_{i=1}^{r_3}  \bar{\M{G}}_i^{\rm H}\M{A}\bar{\M{F}}_i.
        \end{equation}
  \item ($K\geq M$) $\M{\Xi}$ has full column rank;
  \item ($K=1$) The eigenspace corresponding to the maximum eigenvalue of the following matrix
  \begin{equation}
  \M{B}_1:=\sum_{i=1}^{r_1} (\widetilde{\M{F}}_i^{\rm H} \widetilde{\M{F}}_i)^{\rm T} + \sum_{j=1}^{r_2} \widetilde{\M{G}}_j^{\rm H} \widetilde{\M{G}}_j
  \end{equation}
  has dimensionality $1$.
\end{enumerate}
\begin{IEEEproof}
(sketch) For the $K\geq M$ case, it is straightforward that when $\M{\Xi}$ has full column rank, the map $\M{\Phi}(\cdot)$ is injective. For the general case, the problem \eqref{dual_problem} is the dual of the original problem \eqref{opt_problem}. Since strong duality holds, these two problems have identical optimal solutions, but the uniqueness condition derived from the dual problem is more convenient as it does not involve the maximum-rank solution. The condition for the $K=1$ case follows from a simplification of the general condition. For a detailed proof, please refer to Appendix \ref{sec:proof_sufficient}.
\end{IEEEproof}
\end{proposition}

When the sensing-optimal covariance matrix is unique, we have the following result.
\begin{corollary}\label{coro:psc_rank}
If the sensing-optimal $\widetilde{\M{R}}_{\RM{X}}$ is unique, its rank is upper-bounded by
\begin{equation}
{\rm rank}(\widetilde{\M{R}}_{\RM{X}}) \leq \min\{K,M\}.
\end{equation}
\begin{IEEEproof}
Please refer to Appendix \ref{sec:proof_psc_rank}.
\end{IEEEproof}
\end{corollary}

\subsection{Point $P_{\rm SC}$--Achieving Strategy}\label{ssec:psc}
In general, the optimization problem \eqref{opt_problem} has to be solved on a case-by-case basis. Considering the limited scope of this paper, we defer the discussion of the detailed structure of the optimal solution to our future works, and simply denote the optimal sample covariance matrices as $\RM{R}_{\RM{X}}^{\rm SC}$. Note that
\begin{equation}
I(\RM{Y}_{\rm c};\RM{X}|\RM{H}_{\rm c}) = I(\RM{Y}_{\rm c};\RM{X},\RM{R}_{\RM{X}}^{\rm SC}|\RM{H}_{\rm c}),
\end{equation}
since $\RM{R}_{\RM{X}}^{\rm SC}$ is a deterministic function of $\RM{X}$, and thus using the chain rule of mutual information, we have
\begin{equation}
I(\RM{Y}_{\rm c};\RM{X}|\RM{H}_{\rm c}) = I(\RM{Y}_{\rm c};\RM{R}_{\RM{X}}^{\rm SC}|\RM{H}_{\rm c})+I(\RM{Y}_{\rm c};\RM{X}|\RM{H}_{\rm c},\RM{R}_{\RM{X}}^{\rm SC}).
\end{equation}
The rate $R_{\rm SC}$ can now be expressed as
\begin{subequations}
\begin{align}
R_{\rm SC}&\!=\!\!\max_{p_{\RM{X}}(\M{X})}\frac{1}{T}\Big(\!I(\RM{Y}_{\rm c};\!\RM{R}_{\RM{X}}^{\rm SC}|\RM{H}_{\rm c})\!+\!I(\RM{Y}_{\rm c};\!\RM{X}|\RM{H}_{\rm c},\!\RM{R}_{\RM{X}}^{\rm SC})\Big), \label{R1_decomposition}\\
&\hspace{6mm}{\rm s.t.}~\RM{R}_{\RM{X}}^{\rm SC} \in \Set{R}_{\rm SC},
\end{align}
\end{subequations}
where $\Set{R}_{\rm SC}$ is the set of optimal solutions of problem \eqref{opt_problem}, given by
$$
\Set{R}_{\rm SC}=\Big\{\RM{R}|\tr{(\M{\Phi}_{P_{\rm T}M}(\RM{R}))^{-1}}=\mathop{\rm min}_{\RM{R}\in\Set{D}}\tr{(\M{\Phi}_{P_{\rm T}M}(\RM{R}))^{-1}}\Big\}
$$
where $\Set{D}$ is constituted by all feasible $\RM{R}$'s satisfying \eqref{opt_constraints}. Especially, we observe that when the sensing-optimal sample covariance matrix is unique, i.e.
\begin{equation}\label{unique_rx}
\RM{R}_{\RM{X}}^{\rm SC}=\widetilde{\M{R}}_{\RM{X}}^{\rm SC}~\mathrm{with~probability~}1,
\end{equation}
the term $I(\RM{Y}_{\rm c}|\RM{H}_{\rm c};\RM{R}_{\RM{X}}^{\rm SC})$ in \eqref{R1_decomposition} is zero, and hence we have
\begin{subequations}
\begin{align}
R_{\rm SC}&=\max_{p_{\RM{X}}(\M{X})}~\frac{1}{T}I\big(\RM{Y}_{\rm c};\RM{X}\big|\RM{H}_{\rm c},\RM{R}_{\RM{X}}^{\rm SC}\!=\!\widetilde{\M{R}}_{\RM{X}}^{\rm SC}\big),\\
&\hspace{7mm}{\rm s.t.}~\widetilde{\M{R}}_{\RM{X}}^{\rm SC} = \mathop{\rm argmin}_{\M{R}\in\Set{D}}\tr{(\M{\Phi}_{P_{\rm T}M}(\M{R}))^{-1}}.
\end{align}
\end{subequations}

When \eqref{unique_rx} holds, we may provide a generic characterization of the point $P_{\rm SC}$ in the high-\ac{snr} regime, as detailed in the following proposition.
\begin{theorem}[Sensing-limited High-\ac{snr} Ergodic Capacity]\label{thm:sensing_limited}
When the sensing-optimal sample covariance matrix is unique, in the high-\ac{snr} regime, namely when $P_{\rm T}/\sigma_{\rm c}^2\rightarrow \infty$, the rate $R_{\rm SC}$ can be expressed as
\begin{equation}\label{r1_formula}
R_{\rm SC} \!=\! \mathbb{E}\Big\{\Big(1\!-\!\frac{\rv{M}_{\rm SC}}{2T}\Big)\log |\sigma_{\rm c}^{-2}\RM{H}_{\rm c}\widetilde{\M{R}}_{\RM{X}}^{\rm SC}\RM{H}_{\rm c}^{\rm H}| \!+\! \rv{c}_0\Big\} \!+\! O(\sigma_{\rm c}^2).
\end{equation}
where $\rv{M}_{\rm SC}$ denotes the rank of $\RM{H}_{\rm c}\widetilde{\M{R}}_{\RM{X}}^{\rm SC}\RM{H}_{\rm c}^{\rm H}$, and the term
\begin{equation}\label{c0_r1}
\rv{c}_0=\frac{\rv{M}_{\rm SC}}{T}\Big[\Big(T\!-\!\frac{\rv{M}_{\rm SC}}{2}\Big)\log\frac{T}{e}\!-\!\log \Gamma(T)\!+\!\log(2\sqrt{\pi})\Big]
\end{equation}
converges to zero as $T\rightarrow \infty$.
\begin{IEEEproof}
Please refer to Appendix \ref{sec:proof_sensing_limited}.
\end{IEEEproof}
\end{theorem}

Theorem \ref{thm:sensing_limited} also enables us to propose a scheme that asymptotically achieves the point $P_{\rm SC}$ in the high-\ac{snr} regime, as given in the following corollary.
\begin{corollary}[Point $P_{\rm SC}$--Achieving Strategy]\label{coro:shc}
In the high-\ac{snr} regime, when \eqref{unique_rx} holds, the rate $R_{\rm SC}$ may be asymptotically achieved by a waveform $\RM{X}_{\rm SC}$ generated as follows:
\begin{equation}\label{shc_waveform}
\RM{X}_{\rm SC}=\sqrt{T}\RM{H}_{\rm c}^\dagger\widetilde{\RM{U}}\widetilde{\RM{\Sigma}}\RM{Q}+\RM{X}_{\perp},
\end{equation}
where $\RM{H}_{\rm c}^\dagger$ denotes the Moore-Penrose pseudo-inverse of $\RM{H}_{\rm c}$, $\widetilde{\RM{U}}$ contains the first $\rv{M}_{\rm SC}$ columns of $\RM{U}$ given by the eigendecomposition of $\RM{H}_{\rm c}\widetilde{\M{R}}_{\RM{X}}^{\rm SC}\RM{H}_{\rm c}^{\rm H}$
\begin{equation}
\RM{H}_{\rm c}\widetilde{\M{R}}_{\RM{X}}^{\rm SC}\RM{H}_{\rm c}^{\rm H} = \RM{U}\RM{\Lambda}\RM{U}^{\rm H},
\end{equation}
$\widetilde{\RM{\Sigma}}=\diag{\rv{\sigma}_1,\dotsc,\rv{\sigma}_{\rv{M}_{\rm SC}}}$, $\rv{\sigma}_i=\sqrt{\rv{\lambda}}_i$, $\rv{\lambda}_i$ is the $i$-th largest eigenvalue of the matrix $\RM{H}_{\rm c}\widetilde{\M{R}}_{\RM{X}}^{\rm SC}\RM{H}_{\rm c}^{\rm H}$, and $\RM{X}_{\perp}\in\mathbb{C}^{M\times T}$ is any matrix satisfying
\begin{subequations}\label{x_perp_conditions}
\begin{align}
\RM{X}_{\perp}(\RM{H}_{\rm c}^\dagger\widetilde{\RM{U}}\widetilde{\RM{\Sigma}}\RM{Q})^{\rm H}&=\M{0},\\
T^{-1}\RM{X}_{\perp}\RM{X}_{\perp}^{\rm H}+\RM{H}_{\rm c}^{\dagger}\RM{H}_{\rm c}\widetilde{\M{R}}_{\RM{X}}^{\rm SC}\RM{H}_{\rm c}^{\dagger}\RM{H}_{\rm c}&=\widetilde{\M{R}}_{\RM{X}}^{\rm SC}.
\end{align}
\end{subequations}
The matrix $\RM{Q}\in\mathbb{C}^{\rv{M}_{\rm SC}\times T}$ contains the modulated data, uniformly sampled from the following rescaled Stiefel manifold
\begin{equation}\label{stiefel_complex_definition}
\Set{V} = \big\{\M{Q}\in\mathbb{C}^{\rv{M}_{\rm SC}\times T}|\M{Q}\M{Q}^{\rm H}=\M{I}_{\rv{M}_{\rm SC}}\big\}.
\end{equation}
Moreover, the uniform sampling over $\Set{V}$ can be implemented as follows:
\begin{enumerate}
  \item Construct $\RM{A}\in\mathbb{C}^{\rv{M}_{\rm SC}\times T}$, in which each column is independently drawn from the standard circular complex Gaussian distribution $\Set{C}\Set{N}(\V{0},\M{I}_{\rv{M}_{\rm SC}})$;
  \item Obtain $\RM{Q}$ by performing the LQ decomposition of $\RM{A}$, namely $\RM{A}=\RM{L}\RM{Q}$, where $\RM{L}$ is a lower-triangular matrix with real diagonal entries, and $\RM{Q}$ is a semi-unitary matrix satisfying $\RM{Q}\RM{Q}^{\rm H}=\M{I}_{\rv{M}_{\rm SC}}$.
\end{enumerate}
\begin{IEEEproof}
Please refer to Appendix \ref{sec:proof_shc}.
\end{IEEEproof}
\end{corollary}

A $P_{\rm SC}$--achieving strategy in the form of \eqref{shc_waveform} seems to rely on the knowledge of $\RM{H}_{\rm c}$. Next, we show that this knowledge is not necessary, by providing a specific form of $\RM{X}_{\rm SC}$ which does not incorporate $\RM{H}_{\rm c}$.
\begin{corollary}\label{coro:zero_knowledge}
One of the possible $\RM{X}_{\rm SC}$ satisfying \eqref{shc_waveform} is given by
\begin{equation}
    \RM{X}_{\rm SC,1} = \sqrt{T}\M{U}_{\rm s}\M{\Lambda}_{\rm s}^{\frac{1}{2}}\RM{Q},
\end{equation}
where $\M{U}_{\rm s}$ and $\M{\Lambda}_{\rm s}$ are obtained from the eigendecomposition of $\widetilde{\M{R}}_{\RM{X}}^{\rm SC}$ as
$$
\widetilde{\M{R}}_{\RM{X}}^{\rm SC}=\M{U}_{\rm s}\M{\Lambda}_{\rm s}\M{U}_{\rm s}^{\rm H},
$$
and $\RM{Q}$ is uniformly sampled from the set $\Set{V}$ given in \eqref{stiefel_complex_definition}.
\begin{IEEEproof}
Please refer to Appendix \ref{sec:proof_shc}.
\end{IEEEproof}
\end{corollary}

\subsection{Point $P_{\rm CS}$--Achieving Strategy}\label{ssec:pcs}
In contrast to that of achieving point $P_{\rm SC}$, the strategy of achieving point $P_{\rm CS}$ is well-known in the literature \cite{goldsmith}. Specifically, we may simply employ the capacity-achieving strategy in the communication-only scenario, which gives the maximum achievable rate
\begin{align}\label{rmax_pcs}
R_{\rm max} &= \max_{\widetilde{\M{R}}_{\RM{X}}} ~ \mathbb{E}\{\log \big|\M{I}+\sigma_{\rm c}^{-2}\RM{H}_{\rm c}\widetilde{\M{R}}_{\RM{X}}\RM{H}_{\rm c}^{\rm H}\big|\},~{\rm s.t.}~\eqref{power_constraint}, \nonumber \\
&= \mathbb{E}\{\log \big|\M{I}+\sigma_{\rm c}^{-2}\RM{H}_{\rm c}\widetilde{\M{R}}_{\RM{X}}^{\rm CS}\RM{H}_{\rm c}^{\rm H}\big|\},
\end{align}
where the maximization with respect to $\widetilde{\M{R}}_{\RM{X}}$ can be carried out using water filling for each realization of $\RM{H}_{\rm c}$ \cite[Sec.~9.4]{cover}, and each column $\RV{x}_i$ in the $P_{\rm CS}$--achieving $\RM{X}$ (denoted by $\RM{X}_{\rm CS}$ following the circularly symmetric complex Gaussian distribution $\mathcal{CN}(\V{0},\widetilde{\M{R}}_{\RM{X}}^{\rm CS})$. Then, the communication-limited minimum \ac{crb} is given by
\begin{equation}\label{j1_formula}
\epsilon_{\rm CS} = \frac{\sigma_{\rm s}^2}{T}\mathbb{E}\bigg\{{\rm tr}\bigg[\Big(\M{\Phi}(\RM{R}_{\RM{X}}^{\rm CS})\Big)^{-1}\bigg]\bigg\}.
\end{equation}
Next, we provide an upper bound for $\epsilon_{\rm CS}$ at the high-\ac{snr} limit.

\begin{theorem}[Sensing \ac{dof} loss]\label{thm:sensing_dofloss}
When the columns in $\RM{X}$ are independent of one another, and identically follow a circularly symmetric complex Gaussian distribution $\mathcal{CN}(\V{0},\widetilde{\M{R}}_{\RM{X}})$, if $[\M{\Phi}(\RM{R}_{\RM{X}})-\widetilde{\M{J}}_{\rm P}]^{-1}$ exists almost surely and $\widetilde{\M{J}}_{\rm P}$ is invertible, we may obtain the following bound:

\begin{equation}\label{sensing_dofloss}
\begin{aligned}
\tr{\M{\Phi}(\widetilde{\M{R}}_{\RM{X}})^{-1}}&\leq \mathbb{E}\left\{\tr{\M{\Phi}(\RM{R}_{\RM{X}})^{-1}}\right\}\\
&\leq \frac{T\cdot \tr{\M{\Phi}(\widetilde{\M{R}}_{\RM{X}})^{-1}}}{T-\min\{K,{\rm rank}(\widetilde{\M{R}}_{\RM{X}})\}}.
\end{aligned}
\end{equation}
\begin{IEEEproof}
Please refer to Appendix \ref{sec:proof_sensing_dofloss}.
\end{IEEEproof}
\end{theorem}

To fully unveil the implications of Theorem \ref{thm:sensing_dofloss}, let us formally define the concept of sensing \ac{dof}. Upon denoting the \ac{snr} as $\gamma$, the communication \ac{dof} is known to be defined as the asymptotic ratio \cite{jafar_dof}
\begin{equation}
\nu_{\rm c} = \lim_{\gamma \rightarrow \infty} \frac{R(\gamma)}{\log(1+\gamma)},
\end{equation}
where $R(\gamma)$ denotes the maximum achievable rate given the \ac{snr} $\gamma$. Similar to the communication \ac{dof}, we define the sensing \ac{dof} as the following high-\ac{snr} limit
\begin{equation}
\nu_{\rm s} = \lim_{\sigma_{\rm s}\rightarrow 0} \frac{\tr{[\M{\Phi}(\widetilde{\M{R}}_{\RM{X}})]^{-1}}}{T^{-1}\mathbb{E}\{\tr{[\M{\Phi}(\RM{R}_{\RM{X}})]^{-1}}\}}.
\end{equation}
Using this definition, we may now interpret the quantity $\min\{K,{\rm rank}(\widetilde{\M{R}}_{\RM{X}})\}$ in \eqref{sensing_dofloss} as an upper bound for the \emph{sensing \ac{dof} loss} induced by employing the communication-optimal transmission scheme. Intuitively, when $\RM{R}_{\RM{X}}=\widetilde{\M{R}}_{\RM{X}}$, the sensing \ac{dof} is $T$, which may be viewed as the \textit{number of independent observations}. In general, for a random $\RM{R}_{\RM{X}}$ corresponding to a sensing \ac{dof} of $\nu_{\rm s}$, the sensing \ac{crb} equals to that of the deterministic covariance matrix scenario (i.e. $\RM{R}_{\RM{X}}=\widetilde{\M{R}}_{\RM{X}}$) with $T=\nu_{\rm s}$. This suggests that $\nu_{\rm s}$ may be interpreted as the \textit{effective number of independent observations}. The \textit{a priori} knowledge about the sensing parameters clearly improves the sensing performance, but it does not contribute to the sensing \ac{dof} in general. This can be observed from \eqref{phi_def}, where the term $\widetilde{\M{J}}_{\rm P}$ converges to zeros as $\sigma_{\rm s}\rightarrow 0$.

The first equality (the best sensing \ac{dof}) in \eqref{sensing_dofloss} is achieved when $\RM{R}_{\RM{X}}$ is deterministic. For the second equality (the worst case) to be achieved, some sufficient conditions are summarized as follows.

\begin{corollary}[Sufficient conditions for maximum sensing \ac{dof} loss]\label{coro:equality_dof}
The second equality in \eqref{sensing_dofloss} is achieved if either of the following conditions holds:
\begin{enumerate}
\item $K\leq {\rm rank}(\widetilde{\M{R}}_{\RM{X}})$, $r_1+r_2=1$;
\item The covariance matrix $\widetilde{\M{R}}_{\RM{X}}$ is invertible, and the matrix
$$
\M{F}^{\rm st} = [\widetilde{\M{F}}_1,~\dotsc,~\widetilde{\M{F}}_{r_1},~\widetilde{\M{G}}_1,~\dotsc,~\widetilde{\M{G}}_{r_2}]
$$
is unitary.
\end{enumerate}
\begin{IEEEproof}
Please refer to Appendix \ref{sec:proof_equality_dof}.
\end{IEEEproof}
\end{corollary}

By presenting Corollary \ref{coro:equality_dof}, we hope that it may provide further intuitions regarding the sensing \ac{dof}, since in the worst case we have precise knowledge about the sensing \ac{dof} loss. In Section \ref{ssec:channel_estimation} we will see that, the task of target response matrix estimation is a typical example of scenarios where the condition 2) in Corollary \ref{coro:equality_dof} is applicable.

\section{Discussions}\label{sec:discussions}
In this section, we discuss the intuitions and implications of the results presented in Section \ref{sec:results}.

\subsection{\ac{snc} Tradeoff as a Two-fold Tradeoff: ST and DRT}\label{ssec:two_fold}
A central implication of the results in Section \ref{sec:results} is that the \ac{snc} tradeoff is two-fold, detailed as follows.

\subsubsection{\Acf{st}}\label{ssec:inf_flow}
According to the discussions in Section~\ref{ssec:psc} and Section~\ref{ssec:pcs}, at point $P_{\rm SC}$, any sensing-optimal $\widetilde{\M{R}}_{\RM{X}}$ has the following structure
\begin{equation}\label{sensing_optimal_rx}
\widetilde{\M{R}}_{\RM{X}} = \M{U}_{\rm s}\M{\Lambda}_{\rm s}\M{U}_{\rm s}^{\rm H},
\end{equation}
where $\M{U}_{\rm s}\in\mathbb{C}^{M\times r}$ contains the eigenvectors of the maximum-rank ($=r$) sensing-optimal $\M{R}_{\RM{X}}$ corresponding to the non-zero eigenvalues, while $\M{\Lambda}_{\rm s}$ is a positive semidefinite Hermitian matrix that ensures $\widetilde{\M{R}}_{\RM{X}}$ to be an optimal solution of the problem \eqref{opt_problem}. In particular, when the sensing optimal $\widetilde{\M{R}}_{\RM{X}}$ is unique, $\M{\Lambda}_{\rm s}$ is a diagonal matrix containing the non-zero eigenvalues of $\widetilde{\M{R}}_{\RM{X}}$. By contrast, at point $P_{\rm CS}$, the communication-optimal $\widetilde{\M{R}}_{\RM{X}}$ takes the following form
\begin{equation}
\widetilde{\M{R}}_{\RM{X}}= \M{U}_{\rm c}\M{\Lambda}_{\rm c}\M{U}_{\rm c}^{\rm H},
\end{equation}
where $\M{U}_{\rm c}\in\mathbb{C}^{M\times {\rm rank}(\RM{H}_{\rm c})}$ contains the left-singular vectors of the communication channel matrix $\RM{H}_{\rm c}$ corresponding to the non-zero singular values, while $\M{\Lambda}_{\rm c}$ is a diagonal matrix determined by the water-filling power allocation strategy.

From the discussion above, we may observe that the matrix $\M{U}_{\rm s}$ characterizes the \emph{sensing subspace}, while $\M{U}_{\rm c}$ characterizes the \emph{communication subspace}. Given a specific statistical covariance matrix $\widetilde{\M{R}}_{\RM{X}}$, when its column space is more closely aligned with the sensing subspace, we may achieve a more favorable sensing performance at the cost of degraded communication performance, and vice versa. This depicts a part of the entire picture of the \ac{snc} tradeoff, which will be referred to as the \acf{st}.

\subsubsection{\Acf{drt}}
Given a specific $\widetilde{\M{R}}_{\RM{X}}$, the \ac{snc} tradeoff can be further adjusted by controlling the ``degree of randomness'' of the signal $\RM{X}$. To obtain a higher communication rate, one should modulate as much information as possible onto the carrier signal, hence the transmitted waveform should be ``as random as possible''. However, sensing systems prefer deterministic signals for achieving a stable sensing performance. This has been an intuitive insight consistent with both the engineers' experience and \ac{snc} signal processing theory developed during the past few decades \cite{9839260}. Indeed, Proposition \ref{prop:max_icrb} tells us that to achieve the optimal sensing performance, the transmitted waveform has to be deterministic to a degree, in the sense that its sample covariance matrix $\RM{R}_{\RM{X}}$ should be deterministic. Consequently, following the notations in \eqref{sensing_optimal_rx}, any sensing-optimal signal should take the following form
\begin{equation}\label{sensing_optimal_x}
\RM{X} = \sqrt{T}\M{U}_{\rm s}\M{\Lambda}_{\rm s}^{\frac{1}{2}}\RM{Q},
\end{equation}
where $\RM{Q}\in\mathbb{C}^{r\times T}$ is a semi-unitary matrix satisfying $\RM{Q}\RM{Q}^{\rm H}=\M{I}$. Among these sensing-optimal signals, those achieving the highest communication rate are given by Corollary \ref{coro:shc}. By contrast, at point $P_{\rm CS}$, the communication-optimal signal is given by
\begin{equation}\label{comms_optimal_x}
\RM{X} = \M{U}_{\rm c}\M{\Lambda}_{\rm c}^{\frac{1}{2}} \RM{D},
\end{equation}
where $\RM{D}\in\mathbb{C}^{{\rm rank}(\RM{H}_{\rm c})\times T}$ contains \ac{iid} standard circularly symmetric complex Gaussian entries.

Comparing \eqref{sensing_optimal_x} with \eqref{comms_optimal_x}, it can be observed that the sensing-optimal signal $\RM{X}$ sacrifices some communication \acp{dof}, in the sense that the rows in $\RM{Q}$ are forced to have unit norms and to be orthogonal to one another. On the other hand, as indicated by Theorem \ref{thm:sensing_dofloss}, the communication-optimal $\RM{X}$ also sacrifices some sensing \acp{dof}. We refer to this part of the \ac{snc} tradeoff as the \acf{drt}.
A noteworthy issue is that, when the coherent sensing period $T$ is sufficiently long compared to $\rv{M}_{\rm SC}$, the \ac{drt} becomes less prominent, since even a communication-optimal Gaussian waveform $\RM{X}$ would have asymptotically orthogonal rows as $T/\rv{M}_{\rm SC}\rightarrow \infty$ \cite{asymptotic_orthogonal}. A related fact is that most existing contributions on the joint design of \ac{isac} systems treat the sample covariance matrix $\RM{R}_{\RM{X}}$ as a deterministic matrix\cite{crb_radcom,deterministic_rx_1,deterministic_rx_2,deterministic_rx_3}. Such designs may now be interpreted as neglecting the \ac{drt}, and hence can be viewed as \emph{infinite-$T$ approximations} to the original \ac{isac} design problem.

\subsubsection{\ac{st} and \ac{drt} in terms of \ac{dof}}
Next, we provide some quantified characterizations of the \ac{st} and \ac{drt}. Let us first take the point $P_{\rm SC}$ as an example. According to \eqref{r1_formula}, the sensing-limited rate $R_{\rm SC}$ in the high-\ac{snr} regime is given by
\begin{equation}
R_{\rm SC} = \mathbb{E}\Big\{\Big(1-\frac{\rv{M}_{\rm SC}}{2T}\Big)\log |\sigma_{\rm c}^{-2}\RM{H}_{\rm c}\widetilde{\M{R}}_{\RM{X}}^{\rm SC}\RM{H}_{\rm c}^{\rm H}|+\rv{c}_0\Big\} + O(\sigma_{\rm c}^2),
\end{equation}
which implies that the number of communication \acp{dof} at the point $P_{\rm SC}$ is $\rv{M}_{\rm SC}\Big(1-\frac{\rv{M}_{\rm SC}}{2T}\Big)$. Hence we may define the \emph{communication \ac{dof} efficiency} at point $P_{\rm SC}$ as
\begin{equation}\label{dof_efficiency}
\zeta_{\rm SC}:= \alpha_{\rm SC}\Big(1-\frac{\rv{M}_{\rm SC}}{2T}\Big),
\end{equation}
where $\alpha_{\rm SC}=\rv{\alpha}(\widetilde{\M{R}}_{\RM{X}}^{\rm SC})$ is the \emph{communication subspace overlap coefficient} at point $P_{\rm SC}$, defined as
\begin{equation}
\rv{\alpha}(\widetilde{\M{R}}_{\RM{X}})=\frac{{\rm rank}(\RM{H}_{\rm c}\widetilde{\M{R}}_{\RM{X}}\RM{H}_{\rm c}^{\rm H})}{{\rm rank}(\RM{H}_{\rm c}\RM{H}_{\rm c}^{\rm H})}.
\end{equation}
With the definition $\rv{M}_{\rm CS}={\rm rank}(\RM{H}_{\rm c}\widetilde{\M{R}}_{\RM{X}}^{\rm CS}\RM{H}_{\rm c}^{\rm H})$, we see that
\begin{equation}
\alpha_{\rm SC} = \rv{M}_{\rm SC}/\rv{M}_{\rm CS}.
\end{equation}
Observe that in general, the communication subspace overlap coefficient $\rv{\alpha}(\widetilde{\M{R}}_{\RM{X}})$ takes value in $[\alpha_{\rm SC},1]$, which may be viewed as an indicator of the \ac{st}. The \ac{drt} is quantified by the term $(1-\rv{M}_{\rm SC}/2T)$ in \eqref{dof_efficiency}, which is less than $1$ as long as $T$ is finite.

Regarding the point $P_{\rm CS}$, we see from \eqref{sensing_dofloss} that
\begin{equation}
\mathbb{E}\{\tr{[\M{\Phi}(\RM{R}_{\RM{X}})]^{-1}}\}\leq \frac{T\tr{[\M{\Phi}(\widetilde{\M{R}}_{\RM{X}})]^{-1}}}{T-\min\{K,{\rm rank}(\widetilde{\M{R}}_{\RM{X}})\}},
\end{equation}
which implies that the sensing \ac{dof} is lower bounded by $T-\min\{K,{\rm rank}(\widetilde{\M{R}}_{\RM{X}})\}$. Note that at $P_{\rm CS}$, we have  ${\rm rank}(\widetilde{\M{R}}_{\RM{X}})={\rm rank}(\RM{H}_{\rm c}\widetilde{\M{R}}_{\RM{X}}\RM{H}_{\rm c}^{\rm H})$. Thus when $K\leq {\rm rank}(\widetilde{\M{R}}_{\RM{X}})$, we may observe an interesting symmetry between $P_{\rm SC}$ and $P_{\rm CS}$: the maximum sensing \ac{dof} loss at $P_{\rm CS}$ is $\rv{M}_{\rm CS}$, while the communication \ac{dof} loss at $P_{\rm SC}$ is determined by $\rv{M}_{\rm SC}$. When $K>{\rm rank}(\widetilde{\M{R}}_{\RM{X}})$, the maximum sensing \ac{dof} loss becomes $K$. Intuitively, we may interpret the ratio $\min\{K,{\rm rank}(\widetilde{\M{R}}_{\RM{X}})\}/T$ as the \emph{sensing load}, which depicts the number of effective sensing parameters per unit time. As the sensing load tends to zero (e.g. $T\rightarrow \infty$), we see that the maximum sensing \ac{dof} loss also converges to zero. This corroborates the intuition obtained from the discussion of $P_{\rm SC}$, namely, the \ac{drt} becomes negligible in the large $T$ regime, where pure Gaussian signals are sufficient for achieving the Pareto-optimal \ac{snc} performance (i.e. the boundary of the \ac{crb}-rate region).

\subsection{Achievable Inner Bounds Connecting $P_{\rm SC}$ and $P_{\rm CS}$}\label{ssec:inner_bounds}
Besides $P_{\rm SC}$ and $P_{\rm CS}$, other points on the boundary of the \ac{crb}-rate region are also of significant practical interest. However, obtaining the genuine boundary would in general require solving the following complicated functional optimization problem:
$$
\begin{aligned}
\min_{p(\RM{X})} &~~\mathbb{E}\left\{\tr{[\M{\Phi}(\RM{R}_{\RM{X}})]^{-1}}\right\}-\lambda \cdot I(\RM{Y}_{\rm c};\RM{X}|\RM{H}_{\rm c})\\
{\rm s.t.}&~~\tr{\RM{R}_{\RM{X}}}=P_{\rm T}M,~\RM{R}_{\RM{X}}\succeq \M{0},~\RM{R}_{\RM{X}}=\RM{R}_{\RM{X}}^{\rm H},
\end{aligned}
$$
which is typically intractable in the vector Gaussian channel scenario considered in this paper, due to the lack of explicit form of the mutual information $I(\M{Y}_{\rm c};\RM{X}|\RM{H}_{\rm c})$ and the complexity of the \ac{crb} term. In light of this, in this subsection, we discuss two strategies that may be applied either separately or jointly, forming inner bounds of the actual boundary.

\subsubsection{Time sharing}
In Section~\ref{sec:results}, the time-sharing strategy has been applied to conceive the pentagon inner bound. It can be generalized to any pair of given signalling schemes corresponding to a pair of points in the \ac{crb}-rate region, by employing one of the schemes with probability $p$ and the other with probability $1-p$. In general, once a specific achievable inner bound is given, its convex envelope is also achievable with the aid of time sharing.

\subsubsection{Statistical covariance shaping}
It is also possible to adjust the \ac{snc} tradeoff by altering the statistical covariance matrix $\widetilde{\M{R}}_{\RM{X}}$. Specifically, given a realization of $\RM{H}_{\rm c}$, a flexible tradeoff may be struck by solving the following optimization problem
\begin{subequations}\label{opt_problem_pareto}
\begin{align}
\min_{\widetilde{\M{R}}_{\RM{X}}} &~~(1-\alpha){\rm tr}\Big\{\Big[\M{\Phi}_{P_{\rm T}M}(\widetilde{\M{R}}_{\RM{X}})\Big]^{-1}\Big\} \nonumber \\
&\hspace{7mm}-\alpha \log \big|\M{I}+\sigma_{\rm c}^{-2}\RM{H}_{\rm c}\widetilde{\M{R}}_{\RM{X}}\RM{H}_{\rm c}^{\rm H}\big|, \\
{\rm s.t.}&~~\tr{\widetilde{\M{R}}_{\RM{X}}}=P_{\rm T}M,~\widetilde{\M{R}}_{\RM{X}}\succeq \V{0},~\widetilde{\M{R}}_{\RM{X}}=\widetilde{\M{R}}_{\RM{X}}^{\rm H},
\end{align}
\end{subequations}
where $\alpha\in[0,1]$ controls the preference between the sensing and the communication performance. This is a convex optimization problem, which may be efficiently solved by off-the-shelf numerical solvers \cite{cvx,gb08}. Note that when $\alpha=0$, the problem \eqref{opt_problem_pareto} degenerates to the sensing-only problem \eqref{opt_problem}. An alternative formulation is given by
\begin{subequations}\label{opt_problem_pareto_alter}
\begin{align}
\max_{\widetilde{\M{R}}_{\RM{X}}} &~~\log \big|\M{I}+\sigma_{\rm c}^{-2}\RM{H}_{\rm c}\widetilde{\M{R}}_{\RM{X}}\RM{H}_{\rm c}^{\rm H}\big|,\\
{\rm s.t.}&~~\tr{\widetilde{\M{R}}_{\RM{X}}}=P_{\rm T}M,~\widetilde{\M{R}}_{\RM{X}}\succeq \V{0},~\widetilde{\M{R}}_{\RM{X}}=\widetilde{\M{R}}_{\RM{X}}^{\rm H},\\
&~~{\rm tr}\Big\{\Big[\M{\Phi}_{P_{\rm T}M}(\widetilde{\M{R}}_{\RM{X}})\Big]^{-1}\Big\}\leq \epsilon_{\alpha},\label{constraint_crb}
\end{align}
\end{subequations}
where $\epsilon_{\alpha}$ is the value of ${\rm tr}\Big\{\Big[\M{\Phi}_{P_{\rm T}M}(\widetilde{\M{R}}_{\RM{X}}(\alpha))\Big]^{-1}\Big\}$, with $\widetilde{\M{R}}_{\RM{X}}(\alpha)$ being the optimal $\widetilde{\M{R}}_{\RM{X}}$ of the problem \eqref{opt_problem_pareto} given a specific value of $\alpha$.

Once a specific statistical covariance matrix $\widetilde{\M{R}}_{\RM{X}}(\alpha)$ is obtained from \eqref{opt_problem_pareto} or \eqref{opt_problem_pareto_alter} for a given $\alpha$, we can first obtain a natural outer bound of the \ac{crb}-rate region as follows:
\begin{subequations}\label{the_outer_bound}
\begin{align}
R_{\rm out}(\alpha)&= \mathbb{E}\big\{\log \big|\M{I}+\sigma_{\rm c}^{-2}\RM{H}_{\rm c}\widetilde{\M{R}}_{\RM{X}}(\alpha)\RM{H}_{\rm c}^{\rm H}\big|\big\}, \\
\epsilon_{\rm out}(\alpha)&= \frac{\sigma_{\rm s}^2}{T}{\rm tr}\Big\{\Big[\M{\Phi}_{P_{\rm T}M}(\widetilde{\M{R}}_{\RM{X}}(\alpha))\Big]^{-1}\Big\}.
\end{align}
\end{subequations}
Next, we may decide the specific signalling strategy. One of the possible strategies is to transmit the following signal
\begin{equation}\label{gaussian_signalling}
\RM{X} = \RM{X}_{\rm G}+\RM{X}_{\rm G,\perp},
\end{equation}
where $\RM{X}_{\rm G}$ has \ac{iid} columns following the circularly symmetric complex Gaussian distribution $\mathcal{CN}(\V{0},\RM{H}_{\rm c}^\dagger\RM{H}_{\rm c}\widetilde{\M{R}}_{\RM{X}}\RM{H}_{\rm c}^\dagger\RM{H}_{\rm c})$, while $\RM{X}_{\rm G,\perp}\in\mathbb{C}^{M\times T}$ is any matrix satisfying
\begin{equation}
T^{-1}\RM{X}_{\rm G,\perp}\RM{X}_{\rm G,\perp}^{\rm H}+\RM{H}_{\rm c}^{\dagger}\RM{H}_{\rm c}\widetilde{\M{R}}_{\RM{X}}\RM{H}_{\rm c}^{\dagger}\RM{H}_{\rm c}=\widetilde{\M{R}}_{\RM{X}}.
\end{equation}
This leads to an achievable inner bound referred to as the ``Gaussian inner bound'', as follows:
\begin{subequations}
\begin{align}
R_{\rm in,G}(\alpha)&=R_{\rm out}(\alpha),\\
\epsilon_{\rm in,G}(\alpha)&=\frac{\sigma_{\rm s}^2}{T}\mathbb{E}\Big\{{\rm tr}\Big[\Big(\M{\Phi}(\RM{R}_{\RM{X}}(\alpha))\Big)^{-1}\Big]\Big\}.
\end{align}
\end{subequations}
Another possible strategy is given by
\begin{equation}
\RM{X} = \sqrt{T} \RM{H}_{\rm c}^{\dagger}\widetilde{\RM{U}}\widetilde{\RM{\Lambda}}^{\frac{1}{2}}\RM{Q} + \RM{X}_{\rm U,\perp},
\end{equation}
where $\RM{Q}\in\mathbb{C}^{\rv{M}_{\rm U}\times T}$ contains the modulated data, which is uniformly sampled from
$$
\Set{V}_{\widetilde{\M{R}}_{\RM{X}}} = \{\M{Q}\in\mathbb{C}^{\rv{M}_{\rm U}\times T}|\M{Q}\M{Q}^{\rm H}=\M{I}_{\rv{M}_{\rm U}}\},
$$
$\rv{M}_{\rm U}={\rm rank}(\RM{H}_{\rm c}\widetilde{\M{R}}_{\RM{X}}\RM{H}_{\rm c}^{\rm H})$, $\widetilde{\RM{\Lambda}}\in\mathbb{C}^{\rv{M}_{\rm U}\times \rv{M}_{\rm U}}$ is a diagonal matrix consisting of all non-zero eigenvalues of $\RM{H}_{\rm c}\widetilde{\M{R}}_{\RM{X}}\RM{H}_{\rm c}^{\rm H}$, and
$$
\RM{H}_{\rm c}\widetilde{\M{R}}_{\RM{X}}\RM{H}_{\rm c}^{\rm H} = \widetilde{\RM{U}}\widetilde{\RM{\Lambda}}\widetilde{\RM{U}}^{\rm H}.
$$
The matrix $\RM{X}_{\rm U,\perp}\in\mathbb{C}^{M\times T}$ is any matrix satisfying
\begin{subequations}
\begin{align}
\RM{X}_{\rm U,\perp}(\RM{H}_{\rm c}^\dagger\widetilde{\RM{U}}\widetilde{\RM{\Lambda}}^{\frac{1}{2}}\RM{Q})^{\rm H}&=\M{0},\\
T^{-1}\RM{X}_{\rm U,\perp}\RM{X}_{\rm U,\perp}^{\rm H}+\RM{H}_{\rm c}^{\dagger}\RM{H}_{\rm c}\widetilde{\M{R}}_{\RM{X}}\RM{H}_{\rm c}^{\dagger}\RM{H}_{\rm c}&=\widetilde{\M{R}}_{\RM{X}}.
\end{align}
\end{subequations}
This corresponds to another inner bound referred to as the ``semi-unitary inner bound'', characterized by
\begin{subequations}
\begin{align}
R_{\rm in,U}(\alpha)&=\mathbb{E}\Big\{\Big(1-\frac{\rv{M}_{\rm U}}{2T}\Big)\log |\sigma_{\rm c}^{-2}\RM{H}_{\rm c}\widetilde{\M{R}}_{\RM{X}}(\alpha)\RM{H}_{\rm c}^{\rm H}| \nonumber \\
&\hspace{5mm}+ \rv{c}_{0,\rm U}\Big\} + O(\sigma_{\rm c}^2),\\
\epsilon_{\rm in,U}(\alpha)&=\epsilon_{\rm out}(\alpha),
\end{align}
\end{subequations}
where
\begin{equation}
\rv{c}_{0,\rm U}=\frac{\rv{M}_{\rm U}}{T}\Big[\Big(T\!-\!\frac{\rv{M}_{\rm U}}{2}\Big)\log\frac{T}{e}\!-\!\log \Gamma(T)\!+\!\log(2\sqrt{\pi})\Big].
\end{equation}

Furthermore, we may obtain a refined inner bound on the basis of the Gaussian inner bound and the semi-unitary inner bound, by applying the time-sharing strategy. Specifically, this inner bound, referred to as the ``semi-unitary--Gaussian inner bound'', is constructed by computing the convex envelope of the union of the Gaussian inner bound and the semi-unitary inner bound. Each point on this refined inner bound is achievable by applying a suitable time-sharing strategy, which interpolates between a signalling scheme adapted from the Gaussian inner bound and another signalling scheme adapted from the semi-unitary inner bound.

Regarding the tightness of the semi-unitary--Gaussian inner bound, we have the following result.
\begin{proposition}\label{prop:ts_sug}
When the optimal objective function value of \eqref{opt_problem_pareto_alter} is not identical for all $\alpha\in[0,1]$, the semi-unitary--Gaussian inner bound is a tighter bound compared to the pentagon inner bound obtained by connecting $P_{\rm SC}$ and $P_{\rm CS}$.
\begin{IEEEproof}
Please refer to Appendix \ref{sec:proof_ts_sug}.
\end{IEEEproof}
\end{proposition}

Proposition \ref{prop:ts_sug} suggests that, under mild technical assumptions, it would be beneficial to combine the time-sharing strategy with statistical covariance shaping, rather than applying them separately.

\subsection{The Connection Between Existing \ac{isac} Schemes and $P_{\rm SC}$-- \& $P_{\rm CS}$--Achieving Strategies}\label{ssec:connections}
In this subsection, we will discuss how the design philosophy of current \ac{isac} systems is related to different \ac{crb}-rate boundary-approaching strategies.

\subsubsection{Sensing-Centric Designs}
Sensing-centric schemes are typically implemented relying on existing sensing infrastructures, such as radars \cite{sensing_centric1,sensing_centric2,majorcom}. They are designed aiming for incorporating communication functionalities into the system, without compromising the sensing performance.

A representative sensing-centric design is based on \ac{im} \cite{sensing_centric2,majorcom}, in which the communication information is encoded into the index of the waveform chosen from a preset codebook. More precisely, the transmitted signal in the \ac{im}-based scheme designed for colocated \ac{mimo} radars takes the following form
\begin{equation}
\RM{X} = \sqrt{P_{\rm T}}\RM{P}\M{U},
\end{equation}
where $\M{U}\in\mathbb{C}^{M\times T}$ represents the radar waveform codebook, which is a deterministic semi-unitary matrix satisfying $\M{U}\M{U}^{\rm H}=\M{I}_M$, while $\RM{P}$ is an $M\times M$ permutation matrix that conveys the communication information. Observe that the sample covariance matrix $\RM{R}_{\RM{X}}$ is given by
\begin{equation}
\RM{R}_{\RM{X}}=\frac{1}{T}\RM{X}\RM{X}^{\rm H} =P_{\rm T}\M{I}_M,
\end{equation}
which is deterministic. As will be discussed in Section \ref{ssec:channel_estimation}, this choice yields the optimal sensing performance when the sensing objective is the entire channel $\RM{H}_{\rm s}$. By contrast, the maximum communication throughput of this scheme is $\log_2 M!$ bits per transmission (i.e. each length-$T$ block), which is typically lower than the sensing-optimal capacity that is asymptotically achieved by the uniform distribution over the entire ensemble of $M\times T$ semi-unitary matrices, as detailed in Corollary \ref{coro:shc}. In light of this, we may conclude that the \ac{im}-based scheme follows a design philosophy that aims for achieving the $P_{\rm SC}$ point. However, due to its suboptimal communication throughput, the scheme would actually achieve a point right below $P_{\rm SC}$ in the \ac{crb}-rate region.

\subsubsection{Communication-Centric Designs}
When an \ac{isac} system is built upon commercialized communication networks, communication-centric designs are practically more attractive, whose primary objective is to guarantee the communication performance. Such designs typically rely on existing communication-oriented protocol stacks, for example, Wi-Fi and 5G new radio (NR).

A representative approach to communication-centric designs is to exploit the widely-used \ac{ofdm}-based waveform \cite{sturm2011waveform}. In the absence of the \ac{cp}, both the communication channel and the target response matrix can be represented by Toeplitz matrices in the following form:
\begin{equation}
\RM{H} = \left[
           \begin{array}{cccc}
             h_1 & 0 & \dotsc & 0 \\
             h_2 & h_1 & \ddots & \vdots \\
             \vdots & h_2 & \ddots & 0 \\
             h_{T_{\rm CIR}} & \vdots & \ddots & h_1 \\
             0 &  h_{T_{\rm CIR}} & \ddots & h_2 \\
             \vdots & \ddots & \ddots & \vdots \\
             0 & \dotsc & 0 & h_{T_{\rm CIR}} \\
           \end{array}
         \right],
\end{equation}
where $T_{\rm CIR}$ is the duration of the \ac{cir}. When the \ac{cp} is attached to the data symbols, and it is longer than the duration of the \ac{cir}, the channels can also be expressed in the form of circulant matrices (with $M=N_{\rm c}=N_{\rm s}$), and hence could be diagonalized by the \ac{dft} matrices as follows
\begin{subequations}
\begin{align}
\RM{H}_{\rm c} = \M{U}_{{\rm F},M}^{\rm H}\RM{D}_{\rm c}\M{U}_{{\rm F},M}, \\
\RM{H}_{\rm s} = \M{U}_{{\rm F},M}^{\rm H}\RM{D}_{\rm s}\M{U}_{{\rm F},M},
\end{align}
\end{subequations}
where $\RM{D}_{\rm c}$ and $\RM{D}_{\rm s}$ are diagonal matrices representing the frequency-domain communication channel and the target response, respectively, while $\M{U}_{{\rm F},M}$ denotes the $M$-point \ac{dft} matrix. Before the transmission, the raw data symbols are pre-processed by the inverse \ac{dft}, and thus can be expressed as
\begin{equation}
\RM{X} = \M{U}_{{\rm F},M}^{\rm H}\RM{X}_{\rm F},
\end{equation}
where each row of $\RM{X}_{\rm F}$ contains the data symbols modulated on a single subcarrier. Each data symbol is typically chosen from a fixed constellation, for example, \ac{psk} and \ac{qam}, according to the requirement of communication throughput.

Let us illustrate the \ac{drt} of the \ac{ofdm}-based scheme by considering a specific sensing task, namely that of estimating the \ac{cir}, which is a widely-used pre-processing procedure for delay estimation \cite{pieee_ofdm_radar}. In this example, the sensing parameter is $\RV{\eta}=\M{U}_{\rm F,M}^{\rm H}{\rm mdiag}(\RM{D}_{\rm s})$. Note that
\begin{equation}
\RV{\eta} = \M{U}_{\rm F,M}^{\rm H}\M{P}\M{K}_{\rm U}{\rm vec}(\RM{H}_{\rm s}^{\rm T}),
\end{equation}
where $\M{K}_{\rm U}=\M{U}_{{\rm F},M}\otimes \M{U}_{{\rm F},M}^\ast$, and $\M{P}\in\mathbb{R}^{M^2\times M^2}$ is a matrix satisfying $\M{P}{\rm vec}(\M{A}) = {\rm mdiag}(\M{A})$. Thus the linear map $\M{\Phi}(\RM{R}_{\RM{X}})$ is given by

\begin{align}
\M{\Phi}(\RM{R}_{\RM{X}})&\!=\!\M{U}_{\rm F,M}\M{P}\M{K}_{\rm U}^{\rm H}(\M{I}\otimes \RM{R}_{\RM{X}}^{\rm T})\M{K}_{\rm U}\M{P}^{\rm H}\M{U}_{\rm F,M}^{\rm H} \nonumber \\
&\!=\!\M{U}_{\rm F,M}\M{P}\left(\M{I}\otimes \M{U}_{{\rm F},M}^{\rm T}\RM{R}_{\RM{X}}^{\rm T}\M{U}_{{\rm F},M}^{\ast}\right)\M{P}^{\rm H}\M{U}_{\rm F,M}^{\rm H} \nonumber \\
&\!=\!\M{U}_{\rm F,M}{\rm mat}\left({\rm mdiag}\left(\M{U}_{{\rm F},M}^{\rm T}\RM{R}_{\RM{X}}^{\rm T}\M{U}_{{\rm F},M}^{\ast}\right)\right)\M{U}_{\rm F,M}^{\rm H},
\end{align}

hence the sensing-optimal sample covariance matrix should satisfy
\begin{equation}
{\rm mat}\left({\rm mdiag}\left(\M{U}_{{\rm F},M}^{\rm T}\RM{R}_{\RM{X}}^{\rm T}\M{U}_{{\rm F},M}^{\ast}\right)\right) = P_{\rm T}\M{I}_M,
\end{equation}
which implies that
\begin{equation}
\|[\RM{X}_{\rm F}]_{i,:}\|^2=TP_{\rm T},~\forall i=1,\dotsc, M.
\end{equation}
In other words, the transmit power should be equally allocated to each subcarrier for achieving the optimal \ac{cir} estimation performance. In a typical communication-oriented \ac{ofdm} system, the statistical covariance matrix would in general be a diagonal matrix corresponding to the specific power allocation strategy. When the power is equally allocated to each subcarrier, we have $\widetilde{\M{R}}_{\RM{X}}=P_{\rm T}\M{I}_M$, since the data symbols are independent of one another. However, the sample covariance matrix is not necessarily the same as $\widetilde{\M{R}}_{\RM{X}}$ due to the randomness of the data. The issue could be resolved by applying \ac{psk} constellations, at the cost of a lower communication throughput, since the amplitude of the symbols does not carry information in \ac{psk}. By varying the power allocation strategy as well as the constellation (possibly accompanied by error-correcting codes for rate adaptation), we may strike a flexible \ac{snc} tradeoff from point $P_{\rm SC}$ to point $P_{\rm CS}$.

\subsection{Semi-unitary Signalling: The $P_{\rm SC}$--Achieving Strategy vs. Non-coherent Communication}
The $P_{\rm SC}$--achieving strategy discussed in Section \ref{ssec:psc} bears some resemblance to the \ac{ustm} \cite{ustm_marzetta} designed for communicating over non-coherent channels. To elaborate, let us consider the scenario where the communication channel $\RM{H}_{\rm c}$ follows an entrywise \ac{iid} zero-mean, unit-variance complex Gaussian distribution, namely ${\rm vec}(\M{H}_{\rm c})\sim\mathcal{CN}(\V{0},\M{I}_{MN_{\rm c}})$, and is not known to the either the transmitter or the receiver. In this context, it has been shown that the capacity is achieved by signals satisfying\footnote{For simplicity of discussion, in the rest of this subsection we assume $T\geq 2M$. When this is not the case, the analysis becomes tedious and less relevant to this paper. Interested readers are referred to \cite{capacity_marzetta,zheng_tse}.}
\begin{equation}\label{ustm}
\RM{X} = \RM{A}\RM{Q},
\end{equation}
where $\RM{Q}\in\mathbb{C}^{M\times T}$ is a semi-unitary matrix and $\RM{A}\in\mathbb{C}^{M\times M}$ is a diagonal matrix being invariant under permutation of its diagonal entries, which is asymptotically proportional to an identity matrix in the high-\ac{snr} regime \cite{capacity_marzetta}. The specific designs of the semi-unitary matrix have then been referred to as \ac{ustm} schemes, which are related to coding on the Grassmann manifold $\Set{G}_{T,M}$ defined as the quotient space between two Stiefel manifolds $\Set{S}_{T,M}$ and $\Set{S}_{M,M}$ under the equivalence relationship
$$
\M{P}=\M{U}\M{Q}\Longleftrightarrow \M{P}~{\rm is~equivalent~to}~\M{Q},~\M{P},~\M{Q}\in\Set{S}_{T,M},
$$
where $\M{U}\in\Set{S}_{M,M}$ is a unitary matrix, and the Stiefel manifold $\Set{S}_{T,M}$ is defined as
$$
\Set{S}_{T,M} = \left\{\M{Q}\in\mathbb{C}^{M\times T}|\M{Q}\M{Q}^{\rm H}=\M{I}\right\}.
$$

Comparing \eqref{ustm} with \eqref{sensing_optimal_x}, we see that the \ac{ustm} and the $P_{\rm SC}$--achieving signals are similar in form. Nevertheless, it is noteworthy that they are substantially different in the following aspects:
\begin{itemize}
\item \textit{Precoding}: In the non-coherent communication scenario, the communication channel $\RM{H}_{\rm c}$ is isotropic. Consequently, the precoding matrix $\RM{A}$ in \eqref{ustm} is also isotropic in terms of its diagonal entries. By contrast, in the $P_{\rm SC}$--achieving strategy, the precoding matrix $ \sqrt{T}\M{U}_{\rm s}\M{\Lambda}_{\rm s}^{\frac{1}{2}}$ is determined by the sensing-optimality constraint, which aligns the signal with the sensing subspace;
\item \textit{Reason of optimality}: An intuitive interpretation for the optimality of \ac{ustm} is provided in \cite{zheng_tse}, which states that in the high-\ac{snr} regime, $\RM{H}_{\rm c}\RM{X}$ should be entrywise i.i.d. Gaussian distributed for achieving the non-coherent channel capacity, which implies that $\RM{X}$ should take the form of \eqref{ustm}. By contrast, in this paper, the reason of using semi-unitary signals follows from the sensing-optimality constraint, while the communication-optimality is achieved by applying uniform sampling over Stiefel manifolds instead of coding over Grassmann manifolds.
\item \textit{Communication \ac{dof}}: In the non-coherent communication scenario, it has been shown \cite{zheng_tse} that an alternative optimal strategy is to estimate the channel using pilot symbols before transmitting data symbols. This would impose a pilot cost of $MN_{\rm s}$ symbols since $\RM{H}_{\rm c}$ has $MN_{\rm s}$ unknown entries, and hence results in a communication \ac{dof} loss of $M^2/T$. This may also be inferred from the fact that the dimensionality of the Grassmann manifold is $M(T-M)$. By contrast, in this paper, the communication \ac{dof} is determined by the dimensionality of the Stiefel manifold, which is $\rv{M}_{\rm SC}(T-\frac{1}{2}\rv{M}_{\rm SC})$. Consequently, the communication \ac{dof} loss of $\frac{\rv{M}_{\rm SC}^2}{2T}$, which is half of the communication \ac{dof} loss in the non-coherent communication scenario when $\rv{M}_{\rm SC}=M$.
\end{itemize}

\section{Case Study}\label{sec:case_study}

\subsection{Target Angle Estimation}\label{ssec:angle}
Let us first consider the scenario, where the sensing task is to estimate the angles of targets using a \ac{mimo} radar equipped with co-located antennas. In this scenario, the target response matrix $\RM{H}_{\rm s}$ admits the following parametrization \cite{jian_li}:
\begin{equation}
\RM{H}_{\rm s}=\sum_{n=1}^{N_{\rm T}} \rv{\alpha}_n \RV{a}(\rv{\theta}_n)\RV{v}^{\rm T}(\rv{\theta}_n),
\end{equation}
where $N_{\rm T}$ represents the number of targets, $\rv{\theta}_n$ denotes the angle of the $n$-th target relative to the radar, $\rv{\alpha}_n$ denotes the complex amplitude of the echo received from the $n$-th target, while $\RV{v}(\cdot)$ and $\RV{a}(\cdot)$ denote the mappings from the angle to the steering vectors of the transmitting and receiving antennas, respectively. We assume the transmitting and receiving antenna arrays are conjugate symmetric. For the simplicity of discussion, we consider the case of $N_{\rm T}=1$, which corresponds to the parameter vector
$$
\RV{\eta}=[\rv{\theta},~{\rm Re}\{\rv{\alpha}\},~{\rm Im}\{\rv{\alpha}\}]^{\rm T},
$$
where $\rv{\theta}=\rv{\theta}_1$, $\rv{\alpha}=\rv{\alpha}_1$. According to \cite{jian_li}, we have
\begin{equation}
\RM{J}_{\RV{\eta}|\RM{X}}=2\left[
                             \begin{array}{ccc}
                               {\rm Re}\{\rv{f}_{11}\} & {\rm Re}\{\rv{f}_{12}\} & -{\rm Im}\{\rv{f}_{12}\}\\
                               {\rm Re}\{\rv{f}_{12}\} & {\rm Re}\{\rv{f}_{22}\} & -{\rm Im}\{\rv{f}_{22}\}\\
                               -{\rm Im}\{\rv{f}_{12}\} & -{\rm Im}\{\rv{f}_{22}\} & {\rm Re}\{\rv{f}_{22}\} \\
                             \end{array}
                           \right]+\M{J}_{\rm P},
\end{equation}
where
\begin{align}
\rv{f}_{11}&=\frac{T|\rv{\alpha}|^2}{\sigma_{\rm s}^2}\left(\|\dot{\RV{a}}(\rv{\theta})\|^2\tr{\RV{v}(\rv{\theta})\RV{v}^{\rm H}(\rv{\theta})\RM{R}_{\RM{X}}^{\rm T}}\right. \nonumber \\
&\hspace{5mm}\left.+2{\rm Re}\left\{\dot{\RV{a}}^{\rm H}(\rv{\theta})\RV{a}(\rv{\theta})\tr{\dot{\RV{v}}(\rv{\theta})\RV{v}^{\rm H}(\rv{\theta})\RM{R}_{\RM{X}}^{\rm T}}\right\}\right. \nonumber \\
&\hspace{5mm}\left.+\|\RV{a}(\rv{\theta})\|^2\tr{\dot{\RV{v}}(\rv{\theta})\dot{\RV{v}}^{\rm H}(\rv{\theta})\RM{R}_{\RM{X}}^{\rm T}}\right),\\
\rv{f}_{12} &= \frac{T\rv{\alpha}^*}{\sigma_{\rm s}^2}\left(\dot{\RV{a}}^{\rm H}(\rv{\theta})\RV{a}(\rv{\theta})\tr{\RV{v}(\rv{\theta})\RV{v}^{\rm H}(\rv{\theta})\RM{R}_{\RM{X}}^{\rm T}}\right.\nonumber \\
&\hspace{5mm}\left.+ \|\RV{a}(\rv{\theta})\|^2\tr{\RV{v}(\rv{\theta})\dot{\RV{v}}^{\rm H}(\rv{\theta})\RM{R}_{\RM{X}}^{\rm T}}\right),\\
\rv{f}_{22}&=\frac{T}{\sigma_{\rm s}^2}\|\RV{a}(\rv{\theta})\|^2\tr{\RV{v}(\rv{\theta})\RV{v}^{\rm H}(\rv{\theta})\RM{R}_{\RM{X}}^{\rm T}},
\end{align}
where $\dot{\RV{a}}(\rv{\theta})=\partial \RV{a}(\rv{\theta})/\partial \rv{\theta}$, and $\dot{\RV{v}}(\rv{\theta})=\partial \RV{v}(\rv{\theta})/\partial \rv{\theta}$. Let us assume that the prior distributions of $\rv{\theta}$, ${\rm Re}\{\rv{\alpha}\}$ and ${\rm Im}\{\rv{\alpha}\}$ are independent of one another, and hence
\begin{equation}
\M{J}_{\rm P} = {\rm diag}\left(J_{\rv{\theta}}^{\rm P},J_{{\rm Re}\{\rv{\alpha}\}}^{\rm P},J_{{\rm Im}\{\rv{\alpha}\}}^{\rm P}\right).
\end{equation}
Furthermore, we assume that the complex amplitude $\rv{\alpha}$ is circularly symmetric, which implies that $\mathbb{E}\{\rv{\alpha}\}=\mathbb{E}\{\rv{\alpha}^*\}=0$, and that $J_{{\rm Re}\{\rv{\alpha}\}}^{\rm P}=J_{{\rm Im}\{\rv{\alpha}\}}^{\rm P}=J_{\rv{\alpha}}^{\rm P}$. Since we are only interested in the angle $\rv{\theta}$, we consider its equivalent \ac{bfim} (by treating $\rv{\alpha}$ as a nuisance parameter) given by \cite{shen_fundamental}
\begin{equation}
\begin{aligned}
J_{\rm e}(\rv{\theta}) &= 2\mathbb{E}\{\rv{f}_{11}\} + J_{\rv{\theta}}^{\rm P}\\
&\hspace{3mm} - 4\mathbb{E}\{\rv{f}_{12}^*\}(2\mathbb{E}\{\rv{f}_{22}\}+J_{\rv{\alpha}}^{\rm P})^{-1}\mathbb{E}\{\rv{f}_{12}\},
\end{aligned}
\end{equation}
which equals to the first diagonal entry (corresponding to $\rv{\theta}$) in $\RM{J}_{\RV{\eta}|\RM{X}}^{-1}$. Using the assumption that $\rv{\alpha}$ is circularly symmetric, we obtain that $\mathbb{E}\{f_{12}\}=0$, and thus
\begin{align}\label{angle_fim}
\rv{J}_{\rv{\theta}|\RM{X}}^{\rm e} &= 2\mathbb{E}\{\rv{f}_{11}\} + J_{\rv{\theta}}^{\rm P} \nonumber \\
&=\frac{2T\mathbb{E}\{|\rv{\alpha}|^2\}}{\sigma_{\rm s}^2}\tr{\overline{\M{M}}\RM{R}_{\RM{X}}} + J_{\rv{\theta}}^{\rm P},
\end{align}
where $\overline{\M{M}}=\mathbb{E}\{\RM{M}^\ast(\rv{\theta})\}$, and
\begin{equation}
\begin{aligned}
\RM{M}(\rv{\theta})&=\|\dot{\RV{a}}(\rv{\theta})\|^2\RV{v}(\rv{\theta})\RV{v}^{\rm H}(\rv{\theta})+\dot{\RV{a}}^{\rm H}(\rv{\theta})\RV{a}(\rv{\theta})\dot{\RV{v}}(\rv{\theta})\RV{v}^{\rm H}(\rv{\theta})\\
&\hspace{3mm}+\RV{a}^{\rm H}(\rv{\theta})\dot{\RV{a}}(\rv{\theta})\RV{v}(\rv{\theta})\dot{\RV{v}}^{\rm H}(\rv{\theta})+\|\RV{a}(\rv{\theta})\|^2\dot{\RV{v}}(\rv{\theta})\dot{\RV{v}}^{\rm H}(\rv{\theta}).
\end{aligned}
\end{equation}
We may simplify the expression of $\RM{M}(\rv{\theta})$ by choosing the phase reference point of the transmitting and receiving arrays such that $\dot{\RV{v}}^{\rm H}(\rv{\theta})\RV{v}(\rv{\theta})=0$ and $\dot{\RV{a}}^{\rm H}(\rv{\theta})\RV{a}(\rv{\theta})=0$, and hence
\begin{equation}
\RM{M}(\rv{\theta})=\|\dot{\RV{a}}(\rv{\theta})\|^2\RV{v}(\rv{\theta})\RV{v}^{\rm H}(\rv{\theta})+\|\RV{a}(\rv{\theta})\|^2\dot{\RV{v}}(\rv{\theta})\dot{\RV{v}}^{\rm H}(\rv{\theta}).
\end{equation}

\subsubsection{Sensing \ac{dof}}
According to Theorem \ref{thm:sensing_dofloss}, the sensing \ac{dof} in the single-target angle estimation scenario at point $P_{\rm CS}$ is lower-bounded by
\begin{equation}\label{target_sensing_dof_lower}
\nu_{\rm s} \geq T-K = T-1.
\end{equation}
In particular, when the communication channel $\RM{H}_{\rm c}$ is rank-1 (e.g. \ac{los} \ac{mimo} channel), the lower bound in \eqref{target_sensing_dof_lower} is achieved. To elaborate, note that when $\RM{H}_{\rm c}$ is rank-1, the communication-optimal covariance matrix $\widetilde{\M{R}}_{\RM{X}}^{\rm CS}$ is also rank-1. In this case, since we have $\RM{R}_{\RM{X}}=\RM{R}_{\RM{X}}^{\rm CS}$, the term $\tr{\overline{\M{M}}\RM{R}_{\RM{X}}}$ in \eqref{angle_fim} is given by
\begin{equation}
\tr{\overline{\M{M}}\RM{R}_{\RM{X}}} \!=\! \frac{P_{\rm T}M\V{r}^{\rm H}\overline{\M{M}}\V{r}}{T}\sum_{i=1}^T |\rv{n}_i|^2,
\end{equation}
where $\V{r}$ is the eigenvector corresponding to the maximum eigenvalue of $\widetilde{\M{R}}_{\RM{X}}^{\rm CS}$, and $\rv{n}_i$'s are mutually independent zero-mean circularly symmetric complex Gaussian random variables with unit variance. Hence $\tr{\overline{\M{M}}\RM{R}_{\RM{X}}}$ is proportional to a chi-squared distributed random variable having \ac{dof} $2T$, which implies that
\begin{equation}
\mathbb{E}\bigg[\Big(\tr{\overline{\M{M}}\RM{R}_{\RM{X}}^{\rm CS}}\Big)^{-1}\bigg]= \frac{T}{(T-1)\tr{\overline{\M{M}}\widetilde{\M{R}}_{\RM{X}}^{\rm CS}}}.
\end{equation}
Thus the sensing \ac{dof} can be calculated as
\begin{align}
\nu_{\rm s}&=\lim_{\sigma_{\rm s}\rightarrow 0}\frac{T\mathbb{E}\left\{\left(\tr{\overline{\M{M}}\RM{R}_{\RM{X}}^{\rm CS}}+\frac{\sigma_{\rm s}^2J_{\rv{\theta}}^{\rm P}}{T\mathbb{E}\{|\rv{\alpha}|^2\}}\right)^{-1}\right\}}{\left(\tr{\overline{\M{M}}\widetilde{\M{R}}_{\RM{X}}^{\rm CS}}+\frac{\sigma_{\rm s}^2J_{\rv{\theta}}^{\rm P}}{T\mathbb{E}\{|\rv{\alpha}|^2\}}\right)^{-1}} \nonumber \\
&=T-1,
\end{align}
which indeed achieves the lower bound in \eqref{target_sensing_dof_lower}.

\subsubsection{Communication \ac{dof}}
Observe from \eqref{angle_fim} that, at point $P_{\rm SC}$, the columns of $\RM{R}_{\RM{X}}^{\rm SC}$ should be spanned by the subspace corresponding to the largest eigenvalue of $\overline{\M{M}}$. Therefore, when the largest eigenvalue of $\overline{\M{M}}$ has multiplicity $1$, the sensing-optimal sample covariance matrix $\RM{R}_{\RM{X}}^{\rm SC}$ is unique and rank-1. In this case, the high-\ac{snr} sensing-limited capacity is asymptotically given by
\begin{equation}\label{target_capacity}
R_{\rm SC} = \mathbb{E}\bigg\{\frac{2T-1}{2T}\log |\sigma_{\rm c}^{-2}\RM{H}_{\rm c}\widetilde{\M{R}}_{\RM{X}}^{\rm SC}\RM{H}_{\rm c}^{\rm H}|+\rv{c}_0\bigg\} + O(\sigma_{\rm c}^2),
\end{equation}
when the column space of $\RM{R}_{\RM{X}}^{\rm SC}$ ($=\widetilde{\M{R}}_{\RM{X}}^{\rm SC}$) is not orthogonal to the column space of $\RM{H}_{\rm c}$, otherwise the capacity is zero. We see from \eqref{target_capacity} that the communication \ac{dof} is $(2T-1)/2T$, and hence the communication \ac{dof} loss induced by the DRT is $1/2T$.

To further understand the \ac{snc} tradeoff in the task of target angle estimation, let us consider a simplistic scenario, where the communication receiver is equipped with a single antenna. We further assume that the communication channel is deterministic and denoted as $\V{h}_{\rm c}$, hence the (per antenna) maximum sensing and communication receiving \ac{snr} may be computed as
\begin{subequations}
\begin{align}
{\rm SNR}_{\rm s} &= MP_{\rm T}\mathbb{E}\{|\rv{\alpha}|^2\}\sigma_{\rm s}^{-2},\\
{\rm SNR}_{\rm c} &= MP_{\rm T} \|\V{h}_{\rm c}\|^2\sigma_{\rm c}^{-2},
\end{align}
\end{subequations}
respectively. Furthermore, we assume that the maximum eigenvalue of $\overline{\M{M}}$ has multiplicity $1$. Since $N_{\rm c}=1$, we now see that both the sensing-optimal and the communication-optimal sample covariance matrices are rank-1. Therefore, the communication subspace is spanned by $\V{h}_{\rm c}$, while the sensing subspace is spanned by the sensing-optimal steering vector $\V{u}_{\rm s}$ characterized by
\begin{equation}
\overline{\M{M}}\V{u}_{\rm s} = \lambda_1\{\overline{\M{M}}\}\V{u}_{\rm s},
\end{equation}
where $\lambda_1(\cdot)$ denotes the maximum eigenvalue of its argument. In other words, $\V{u}_{\rm s}$ is the eigenvector corresponding to the maximum eigenvalue of $\overline{\M{M}}$. In this scenario, the overlap between the communication subspace and the sensing subspace can be depicted by the following quantity
\begin{equation}
\rho = \frac{\V{h}_{\rm c}^{\rm H}\overline{\M{M}}\V{h}_{\rm c}}{\|\V{h}_{\rm c}\|^2\lambda_1\{\overline{\M{M}}\}}.
\end{equation}

Under the aforementioned assumptions, besides the pentagon inner bound obtained by employing time-sharing strategies between $P_{\rm SC}$ and $P_{\rm CS}$, we can also obtain some refined inner bounds of the \ac{crb}-rate region using the statistical covariance shaping method discussed in Section \ref{ssec:inner_bounds}, as well as an outer bound. Specifically, the statistical covariance shaping problem now takes the following form
\begin{equation}\label{infinite_T_optimization}
\begin{aligned}
\max_{\widetilde{\M{R}}_{\RM{X}}}&~~\tr{\overline{\M{M}}\widetilde{\M{R}}_{\RM{X}}}+\lambda \V{h}_{\rm c}^{\rm H}\widetilde{\M{R}}_{\RM{X}}\V{h}_{\rm c}\\
{\rm s.t.}&~~\tr{\widetilde{\M{R}}_{\RM{X}}}=P_{\rm T}M,~\widetilde{\M{R}}_{\RM{X}}\succeq \V{0},~\widetilde{\M{R}}_{\RM{X}}=\widetilde{\M{R}}_{\RM{X}}^{\rm H},
\end{aligned}
\end{equation}
where $\lambda\in[0,+\infty)$ controls the preference between the sensing and the communication performance. The solution to this problem can be expressed as follows:
\begin{equation}
\widetilde{\M{R}}_{\RM{X}}(\lambda) = P_{\rm T}M\V{r}(\lambda)\V{r}^{\rm H}(\lambda),
\end{equation}
where $\V{r}(\lambda)$ is the eigenvector corresponding to the largest eigenvalue of the matrix
\begin{equation}
\M{M}(\lambda) = \overline{\M{M}}+\lambda \V{h}_{\rm c}\V{h}_{\rm c}^{\rm H}.
\end{equation}
By employing this Pareto-optimal power allocation strategy and ignoring the \ac{drt}, according to \eqref{angle_fim}, we obtain an outer bound characterized as follows:
\begin{subequations}\label{outerbound1}
\begin{align}
\lambda &\in [0,+\infty),\\
R_{\rm out}(\lambda) &= \log_2(1+\|\V{h}_{\rm c}\|^{-2}N_{\rm c}|\V{r}^{\rm H}(\lambda)\V{h}_{\rm c}|^2{\rm SNR}_{\rm c}),\\
\epsilon_{\rm out}(\lambda) &= \Big(2T{\rm SNR_{\rm s}}\V{r}^{\rm H}(\lambda)\overline{\M{M}}\V{r}(\lambda) + J_{\rv{\theta}}^{\rm P}\Big)^{-1},
\end{align}
\end{subequations}
where we have used the base-2 logarithm to ensure that the rate is in the unit of bit per channel use (bpcu). By contrast, when we apply the power allocation strategy to the Gaussian signal, we obtain the following inner bound
\begin{subequations}\label{innerboundG}
\begin{align}
R_{\rm in,G}(\lambda)&\!=\!\log_2(1+\|\V{h}_{\rm c}\|^{-2}N_{\rm c}|\V{r}^{\rm H}(\lambda)\V{h}_{\rm c}|^2{\rm SNR}_{\rm c}),\\
\epsilon_{\rm in,G}(\lambda)&\!=\!\mathbb{E}\Big\{\Big(\rv{\chi}^2_{2T}{\rm SNR_{\rm s}}\V{r}^{\rm H}(\lambda)\overline{\M{M}}\V{r}(\lambda) + J_{\rv{\theta}}^{\rm P}\Big)^{-1}\Big\}\nonumber \\
&\!=\!(2{\rm SNR_{\rm s}}\V{r}^{\rm H}(\lambda)\overline{\M{M}}\V{r}(\lambda))^{-1}\zeta^{T-1}e^\zeta\Gamma(1-T,\zeta)\nonumber \\
&\!=\!\Big(2(T\!-\!1){\rm SNR_{\rm s}}\V{r}^{\rm H}(\lambda)\overline{\M{M}}\V{r}(\lambda)\Big)^{-1}(1\!+\!r_{\zeta}),
\end{align}
\end{subequations}
where $\rv{\chi}^2_{2T}$ is a chi-squared distributed random variable with \ac{dof} $2T$, $\Gamma(a,x)=\int_x^{\infty} t^{a-1}e^{-t} {\rm d}t$ denotes the incomplete Gamma function \cite{tables}, and $\zeta = J_{\rv{\theta}}^{\rm P}(2{\rm SNR_{\rm s}}\V{r}^{\rm H}(\lambda)\overline{\M{M}}\V{r}(\lambda))^{-1}$. The correction term $r_{\zeta}$ is on the order of $O(\zeta)$, given by
$$
r_{\zeta}=\sum_{n=1}^{T-2}\frac{(-1)^n\zeta^n}{\prod_{i=1}^n (T-i-1)}+\underbrace{(-1)^{T-1}\cdot \frac{e^{\zeta}\zeta^{T-1}\Gamma(0,\zeta)}{\Gamma(T-1)}}_{O(\zeta^{T-1}\log \zeta)},
$$
which can be derived from \cite[Sec.~8.352]{tables}. We can also obtain another inner bound by applying the power allocation strategy to the semi-unitary signal (a constant-norm vector in this scenario, since the sensing-optimal sample covariance matrix is rank-1)
\begin{subequations}\label{innerboundU}
\begin{align}
R_{\rm in,U}(\lambda)&=\frac{2T-1}{2T}\log_2(N_{\rm c}\|\V{h}_{\rm c}\|^{-2}|\V{r}^{\rm H}(\lambda)\V{h}_{\rm c}|^2{\rm SNR}_{\rm c}) \nonumber \\
&\hspace{3mm}+ c_0+O(\sigma_{\rm c}^2),\\
\epsilon_{\rm in,U}(\lambda)&=\Big(2T{\rm SNR_{\rm s}}\V{r}^{\rm H}(\lambda)\overline{\M{M}}\V{r}(\lambda)\} +  J_{\rv{\theta}}^{\rm P}\Big)^{-1},
\end{align}
\end{subequations}
where the correction term $c_0$ is given by
$$
c_0=\frac{1}{T}\Big[\Big(T\!-\!\frac{1}{2}\Big)\log\frac{T}{e}\!-\!\log \Gamma(T)\!+\!\log(2\sqrt{\pi})\Big],
$$
and the $O(\sigma_{\rm c}^2)$ residual term can be computed numerically.

Moreover, we can obtain a tighter inner bound, by employing time-sharing strategies between the Gaussian signal and the semi-unitary signal. This inner bound can be obtained by numerically computing the convex envelope of the union of the regions determined by \eqref{innerboundG} and \eqref{innerboundU}.

\subsection{Target Response Matrix Estimation}\label{ssec:channel_estimation}
Next, let us consider the task of target response matrix estimation commonly seen in the literature of statistical \ac{mimo} radar \cite{trm}, where the sensing objective is to estimate the entire matrix $\RM{H}_{\rm s}$. More precisely, we have
\begin{equation}
\RV{\eta} = \left[\mathrm{Re}\{\mathrm{vec}(\RM{H}_{\rm s}^{\rm T})\}^{\rm T},~\mathrm{Im}\{\mathrm{vec}(\RM{H}_{\rm s}^{\rm T})\}^{\rm T}\right]^{\rm T}.
\end{equation}
We further assume that the \textit{a priori} distribution of each entry in $\RM{H}_{\rm s}$ is identically $\mathcal{CN}(0,\sigma_{\rm p}^2)$, and hence
$$
\M{J}_{\rm P}=\sigma_{\rm p}^{-2}\M{I}_{2N_{\rm s}M},~\widetilde{\M{J}}_{\rm P}=\frac{\sigma_{\rm s}^2}{\sigma_{\rm p}^2T}\M{I}_{2N_{\rm s}M}.
$$
Note that
\begin{equation}
\RV{\eta}=\frac{1}{\sqrt{2}}\M{U}_{\rm Had}\RV{h}_{\rm s},
\end{equation}
where $\RV{h}_{\rm s}:=[{\rm vec}(\RM{H}_{\rm s}^{\rm T})^{\rm T},~{\rm vec}(\RM{H}_{\rm s}^{\rm T})^{\rm H}]^{\rm T}$, and $\M{U}_{\rm Had}$ is a unitary matrix given by
$$
\M{U}_{\rm Had}=\frac{1}{\sqrt{2}}\left[
                 \begin{array}{cc}
                   1 & 1 \\
                   \imath & -\imath \\
                 \end{array}
               \right]\otimes \M{I}_{N_{\rm s}M}.
$$
In light of this, the affine map $\M{\Phi}(\cdot)$ can be expressed explicitly as
\begin{equation}
\begin{aligned}
\M{\Phi}(\RM{R}_{\RM{X}})&=2\M{U}_{\rm Had}^{\rm H}\Big(\M{I}_{N_{\rm s}}\otimes \mathrm{blkdiag}\big(\RM{R}_{\RM{X}}^{\rm T},\RM{R}_{\RM{X}}\big)\Big)\M{U}_{\rm Had}\\
&\hspace{3mm}+\frac{2\sigma_{\rm s}^2}{\sigma_{\rm p}^2T}\M{I}_{2N_{\rm s}M}.
\end{aligned}
\end{equation}
Thus  the \ac{crb} of $\RV{\eta}$ is given by
\begin{align}
\epsilon &= \tr{\mathbb{E}\left[\frac{\sigma_{\rm s}^2}{T}[\M{\Phi}(\RM{R}_{\RM{X}})]^{-1}\right]} \nonumber \\
&=\frac{\sigma_{\rm s}^2N_{\rm s}}{T}\tr{ \mathbb{E}\left[\left(\RM{R}_{\RM{X}}+\frac{\sigma_{\rm s}^2}{\sigma_{\rm p}^2T}\M{I}\right)^{-1}\right]},
\end{align}
which follows from the fact that the unitary matrix $\M{U}_{\rm Had}$ preserves trace, and that the transpose operation preserves trace. Next, let us consider the sensing and communication performance at points $P_{\rm SC}$ and $P_{\rm CS}$, respectively.

\subsubsection{Sensing \ac{dof}}
At point $P_{\rm SC}$, the optimal $\widetilde{\M{R}}_{\RM{X}}$ is obtained at $\widetilde{\M{R}}_{\RM{X}}^{\rm SC}=P_{\rm T}\M{I}_M$. Moreover, we have
\begin{align}
\tr{ \mathbb{E}\left[\left(\RM{R}_{\RM{X}}+\frac{\sigma_{\rm s}^2}{\sigma_{\rm p}^2T}\M{I}\right)^{-1}\right]}&=\tr{\left(\widetilde{\M{R}}_{\RM{X}}+\frac{\sigma_{\rm s}^2}{\sigma_{\rm p}^2T}\M{I}\right)^{-1}} \nonumber \\
&=\frac{M}{P_{\rm T}+\sigma_{\rm s}^2(\sigma_{\rm p}^2T)^{-1}},
\end{align}
since the sensing-optimal sample covariance matrix $\RM{R}_{\RM{X}}$ in this scenario is deterministic. Hence the minimum \ac{crb} can be expressed as
\begin{equation}
\epsilon_{\min} = \frac{\sigma_{\rm s}^2 N_{\rm s}M}{TP_{\rm T}+\sigma_{\rm s}^2\sigma_{\rm p}^{-2}}.
\end{equation}
The sensing \ac{dof} at point $P_{\rm SC}$ is clearly $\nu_{\rm s,max}=T$. Observe that the minimum \ac{crb} may be alternatively expressed as
\begin{equation}
\epsilon_{\min} = \frac{N_{\rm s}M}{T} \cdot \frac{1}{P_{\rm T}\sigma_{\rm s}^{-2} +(\sigma_{\rm p}^2T)^{-1}},
\end{equation}
which implies that the \textit{a priori} knowledge contributes to an additional effective \ac{snr} of $(\sigma_{\rm p}^2T)^{-1}$.

\begin{figure}[t]
\centering
\begin{overpic}[width=.45\textwidth]{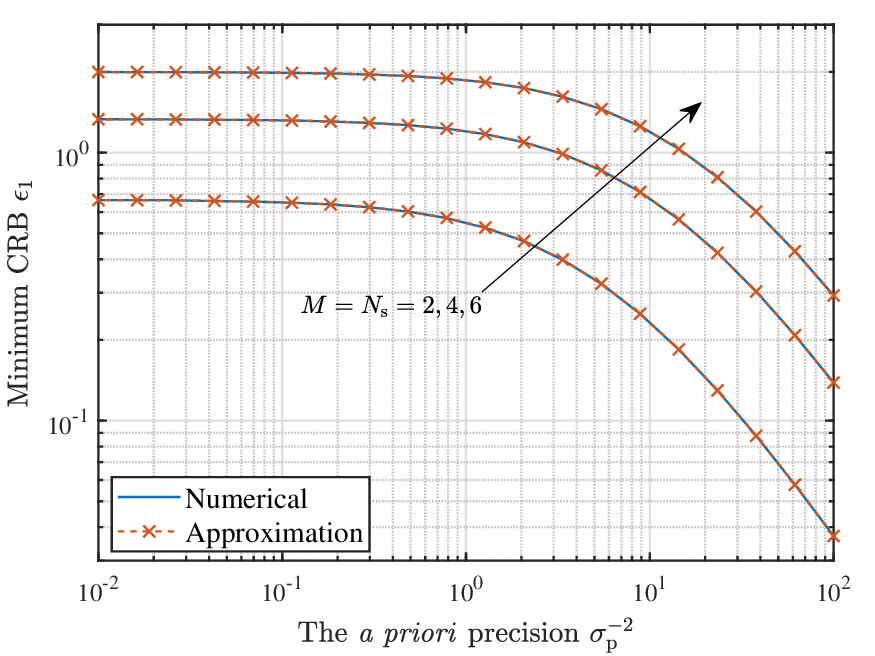}
\end{overpic}
\caption{The minimum achievable \ac{crb} $\epsilon_{\rm CS}$ and its Marchenko-Pastur approximation \eqref{approx_mp1} versus the \textit{a priori} precision $\sigma_{\rm p}^{-2}$, where $\RM{R}_{\RM{X}}^{\rm CS}\sim\mathcal{CW}_M(\M{I},T)$, $\sigma_{\rm s}^2=1$, and $\beta=M/T$ is fixed at $1/4$.}
\label{fig:mp_approx}
\end{figure}

By contrast, at point $P_{\rm CS}$, the minimum achievable \ac{crb} is given by
\begin{equation}
\epsilon_{\rm CS} = \frac{\sigma_{\rm s}^2N_{\rm s}}{T} \tr{\mathbb{E}\left[\left(\RM{R}_{\RM{X}}^{\rm CS}+\frac{\sigma_{\rm s}^2}{\sigma_{\rm p}^2T}\M{I}\right)^{-1}\right]},
\end{equation}
where $\RM{R}_{\RM{X}}^{\rm CS}$ is a complex Wishart distributed matrix having degree of freedom $T$ and scale matrix $\widetilde{\M{R}}_{\RM{X}}^{\rm CS}$. When $\RM{R}_{\RM{X}}^{\rm CS}$ follows the complex Wishart distribution $\mathcal{CW}_M(\M{I},T)$, it is known that the eigenvalue distribution of $\mathcal{CW}_M(\M{I},T)$ converges to the Marchenko-Pastur distribution as $M\rightarrow\infty$, if $\beta=M/T$ is a constant \cite{mp_distribution}. In the same asymptotic limit, according to the Stieltjes transform of the Marchenko-Pastur distribution \cite{mp_distribution}, the \ac{crb} $\epsilon_{\rm CS}$ satisfies
\begin{equation}\label{approx_mp1}
\epsilon_{\rm CS} \rightarrow \frac{\sigma_{\rm s}^2N_{\rm s}M}{2Tz \beta}(\beta-z-1+\sqrt{(z+\beta-1)^2+4z}),
\end{equation}
where $z = \sigma_{\rm s}^2 (\sigma_{\rm p}^2T)^{-1}$. As it can be observed from Fig.~\ref{fig:mp_approx}, this limit serves as an excellent approximation even when $M$ is rather small (e.g. $M=2$). In the more general scenario $\RM{R}_{\RM{X}}^{\rm CS}\sim \mathcal{CW}_M(\M{\Sigma},T)$, if the eigenvalues of $\M{\Sigma}$ are upper and lower bounded by some positive constants as $M\rightarrow \infty$, we have
\begin{equation}\label{approx_mp}
\epsilon_{\rm CS}\rightarrow \frac{\sigma_{\rm s}^2N_{\rm s}}{2Tz \beta}\sum_{i=1}^M \left[\beta\!-\!\frac{z}{\sigma_i}\!-\!1\!+\!\sqrt{(\frac{z}{\sigma_i}\!+\!\beta\!-\!1)^2+4\frac{z}{\sigma_i}}\right],
\end{equation}
where $\sigma_i$ is the $i$-th largest eigenvalue of $\M{\Sigma}$.

As for the sensing \ac{dof}, observe that
\begin{subequations}\label{sensing_dof_channel}
\begin{align}
\nu_{\rm s,CS} &= \lim_{\sigma_{\rm s}\rightarrow 0} \epsilon_{\rm CS}^{-1}\sigma_{\rm s}^2N_{\rm s}\tr{\left[\widetilde{\M{R}}_{\RM{X}}^{\rm CS}+\frac{\sigma_{\rm s}^2}{\sigma_{\rm p}^2T}\M{I}\right]^{-1}}\\
&=T\tr{(\widetilde{\M{R}}_{\RM{X}}^{\rm CS})^{-1}}/\tr{\mathbb{E}\left[(\RM{R}_{\RM{X}}^{\rm CS})^{-1}\right]}\\
&=T-M, \label{3rd_line_dof}
\end{align}
\end{subequations}
when $\widetilde{\M{R}}_{\RM{X}}^{\rm CS}$ has full rank (otherwise the sensing \ac{dof} is zero), where \eqref{3rd_line_dof} follows from the fact that
\begin{equation}
\mathbb{E}\left[(\RM{R}_{\RM{X}}^{\rm CS})^{-1}\right] = \frac{T}{T-M}(\widetilde{\M{R}}_{\RM{X}}^{\rm CS})^{-1}.
\end{equation}
Hence we may conclude that the sensing \ac{dof} loss in the task of target response matrix estimation is
\begin{equation}
\nu_{\rm s,max}-\nu_{\rm s,CS}=M,
\end{equation}
which achieves the upper bound indicated by Theorem \ref{thm:sensing_dofloss}.

\subsubsection{Communication \ac{dof}}
At the point $P_{\rm CS}$, the maximum achievable rate is given by the channel capacity in the absence of sensing constraints, as follows:
\begin{equation}
R_{\rm max}=\mathbb{E}\{\log|\M{I}+\sigma_{\rm c}^{-2}\RM{H}_{\rm c}\widetilde{\M{R}}_{\RM{X}}^{\rm CS}\RM{H}_{\rm c}^{\rm H}|\},
\end{equation}
where $\widetilde{\M{R}}_{\RM{X}}^{\rm CS}$ is determined by the water-filling strategy. The communication \ac{dof} at the point $P_{\rm CS}$ is $\rv{M}_{\rm SC}$.

As for the point $P_{\rm SC}$, according to Proposition \ref{ssec:psc}, when the \ac{snr} is high, the corresponding achievable rate may be asymptotically expressed as
\begin{equation}
R_{\rm SC} = \mathbb{E}\Big\{\Big(1-\frac{\rv{M}_{\rm CS}}{2T}\Big)\log |P_{\rm T}\sigma_{\rm c}^{-2}\RM{H}_{\rm c}\RM{H}_{\rm c}^{\rm H}|+\rv{c}_0\Big\} + O(\sigma_{\rm c}^2),
\end{equation}
where we have used the fact that
$$
\rv{M}_{\rm SC}={\rm rank}(\RM{H}_{\rm c}\widetilde{\M{R}}_{\RM{X}}^{\rm SC}\RM{H}_{\rm c}^{\rm H})={\rm rank}(\RM{H}_{\rm c}\RM{H}_{\rm c}^{\rm H})=\rv{M}_{\rm CS}
$$
holds in the high-\ac{snr} regime. In this case, the communication \ac{dof} is
\begin{equation}
\nu_{\rm c,SC} = \rv{M}_{\rm SC}\Big(1-\frac{\rv{M}_{\rm SC}}{2T}\Big).
\end{equation}
The corresponding communication \ac{dof} loss at point $P_{\rm SC}$ is
\begin{equation}
\nu_{\rm c,max}-\nu_{\rm c,SC}=\frac{\rv{M}_{\rm SC}^2}{2T}.
\end{equation}
Observe that the communication subspace overlap coefficient $\alpha_{\rm SC}$ achieves its maximum $1$, hence the sensing-induced communication performance loss is mainly due to the row-orthogonality of the sensing-optimal coding strategy.

Similar to the target angle estimation problem, we can also conceive an outer bound based on statistical covariance shaping, namely solving the following Pareto optimization problem \cite{deterministic_rx_3}
\begin{equation}\label{pareto_ce}
\begin{aligned}
\min_{\M{R}_{\RM{X}}}&~{\rm tr}\Big\{\Big(\M{R}_{\RM{X}}\!+\!\frac{\sigma_{\rm s}^2}{\sigma_{\rm p}^2T}\M{I}\Big)^{-1}\Big\}\!+\! \alpha \log\left|\M{I}\!+\!\sigma_{\rm c}^{-2}\RM{H}_{\rm c}\M{R}_{\RM{X}}\RM{H}_{\rm c}^{\rm H}\right|\\
{\rm s.t.}&~\tr{\M{R}_{\RM{X}}}\leq P_{\rm T}M,~\M{R}_{\RM{X}}\succeq \M{0},~\M{R}_{\RM{X}}=\M{R}_{\RM{X}}^{\rm H},
\end{aligned}
\end{equation}
for each realization of $\RM{H}_{\rm c}$, where $\alpha\in[0,\infty)$ is a parameter controlling the power allocation strategy. Upon denoting the optimal solution  to \eqref{pareto_ce} as $\widetilde{\M{R}}_{\RM{X}}(\alpha)$ and the corresponding sample covariance matrix as $\RM{R}_{\alpha}$, we could obtain the Gaussian inner bound
\begin{subequations}\label{innerboundG_CE}
\begin{align}
R_{\rm in,G}(\alpha)&=\mathbb{E}\Big\{\log_2\Big|\M{I}+\sigma_{\rm c}^{-2}\RM{H}_{\rm c}\widetilde{\M{R}}_{\RM{X}}(\alpha)\RM{H}_{\rm c}^{\rm H}\Big|\Big\},\\
\epsilon_{\rm in,G}(\alpha)&=\frac{\sigma_{\rm s}^2N_{\rm s}}{T} \tr{\mathbb{E}\left[\left(\RM{R}_{\RM{X}}(\alpha)+\sigma_{\rm s}^2(\sigma_{\rm p}^2T)^{-1}\M{I}\right)^{-1}\right]}, \label{expectation_IBG}
\end{align}
\end{subequations}
the semi-unitary inner bound
\begin{subequations}\label{innerboundU_CE}
\begin{align}
R_{\rm in,U}(\alpha)&=\mathbb{E}\Big\{\Big(1-\frac{\rv{M}_{\alpha}}{2T}\Big)\log_2\Big|\M{I}+\sigma_{\rm c}^{-2}\RM{H}_{\rm c}\widetilde{\M{R}}_{\RM{X}}(\alpha)\RM{H}_{\rm c}^{\rm H}\Big|\nonumber \\
&\hspace{13mm}+\rv{c}_0\Big\} + O(\sigma_{\rm c}^2), \\
\epsilon_{\rm in,U}(\alpha)&=\frac{\sigma_{\rm s}^2N_{\rm s}}{T} \tr{\left(\widetilde{\M{R}}_{\RM{X}}(\alpha)+\sigma_{\rm s}^2(\sigma_{\rm p}^2T)^{-1}\M{I}\right)^{-1}},
\end{align}
\end{subequations}
and the outer bound
\begin{equation}\label{outerbound_CE}
R_{\rm out}(\alpha)=R_{\rm in G}(\alpha),~\epsilon_{\rm out}(\alpha)=\epsilon_{\rm in U}(\alpha),
\end{equation}
following a similar line of reasoning as \eqref{outerbound1}, \eqref{innerboundG} and \eqref{innerboundU}, where $\rv{M}_{\alpha}={\rm rank}(\RM{H}_{\rm c}\widetilde{\M{R}}_{\RM{X}}(\alpha)\RM{H}_{\rm c}^{\rm H})$, and the expectation in \eqref{expectation_IBG} may be approximated using \eqref{approx_mp}.

\begin{remark}[Null space completion by semi-unitary signals]\label{rem:completion}
According to \eqref{gaussian_signalling}, when $\rv{M}_{\rm CS} < M$, the Gaussian inner bound--achieving signalling scheme does not send pure Gaussian signals. To elaborate, in this case, since $\widetilde{\M{R}}_{\RM{X}}^{\rm CS}$ is rank-deficient, the sensing \ac{dof} is zero. In particular, at point $P_{\rm CS}$, the \ac{isac} signal cannot provide any information about the target response matrix in the null space of $\widetilde{\M{R}}_{\RM{X}}^{\rm CS}$, and hence we have to rely entirely on the \textit{a priori} knowledge when estimating this part of the target response matrix. As we move from $P_{\rm CS}$ to $P_{\rm SC}$ on the outer bound, $\widetilde{\M{R}}_{\RM{X}}(\alpha)$ becomes full-rank. However, the power allocated to the null space of $\widetilde{\M{R}}_{\RM{X}}^{\rm CS}$ does not contributed to the communication rate. Therefore, we may transmit semi-unitary signals instead of Gaussian signals in the null space, without eroding the communication rate at all.
\end{remark}

\section{Numerical Results}\label{sec:numerical}

\subsection{Target Angle Estimation}
We first demonstrate the \ac{snc} tradeoff in the task of target angle estimation discussed in Section~\ref{ssec:angle}. We consider the scenario where the sensing Rx and the ISAC Tx are co-located, both equipped with uniform linear arrays. There is a single target with a bearing angle of $\rv{\theta}$, for which the \textit{a priori} distribution is a von Mises distribution with mean $30^\circ$ and standard deviation of $5^\circ$. The communication channel is assumed to be a rank-1 \ac{los} channel. The configurations that are identical across all numerical examples are summarized in Table~\ref{tbl:config_angle}. For all numerical examples in this subsection, we observe that the maximum eigenvalue of $\overline{\M{M}}$ has multiplicity $1$, and thus the discussions in Section~\ref{ssec:angle} are fully applicable.

\begin{table}[t]
\centering
\caption{Configurations for the numerical examples of the target angle estimation problem.}
\label{tbl:config_angle}
\begin{tabular}{|c|c|}
  \hline
  \textbf{Configuration}  & \textbf{Value} \\ \hline
  No. Tx antennas ($M$)  & $10$ \\ \hline
  Tx antenna spacing & $1/2$ wavelength \\ \hline
  No. sensing Rx antennas ($N_{\rm s}$)  & $10$ \\ \hline
  Sensing Rx antenna spacing & $1/2$ wavelength \\ \hline
  No. communication Rx antennas ($N_{\rm c}$)  & $1$ \\ \hline
  Max. sensing receiving \ac{snr} & $20$dB per antenna \\ \hline
  Max. communication receiving \ac{snr} & $33$dB per antenna \\ \hline
\end{tabular}
\end{table}

\begin{figure}[t]
\centering
\begin{overpic}[width=.45\textwidth]{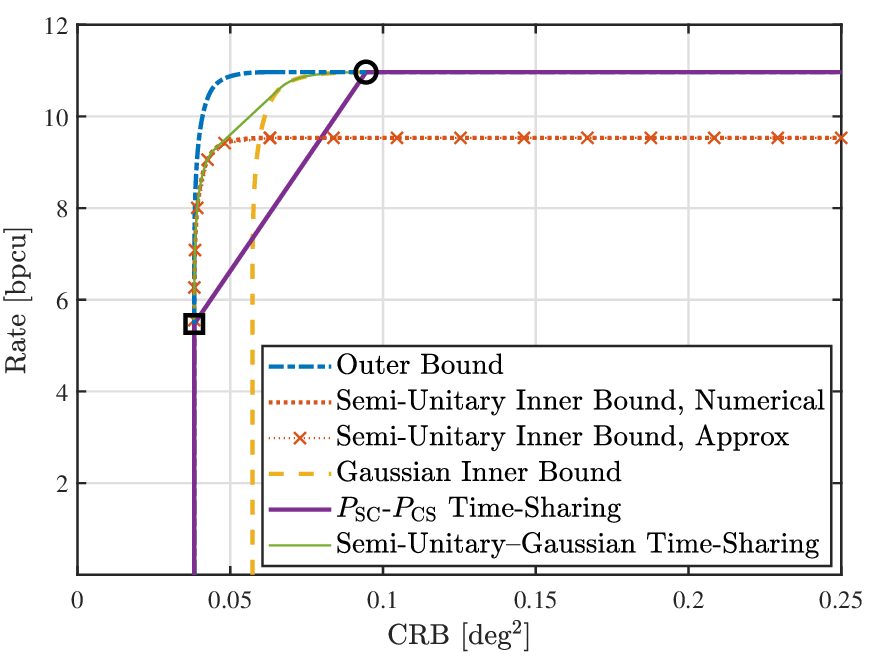}
\put(13,40.5){\footnotesize $P_{\rm SC}$}
\put(42.5,62){\footnotesize $P_{\rm CS}$}
\end{overpic}
\caption{Inner and outer bounds of the \ac{crb}-rate region for the task of single-target angle estimation. The correlation coefficient is $\rho\approx 0.61$.}
\label{fig:cr_basic}
\end{figure}

Let us first fix the bearing angle of the communication Rx at $\theta_{\rm c}=42^\circ$, which corresponds to a correlation coefficient of $\rho\approx 0.61$, and consider the case of $T=3$. The outer bound \eqref{outerbound1}, the Gaussian inner bound \eqref{innerboundG} and the semi-unitary inner bound \eqref{innerboundU}, as well as the refined inner bound obtained by employing the time-sharing strategy between the two inner bounds, are portrayed in Fig.~\ref{fig:cr_basic}. We have numerically computed the rate of the semi-unitary signalling strategy, accompanied by an approximation obtained by neglecting the $O(\sigma_{\rm c}^2)$ term in \eqref{innerboundU}. Observe that the approximation error is rather small, since the \ac{snr} is high. A noteworthy phenomenon is that the semi-unitary--Gaussian inner bound is tighter than the $P_{\rm SC}$--$P_{\rm CS}$ inner bound. An intuitive interpretation is that the semi-unitary--Gaussian inner bound benefits from amalgamating the time-sharing strategy with power allocation, whereas the $P_{\rm SC}$--$P_{\rm CS}$ inner bound only employs the time-sharing strategy. The vertical dashed line and the horizontal dotted line depict the ultimate loss of sensing \ac{dof} and communication \ac{dof} caused by employing the Gaussian and the semi-unitary signalling strategies, respectively.

\begin{figure}[t]
\centering
\begin{overpic}[width=.45\textwidth]{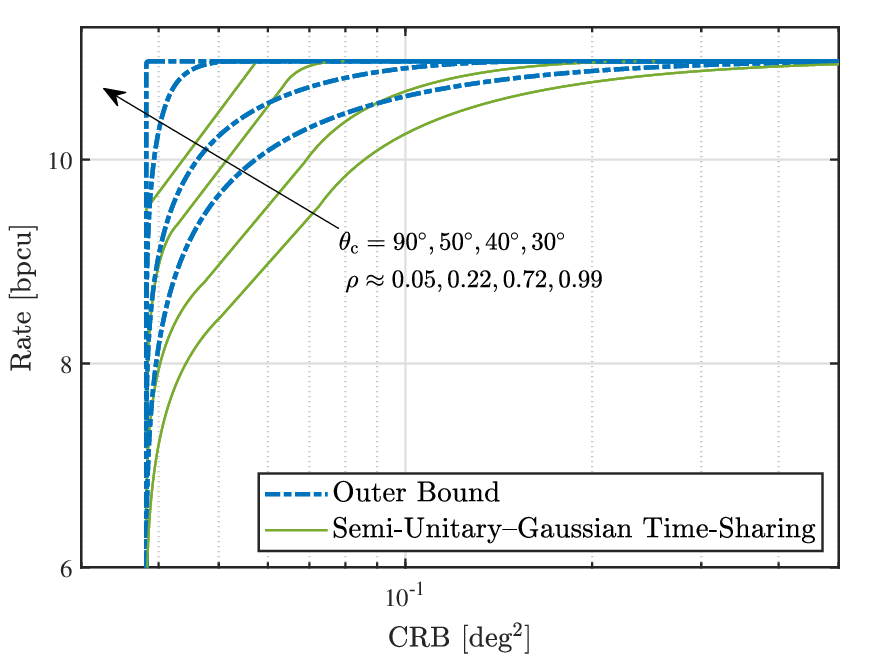}
\end{overpic}
\caption{The outer bound and the unitary-Gaussian time-sharing inner bound of the \ac{crb}-rate region for the task of single-target angle estimation, with various values of $\rho$.}
\label{fig:cr_rho}
\end{figure}

By varying the bearing angle of the communication Rx (hence the value of the correlation coefficient $\rho$), we may observe the dependence of the \ac{isac} integration gain on the overlap between the sensing subspace and the communication subspace, as portrayed in Fig.~\ref{fig:cr_rho}. In this example we also fix the coherent sensing period at $T=3$. We see that as $\rho$ increases, both the semi-unitary--Gaussian inner bound and the outer bound tend to become closer to the rectangular boundary, which corroborates the intuition that the \ac{isac} integration gain should be higher when sensing subspace has a larger overlap with the communication subspace. A noteworthy fact is that the inner bound is not rectangular when $\rho\approx 1$. In this case, the sensing subspace is identical to the communication subspace, and hence the \ac{snc} tradeoff is completely determined by the \ac{drt}. The two corner points of the inner bound correspond to the sensing-optimal and the communication-optimal signalling strategies, respectively.

\begin{figure}[t]
\centering
 \begin{overpic}[width=.45\textwidth]{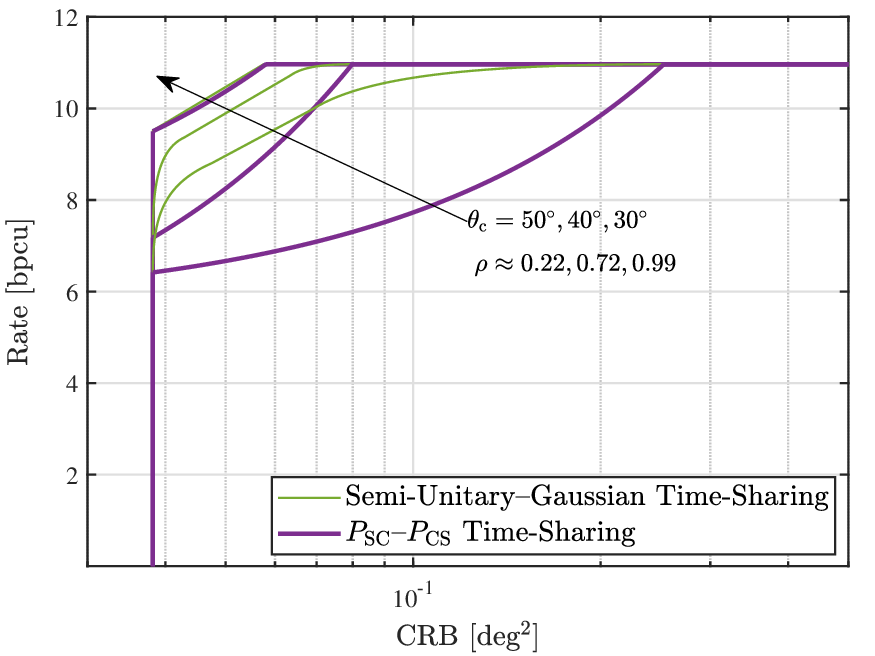}
\end{overpic}
\caption{Comparison between the inner bounds of the \ac{crb}-rate region for the task of single-target angle estimation, with various values of $\rho$.}
\label{fig:cr_rho_pentagon}
\end{figure}

In fact, the gap between the semi-unitary--Gaussian inner bound and the na\"{i}ve $P_{\rm SC}$--$P_{\rm CS}$ time-sharing inner bound may be viewed as the room of improvement upon the na\"{i}ve time-sharing relying on statistical covariance shaping, which corresponds to the optimal adjustment of \ac{st}. This is portrayed in Fig.~\ref{fig:cr_rho_pentagon}, where the curved shape of the na\"{i}ve time-sharing inner bound is due to the logarithmic scale of the abscissa. We may observe from Fig.~\ref{fig:cr_rho_pentagon} that the gap between these two inner bounds grows as the correlation coefficient $\rho$ decreases. This suggests that the performance gain of statistical covariance shaping over na\"{i}ve time-sharing is larger under weaker S\&C subspace correlation.

\begin{figure}[t]
\centering
\begin{overpic}[width=.45\textwidth]{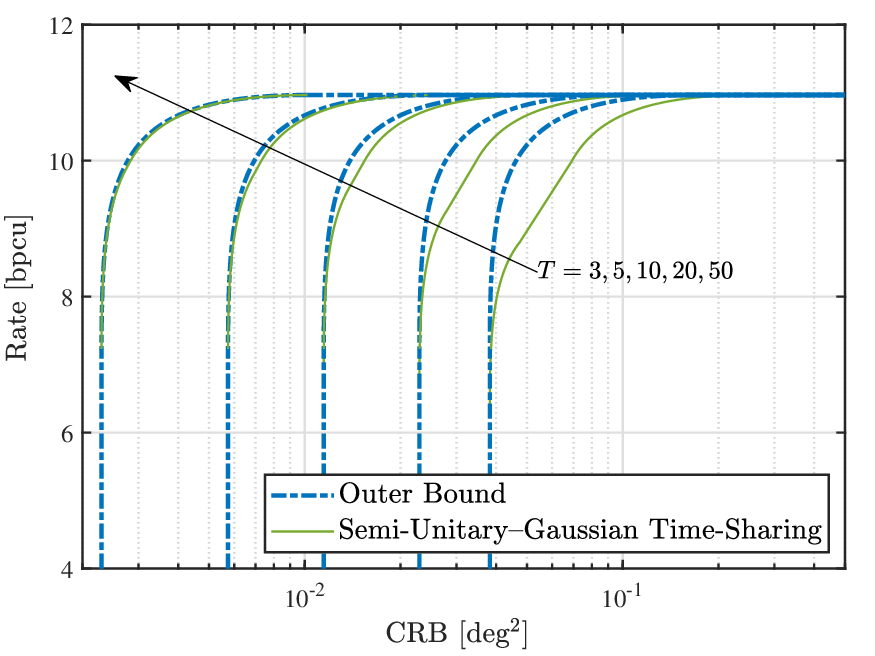}
\end{overpic}
\caption{The outer bound and the unitary-Gaussian time-sharing inner bound of the \ac{crb}-rate region for the task of single-target angle estimation, with various values of $T$.}
\label{fig:cr_T}
\end{figure}

Next we demonstrate the dependence of the \ac{crb}-rate region on the length $T$ of the coherent sensing period. In this example we fix the bearing angle of the communication Rx at $\theta_{\rm c}=50^{\circ}$, which corresponds to a correlation coefficient of $\rho\approx 0.22$. As it can be observed from Fig.~\ref{fig:cr_T}, as $T$ increases, the gap between the semi-unitary--Gaussian inner bound and the outer bound vanishes. This suggests that the \ac{drt} becomes less prominent when the coherent sensing period is long. Ultimately, it would become completely irrelevant to the \ac{snc} tradeoff as $T\rightarrow \infty$.

\begin{figure}[t]
\centering
\begin{overpic}[width=.45\textwidth]{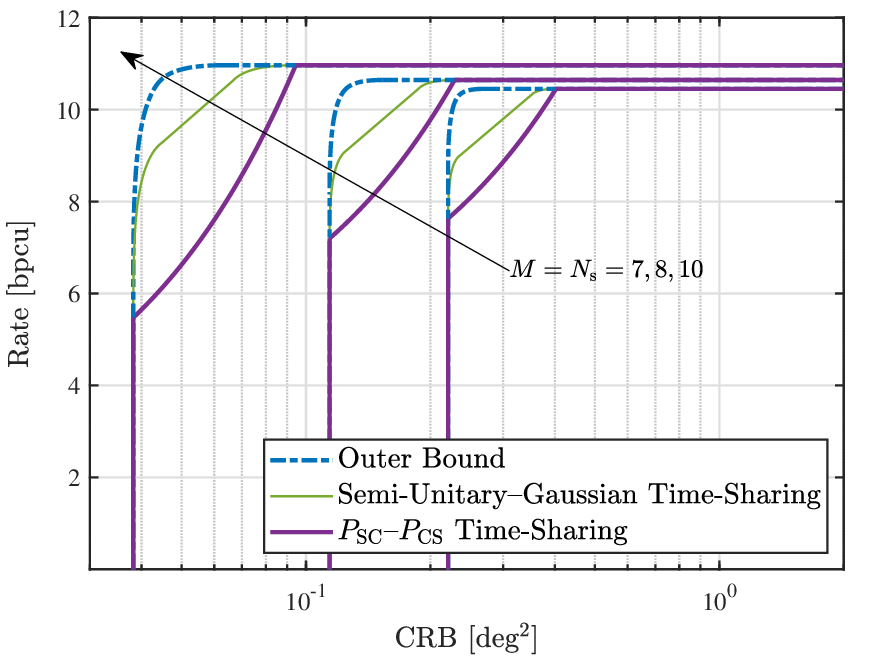}
\end{overpic}
\caption{The inner and outer bounds of the \ac{crb}-rate region for the task of single-target angle estimation, with various values of $M$ and $N_{\rm s}$. The bearing angle of the communication Rx is $\theta_{\rm c}=42^{\circ}$.}
\label{fig:cr_M}
\end{figure}

Finally, let us investigate the relationship between the CRB-rate region and the number of antennas. In this example we choose $\theta_{\rm c}=42^\circ$. The maximum sensing receiving \ac{snr} is $(10+10\log_{10}M)$~dB per antenna, while the maximum communication receiving \ac{snr} is $(23+10\log_{10}M)$~dB per antenna. These parameter values are chosen to be consistent with the example shown in Fig.~\ref{fig:cr_basic}. The inner and outer bounds obtained under various values of $M$ and $N_{\rm s}$ are portrayed in Fig.~\ref{fig:cr_M}. Observe that as $M$ and $N_{\rm s}$ increase, the gap between the semi-unitary--Gaussian inner bound and the na\"{i}ve time-sharing inner bound expands, and the rate at point $P_{\rm SC}$ decreases. These phenomena originate from the fact that the sensing and communication subspaces become asymptotically orthogonal to each other as $M$ and $N_{\rm s}$ tend to infinity, which may be intuitively interpreted as that the resolution of the system improves as the number of antennas increases. This may also be quantitatively depicted by the value of the correlation coefficient $\rho$, which decreases from $\rho\approx 0.82$ to $\rho\approx 0.61$ as $M=N_{\rm s}$ increases from $7$ to $10$. Consequently, the room of improvement due to \ac{st} adjustment also expands as $M$ and $N_{\rm s}$ increase, as suggested by the previous discussion about Fig.~\ref{fig:cr_rho_pentagon}.

\subsection{Target Response Matrix Estimation}
\begin{table}[t]
\centering
\caption{Configurations for the numerical examples of the target response matrix estimation problem.}
\label{tbl:config_hs}
\begin{tabular}{|c|c|}
  \hline
  \textbf{Configuration}  & \textbf{Value} \\ \hline
  No. Tx antennas ($M$)  & $4$ \\ \hline
  No. sensing Rx antennas ($N_{\rm s}$)  & $4$ \\ \hline
  No. communication Rx antennas ($N_{\rm c}$)  & $4$ \\ \hline
  Sensing noise variance ($\sigma_{\rm s}^2$) & $1$ \\ \hline
  Sensing transmit \ac{snr} ($P_{\rm t}/\sigma_{\rm s}^2$) & $24$dB \\ \hline
  Communication transmit \ac{snr} ($P_{\rm t}/\sigma_{\rm c}^2$) & $24$dB \\ \hline
\end{tabular}
\end{table}

Next, let us consider the task of target response matrix estimation discussed in Section \ref{ssec:channel_estimation}. We assume that the \textit{a priori} distributions of each entry in the target response matrix $\RM{H}_{\rm s}$ are independent of one another, and are identically given by $\mathcal{CN}(0,1)$, which implies that $\sigma_{\rm p}^2 = 1$. The configurations of other parameters are summarized in Table~\ref{tbl:config_hs}.

\begin{figure}[t]
\centering
\begin{overpic}[width=.45\textwidth]{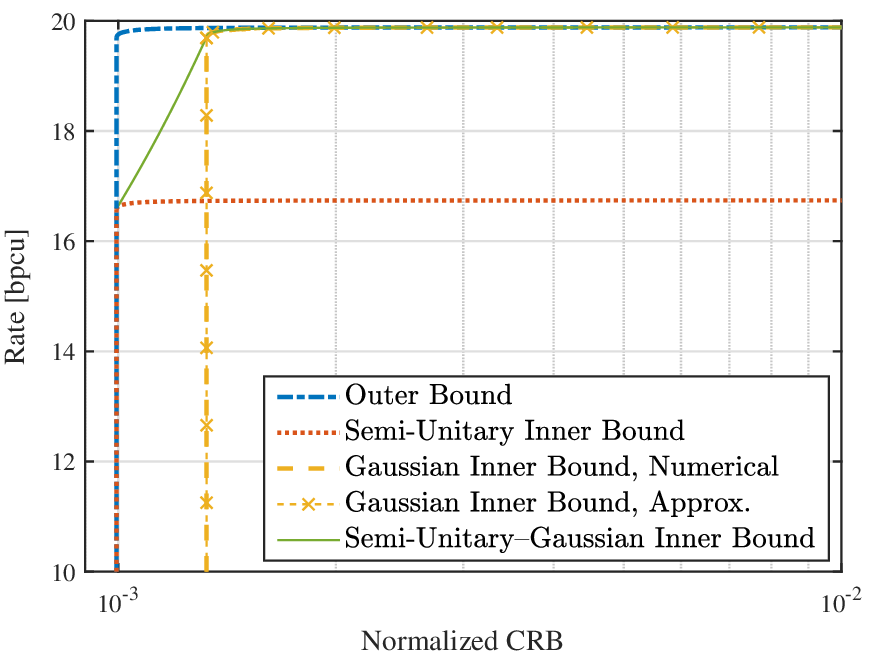}
\end{overpic}
\caption{Inner and outer bounds of the \ac{crb}-rate region for the task of target response matrix estimation, where the communication channel is subject to spatially uncorrelated Rayleigh fading.}
\label{fig:cr_CE_basic}
\end{figure}

We first consider the scenario where the communication channel is subject to spatially uncorrelated Rayleigh fading, namely that each entry in $\RM{H}_{\rm c}$ is independent of one another, and follows the distribution of $\mathcal{CN}(0,M^{-1})$. For the sake of illustration, we normalize the \ac{crb} as follows
$$
\epsilon_{\rm normalized} = \frac{\epsilon}{MN_{\rm s}}.
$$
In this example, we set the coherent sensing period as $T=4M=16$. The outer bound and the inner bounds of the \ac{crb}-rate region are portrayed in Fig.~\ref{fig:cr_CE_basic}, accompanied by the approximated Gaussian inner bound computed according to \eqref{approx_mp}. Observe that the outer bound is rather close to a rectangle in this scenario, which is reminiscent of the $\rho=1$ scenario in the target angle estimation task. We may understand this phenomenon intuitively, by noticing that both the communication channel and the target response matrix have full rank, and that the communication-optimal water-filling strategy becomes similar to the sensing-optimal uniform power allocation strategy in the high \ac{snr} regime.

\begin{figure}[t]
\centering
\begin{overpic}[width=.45\textwidth]{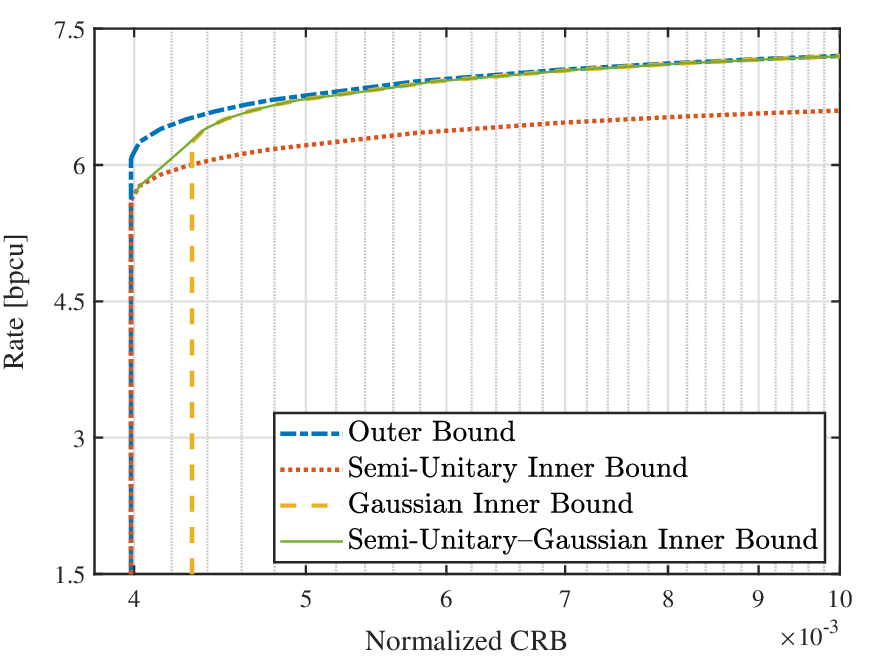}
\end{overpic}
\caption{Inner and outer bounds of the \ac{crb}-rate region for the task of target response matrix estimation, where the communication channel is a rank-1 \ac{los} channel.}
\label{fig:cr_CE_LoS}
\end{figure}

To further validate this intuition, let us consider another scenario, where the communication channel is a rank-1 \ac{los} channel taking the form of
$$
\RM{H}_{\rm c} = \rv{\alpha}\V{a}\V{v}^{\rm T},
$$
where $\V{a}\in\mathbb{C}^{N_{\rm s}}$ and $\V{v}\in\mathbb{C}^M$ are the steering vectors of the communication receiver and the transmitter, respectively, in which every entry has amplitude $1$. Note that in this example, the Gaussian inner bound is achieved by the null space completing technique discussed in Remark \ref{rem:completion}. Indeed, here we have $\rv{M}_{\rm CS}=1<M$. Consequently, the null space has a dimensionality of $M-1$, in which semi-unitary signals are transmitted. We assume that the amplitude $\rv{\alpha}$ is a random variable following $\mathcal{CN}(0,1)$, and set the coherent sensing period as $T=M=4$.  From the \ac{crb}-rate region plotted in Fig.~\ref{fig:cr_CE_LoS} we may see that the boundaries are obviously non-rectangular. This corroborate the aforementioned intuition, since the communication subspace in this scenario is only one-dimensional. Consequently, the overlap between the sensing and the communication subspaces is significantly smaller than the one in the previous scenario of the spatially uncorrelated Rayleigh channel.


\section{Conclusions}\label{sec:conclusions}
In this paper, we have proposed a general framework for the analysis of the \ac{snc} performance tradeoff. In particular, we have introduced the \ac{crb}-rate region for characterizing the \ac{snc} tradeoff, and have shown that it has a pentagon inner bound obtained by connecting the communication-optimal point $P_{\rm CS}$ and the sensing-optimal point $P_{\rm SC}$. Especially for $P_{\rm SC}$, we have shown that it is achieved when the transmitted waveform has a sample covariance matrix with deterministic trace. In particular, when the \ac{crb} minimization problem has a unique solution, the sample covariance matrix is deterministic, and hence the \ac{isac} waveform should be modulated by symbols generated based on the uniform distribution of the Stiefel manifold. For $P_{\rm CS}$, we have shown that the maximum sensing \ac{dof} loss is determined by the number of sensing parameters $K$, as well as the rank of the statistical covariance matrix $\widetilde{\M{R}}_{\RM{X}}$. We have also conceived signalling strategies based on statistical covariance shaping, which are shown to be capable of achieving more favorable \ac{snc} tradeoffs than the pentagon inner bound when applied in conjunction with time sharing.

Based on these analytical results, we have highlighted that the \ac{snc} performance trade-off is two-fold, in the sense that it is determined by the subspace alignment of the transmitted waveform in general, and also by the deterministic-random tradeoff in the finite coherent sensing period regime, which may shed light on the design of practical Pareto-optimal \ac{isac} signalling strategies.

\appendices
\section{Proof of Proposition \ref{prop:fim}}\label{sec:proof_fim}
\begin{IEEEproof}
The sensing model \eqref{sensing_model} may be rewritten as
\begin{equation}
{\rm vec}(\RM{Y}_{\rm s}^{\rm T})=( \M{I}_{N_{\rm s}}\otimes\RM{X}^{\rm T}){\rm vec}(\RM{H}_{\rm s}^{\rm T}) + {\rm vec}(\RM{Z}_{\rm s}^{\rm T}).
\end{equation}
For the moment, let us ignore the contribution of the prior distribution $p_{\RV{\eta}}(\V{\eta})$ to the \ac{bfim}. To facilitate further derivation, we define an extended parameter vector $$\RV{\theta}:=\left[
                                                                                           \begin{array}{c}
                                                                                           (\M{I}_{N_{\rm s}} \otimes\RM{X}^{\rm T} ){\rm vec}(\RM{H}_{\rm s}^{\rm T}) \\
                                                                                            (\M{I}_{N_{\rm s}} \otimes\RM{X}^{\rm H}){\rm vec}(\RM{H}_{\rm s}^{\rm H}) \\
                                                                                           \end{array}
                                                                                         \right],
$$
whose \ac{bfim} can be expressed as
\begin{equation}
\M{J}_{\RV{\theta}} = \sigma_{\rm s}^{-2}\M{I}_{2N_{\rm s}T}.
\end{equation}
Hence the \ac{bfim} of $\RV{h}_{\rm s}:=[{\rm vec}(\RM{H}_{\rm s}^{\rm T})^{\rm T},~{\rm vec}(\RM{H}_{\rm s}^{\rm T})^{\rm H}]^{\rm T}$ can be written as
\begin{align}
\RM{J}_{\RV{h}_{\rm s}|\RM{X}} &= \mathbb{E}_{\RV{h}_{\rm s}}\bigg\{\left(\frac{\partial \RV{\theta}}{\partial \RV{h}_{\rm s}}\right)^{\ast}\M{J}_{\RV{\theta}}\left(\frac{\partial \RV{\theta}}{\partial \RV{h}_{\rm s}}\right)^{\rm T}\bigg\} \nonumber \\
&=\frac{T}{\sigma_{\rm s}^2}\left[
                              \begin{array}{cc}
                                \M{I}_{N_{\rm s}}\otimes \RM{R}_{\RM{X}}^{\rm T} & \M{0} \\
                                \M{0} & \M{I}_{N_{\rm s}}\otimes \RM{R}_{\RM{X}} \\
                              \end{array}
                            \right],
\end{align}
according to the convention in \cite{vandenbos}. Let us denote $\RM{F}:=\frac{\partial \RV{h}_{\rm s}^{\ast}}{\partial \RV{\eta}}\in\mathbb{C}^{K\times 2N_{\rm s}M}$, and partition $\RM{F}$ in the following form
$$
\RM{F}=[\RM{F}_1,~\dotsc,~\RM{F}_{2N_{\rm s}}],
$$
where each $\RM{F}_i$ has a dimensionality of $K\times M$.

Now, taking the contribution of the prior distribution back into account, we have
\begin{align}
\RM{J}_{\RV{\eta}|\RM{X}} &= \frac{T}{\sigma_{\rm s}^2}\mathbb{E}_{\RV{\eta}}\bigg\{\RM{F}\left[
                              \begin{array}{cc}
                               \M{I}_{N_{\rm s}} \otimes \RM{R}_{\RM{X}}^{\rm T} & \M{0} \\
                                \M{0} & \M{I}_{N_{\rm s}} \otimes \RM{R}_{\RM{X}} \\
                              \end{array}
                            \right]\RM{F}^{\rm H}\bigg\} \!+\! \M{J}_{\rm P} \nonumber \\
&=\frac{T}{\sigma_{\rm s}^2}\mathbb{E}_{\RV{\eta}}\bigg\{\sum_{i=1}^{N_{\rm s}}\RM{F}_i\RM{R}_{\RM{X}}^{\rm T} \RM{F}_i^{\rm H}\!+\!\RM{F}_{N_{\rm s}+i}\RM{R}_{\RM{X}} \RM{F}_{N_{\rm s}+i}^{\rm H}\Bigg\} \!+\! \M{J}_{\rm P}.
\end{align}
Let us define a pair of linear super-operators $\RM{\Phi}_1(\cdot)$ and $\RM{\Phi}_2(\cdot)$, characterized by
\begin{subequations}
\begin{align}
\RM{\Phi}_1(\M{A}) &= \sum_{i=1}^{N_{\rm s}} \RM{F}_i \M{A}\RM{F}_i^{\rm H}, \label{phi_1}\\
\RM{\Phi}_2(\M{A}) &= \sum_{i=1}^{N_{\rm s}} \RM{F}_{N_{\rm s}+i} \M{A}\RM{F}_{N_{\rm s}+i}^{\rm H}.
\end{align}
\end{subequations}
Without loss of generality, in the following discussion, we will focus on $\RM{\Phi}_1(\cdot)$. Observe that the following matrix
\begin{equation}
\RM{\Psi}_1 = \sum_{i=1}^M \sum_{j=1}^M \M{E}_{ij} \otimes \RM{\Phi}_1(\M{E}_{ij})
\end{equation}
is a faithful representation of $\RM{\Phi}_1$ (known as the Choi representation, according to the Choi-Jamio{\l}kowski isomorphism \cite{choi,jammy}), where $\M{E}_{ij}$ is an $M\times M$ matrix with its $(i,j)$-th entry equals to $1$, while all other entries are zero. According to \eqref{phi_1}, we have
\begin{equation}
\RM{\Psi}_1 = \sum_{i=1}^{N_{\rm s}} {\rm vec}(\RM{F}_i){\rm vec}(\RM{F}_i)^{\rm H},
\end{equation}
which implies that
$$
\widetilde{\M{\Psi}}_1 := \mathbb{E}\{\RM{\Psi}_1\} =\sum_{i=1}^{N_{\rm s}}\mathbb{E}\big\{{\rm vec}(\RM{F}_i){\rm vec}(\RM{F}_i)^{\rm H}\big\}.
$$
Consider the eigendecomposition of $\widetilde{\M{\Psi}}_1$ as follows
\begin{equation}\label{eigen_choi}
\begin{aligned}
\widetilde{\M{\Psi}}_1 &= \M{U}^{(1)}\M{\Lambda}^{(1)}(\M{U}^{(1)})^{\rm H}\\
&=\sum_{i=1}^{r_1}\Big(\sqrt{\lambda_i^{(1)}}\V{u}_i^{(1)}\Big)\Big(\sqrt{\lambda_i^{(1)}}\V{u}_i^{(1)}\Big)^{\rm H}
\end{aligned}
\end{equation}
where
$$
\M{U}^{(1)}=[\V{u}_1^{(1)},~\dotsc,~\V{u}_{KM}^{(1)}],~\M{\Lambda}^{(1)}={\rm diag}(\lambda_1^{(1)},\dotsc,\lambda_{KM}^{(1)}).
$$
We assume that $\lambda_1^{(1)}\geq \lambda_2^{(1)} \geq \dotsc \geq \lambda_{KM}^{(1)}$, and the number of non-zero eigenvalues is denoted as $r_1$. Since there exist a linear one-to-one correspondence between every realization of $\RM{\Phi}_1$ and its Choi representation $\RM{\Psi}_1$, we see that $\widetilde{\M{\Psi}}_1=\mathbb{E}\{\RM{\Psi}_1\}$ is also the Choi representation of $\mathbb{E}\{\RM{\Phi}_1(\M{A})\}$, and thus
\begin{equation}
\mathbb{E}\{\RM{\Phi}_1(\M{A})\} = \sum_{i=1}^{r_1} \widetilde{\M{F}}_i\M{A}\widetilde{\M{F}}_i^{\rm H},
\end{equation}
where $r_1$ is the rank of $\widetilde{\M{\Psi}}_1$, and
\begin{equation}\label{kraus_operators_F}
\widetilde{\M{F}}_i = \sqrt{\lambda_i^{(1)}} {\rm mat}(\V{u}_i^{(1)}).
\end{equation}
Similar arguments also apply to $\RM{\Phi}_2(\cdot)$, and hence we have
\begin{equation}
\mathbb{E}\{\RM{\Phi}_2(\M{A})\} = \sum_{i=1}^{r_2}\widetilde{\M{G}}_i\M{A}\widetilde{\M{G}}_i^{\rm H},
\end{equation}
where
\begin{equation}\label{kraus_operators_G}
\widetilde{\M{G}}_i = \sqrt{\lambda_i^{(2)}} {\rm mat}(\V{u}_i^{(2)}),
\end{equation}
under the corresponding definitions of $\lambda_i^{(2)}$ and $\V{u}_i^{(2)}$. Consequently, we obtain
\begin{equation}
\RM{J}_{\RV{\eta}|\RM{X}} =  \frac{T}{\sigma_{\rm s}^2}\Big(\sum_{i=1}^{r_1}\widetilde{\M{F}}_i\RM{R}_{\RM{X}}^{\rm T} \widetilde{\M{F}}_i^{\rm H}+\sum_{j=1}^{r_2}\widetilde{\M{G}}_j\RM{R}_{\RM{X}} \widetilde{\M{G}}_j^{\rm H}\Big) + \M{J}_{\rm P},
\end{equation}
Finally, it is obvious that $r_1\leq KM$ and $r_2\leq KM$, since $\widetilde{\M{\Psi}}_1$ and $\widetilde{\M{\Psi}}_2$ are of the size $KM\times KM$. Hence the proof is completed.
\end{IEEEproof}

\section{Proof of Corollary \ref{coro:supermap}}\label{sec:proof_supermap}
\begin{IEEEproof}
First let us consider the following decomposition
\begin{equation}
\M{\Phi}_{\gamma}(\M{A}) = \M{\Phi}_1(\M{\Phi}_{\rm T}(\M{A})) +\M{\Phi}_2(\M{A})+ \M{\Phi}_{\rm P}(\gamma,\M{A}),
\end{equation}
where
\begin{subequations}
\begin{align}
\M{\Phi}_{\rm T}(\M{A})&=\M{A}^{\rm T} = \sum_{i=1}^M \sum_{j=1}^M \M{E}_{ij}\M{A}\M{E}_{ji}^{\rm H}, \label{transpose_map}\\
\M{\Phi}_1(\M{A})&=\sum_{i=1}^{r_1}\widetilde{\M{F}}_i\M{A}\widetilde{\M{F}}_i^{\rm H}, \label{phi_1_det} \\
\M{\Phi}_2(\M{A})&=\sum_{i=1}^{r_2}\widetilde{\M{G}}_i\M{A}\widetilde{\M{G}}_i^{\rm H}, \\
\M{\Phi}_{\rm P}(\gamma,\M{A})&= \frac{{\rm tr}\{\M{A}\}}{\gamma}\M{J}_{\rm P}.
\end{align}
\end{subequations}
From \eqref{transpose_map} and \eqref{phi_1_det}, we have
\begin{equation}\label{phi_1t}
\M{\Phi}_1(\M{\Phi}_{\rm T}(\M{A}))=\sum_{i=1}^{r_1} \sum_{j=1}^M \sum_{k=1}^M \widetilde{\M{F}}_i\M{E}_{jk}\M{A}\M{E}_{kj}^{\rm H}\widetilde{\M{F}}_i^{\rm H},
\end{equation}
which has the following Choi representation
\begin{equation}
\widetilde{\M{\Psi}}_{1{\rm T}} = \sum_{i=1}^{r_1} \sum_{j=1}^M \sum_{k=1}^M {\rm vec}(\widetilde{\M{F}}_i\M{E}_{jk}) {\rm vec}(\widetilde{\M{F}}_i\M{E}_{kj})^{\rm H}.
\end{equation}
As for the term $\M{\Phi}_{\rm P}(\gamma,\M{A})$, we note that its Choi representation is given by
\begin{equation}
\widetilde{\M{\Psi}}_{\rm P}(\gamma) = \frac{1}{\gamma} \M{I}\otimes \M{J}_{\rm P}.
\end{equation}
Now, we may write the Choi representation of $\M{\Phi}_{\gamma}(\M{A})$ as follows
\begin{equation}
\widetilde{\M{\Psi}}_{\gamma} = \widetilde{\M{\Psi}}_{1{\rm T}} + \widetilde{\M{\Psi}}_2 + \widetilde{\M{\Psi}}_{\rm P}(\gamma),
\end{equation}
where $\widetilde{\M{\Psi}}_2$ denotes the Choi representation of $\M{\Phi}_2(\M{A})$, given by
\begin{equation}
\widetilde{\M{\Psi}}_2 = \sum_{j=1}^{r_2}  {\rm vec}(\widetilde{\M{G}}_j) {\rm vec}(\widetilde{\M{G}}_j)^{\rm H}.
\end{equation}
Note that $\widetilde{\M{\Psi}}_{\gamma}$ is not Hermitian, since $\widetilde{\M{\Psi}}_{\rm 1T}$ and $\widetilde{\M{\Psi}}_{\rm P}$ are not. Therefore, the eigendecomposition in \eqref{eigen_choi} is no longer applicable, and we have to resort to the singular value decomposition of $\widetilde{\M{\Psi}}_{\gamma}$
\begin{equation}
\widetilde{\M{\Psi}}_{\gamma} = \M{U}_\gamma \M{\Sigma}_\gamma\M{V}_\gamma^{\rm H},
\end{equation}
where $\M{U}_\gamma=[\V{u}_1,~\dotsc,~\V{u}_{KM}]$, $\M{V}_\gamma=[\V{v}_1,~\dotsc,~\V{v}_{KM}]$, and $\M{\Sigma}_\gamma={\rm diag}(\sigma_1,\dotsc,\sigma_{KM})$. We may then obtain the representation in \eqref{asymmetric_representation}, where
\begin{equation}\label{bar_f_g}
\bar{\M{F}}_i = \sqrt{\sigma_i}{\rm mat}(\V{u}_i),~\bar{\M{G}}_i = \sqrt{\sigma_i}{\rm mat}(\V{v}_i).
\end{equation}
Finally, we have $r_3\leq KM$, since $\widetilde{\M{\Psi}}\in\mathbb{C}^{KM\times KM}$.
\end{IEEEproof}

\section{Proof of Proposition \ref{prop:max_icrb}}\label{sec:proof_max_icrb}
\begin{IEEEproof}
The problem of finding the distribution for the sensing-optimal $\RM{R}_{\RM{X}}$ may be formulated as
\begin{subequations}\label{variational_optimization}
\begin{align}
\min_{p(\RM{R}_{\RM{X}})} &~~\mathbb{E}_{\RM{R}_{\RM{X}}} \{\tr{[\M{\Phi}(\RM{R}_{\RM{X}})]^{-1}}\},\\
{\rm s.t.}&~~ \tr{\mathbb{E}(\RM{R}_{\RM{X}})}=P_{\rm T}M,~\RM{R}_{\RM{X}}\succeq \M{0},~\RM{R}_{\RM{X}}=\RM{R}_{\RM{X}}^{\rm H}, \label{variational_constraints}
\end{align}
\end{subequations}
Observe that
\begin{align}
&\min_{p(\RM{R}_{\RM{X}})} \mathbb{E}\{\tr{[\M{\Phi}(\RM{R}_{\RM{X}})]^{-1}}\} \nonumber \\
&\hspace{3mm} = \min_{p(\RM{R}_{\RM{X}})} \mathbb{E}\Big\{\mathbb{E}\big\{\tr{[\M{\Phi}_{\rv{\gamma}}(\RM{R}_{\RM{X}})]^{-1}}|\tr{\RM{R}_{\RM{X}}}=\rv{\gamma}\big\}\Big\} \nonumber \\
&\hspace{3mm} \geq \min_{p(\rv{\gamma})} \mathbb{E}\Big\{\min_{\tr{\RM{R}_{\RM{X}}} =\rv{\gamma}}\tr{[\M{\Phi}_{\rv{\gamma}}(\RM{R}_{\RM{X}})]^{-1}}\Big\}
\end{align}
holds under the constraints \eqref{variational_constraints}, when the following deterministic optimization problem
\begin{subequations}\label{deterministic_problem}
\begin{align}
\min_{\M{R}} &~~\tr{\Big(\M{\Phi}_{\gamma}(\M{R})\Big)^{-1}} \\
{\rm s.t.}&~~ \tr{\M{R}}=\gamma,~\M{R}\succeq \M{0},~\M{R}=\M{R}^{\rm H} \label{deterministic_constraints}
\end{align}
\end{subequations}
has valid optimal solutions for every positive $\gamma$. Furthermore, we have
\begin{equation}\label{jensen_gamma}
\mathbb{E} f(\rv{\gamma})\geq f(\mathbb{E}\{\rv{\gamma}\}),
\end{equation}
where
\begin{equation}\label{definition_f_gamma}
f(\rv{\gamma}):=\min_{\M{R}\succeq \M{0}} \tr{(\M{\Phi}_{\gamma}(\M{R}))^{-1}},~{\rm s.t.}~\tr{\M{R}}=\rv{\gamma},~\M{R}=\M{R}^{\rm H}.
\end{equation}
The result \eqref{jensen_gamma} follows from the following lemma.
\begin{lemma}
The function $f(\rv{\gamma})$ is convex.
\begin{IEEEproof}
Consider the optimal $\M{R}_1$ corresponding to $f(\gamma_1)$ and the optimal $\M{R}_2$ corresponding to $f(\gamma_2)$, which satisfy
\begin{equation}
\begin{aligned}
\alpha f(\gamma_1) + (1-\alpha) f(\gamma_2) \geq \tr{[\M{\Phi}(\alpha\M{R}_1+(1-\alpha)\M{R}_2)]^{-1}}
\end{aligned}
\end{equation}
for $\alpha\in[0,1]$, since $\tr{[\M{\Phi}_{\gamma}(\M{A})]^{-1}}$ is a convex function of $\M{A}$. Furthermore, we have
\begin{equation}
\tr{[\M{\Phi}_{\gamma}(\alpha\M{R}_1+(1-\alpha)\M{R}_2)]^{-1}} \geq f(\alpha \gamma_1+(1-\alpha)\gamma_2),
\end{equation}
which follows from the definition of $f(\rv{\gamma})$ in \eqref{definition_f_gamma}, and the fact that
$$
\tr{\alpha\M{R}_1+(1-\alpha)\M{R}_2} = \alpha\gamma_1+(1-\alpha)\gamma_2.
$$
Hence the proof is completed.
\end{IEEEproof}
\end{lemma}

The equality in \eqref{jensen_gamma} is achieved when $\rv{\gamma}$ is deterministic, namely when $\tr{\RM{R}_{\RM{X}}}=\tr{\widetilde{\M{R}}_{\RM{X}}}$. Due to the power constraint of $\tr{\widetilde{\M{R}}_{\RM{X}}}=P_{\rm T}M$, we see that the optimal distribution $p(\RM{R}_{\RM{X}})$ may now be simplified as $p(\RM{R}_{\RM{X}}|\tr{\RM{R}_{\RM{X}}}=P_{\rm T}M)$. Furthermore, $p(\RM{R}_{\RM{X}})$ should be defined over the set of optimal solutions of the problem \eqref{deterministic_problem} with $\gamma = P_{\rm T}M$. Thus the proof is completed.
\end{IEEEproof}

\section{Proof of Proposition \ref{prop:uniqueness}}\label{sec:proof_uniqueness}
\begin{IEEEproof}
We will first show the sufficiency, and then show the necessity.
\subsubsection{Sufficiency}
Since $\M{R}_{\rm opt}$ is a maximum-rank solution, we can express any optimal solution as follows
\begin{equation}\label{eigen_R}
\M{R}=\M{U}_{\rm opt}\M{D}\M{U}_{\rm opt}^{\rm H},
\end{equation}
where $\M{D}\in\mathbb{C}^{r \times r}$ is a Hermitian matrix. Otherwise, if we have any other solution $\M{R}^\prime$ that cannot be represented as \eqref{eigen_R}, then $(\M{R}_{\rm opt}+\M{R}^\prime)/2$ is also an optimal solution, but ${\rm rank}\{(\M{R}_{\rm opt}+\M{R}^\prime)/2\}>{\rm rank}\{\M{R}_{\rm opt}\}$, which contradicts to the assumption that $\M{R}_{\rm opt}$ is a maximum-rank optimal solution.

Next, note that the function $\tr{\M{X}^{-1}}$ is strictly convex with respect to $\M{X}$. Therefore, the optimal solution is unique, when the linear map from $\M{D}$ to $\M{\Phi}_{P_{\rm T}M}(\M{U}_{\rm opt}\M{D}\M{U}_{\rm opt}^{\rm H})$ is injective. Observe that
\begin{equation}
\begin{aligned}
&{\rm vec}(\M{\Phi}_{P_{\rm T}M}(\M{U}_{\rm opt}\M{D}\M{U}_{\rm opt}^{\rm H}))\\
&\hspace{3mm}= \M{\Xi}(\M{U}_{\rm opt}^*\otimes \M{U}_{\rm opt}){\rm vec}(\M{D}) + {\rm vec}(\widetilde{\M{J}}_{\rm P}),
\end{aligned}
\end{equation}
hence the map is indeed injective if $\M{\Xi}(\M{U}_{\rm opt}^*\otimes \M{U}_{\rm opt})$ has full column rank.
\subsubsection{Necessity}
If $\M{R}_{\rm opt}$ is the unique optimal solution, there does not exist any $\M{R}$ taking the form of \eqref{eigen_R} corresponding to a zero-trace $\M{D}$ that satisfies
\begin{equation}
\M{\Xi}{\rm vec}(\M{R}) = \V{0}.
\end{equation}
Otherwise, one may construct another optimal solution $\M{R}^\prime=\M{R}_{\rm opt}+\alpha\M{R}$, where $\alpha\in\mathbb{R}$ is chosen to ensure that $\M{R}^\prime$ is positive semidefinite. This implies that the subspace
$$
\Set{V}\!=\!\{(\M{U}_{\rm opt}^*\otimes \M{U}_{\rm opt}){\rm vec}(\M{D})|\M{D}\in\mathbb{C}^{r\times r},\M{D}\!=\!\M{D}^{\rm H},\tr{\M{D}}\!=\!0\}
$$
is orthogonal to the null space of $\M{\Xi}$. Moreover, we see that ${\rm dim}(\Set{V})=r^2-1$, since the complex vector ${\rm vec}(\M{D})$ is linearly independent of its complex conjugate ${\rm vec}(\M{D})^*$, and the zero-trace constraint $\tr{\M{D}}=0$ only reduces the dimensionality by $1$. Furthermore, ${\rm vec}(\M{I})$ is never in the null space of $\M{\Xi}$ (otherwise we would have $\M{\Phi}(\M{I})=\M{0}$ implying that $\RM{J}_{\RV{\eta}|\RM{X}}=\M{0}$ for all $\RM{R}_{\RM{X}}$), hence the column rank of $\M{\Xi}(\M{U}_{\rm opt}^*\otimes \M{U}_{\rm opt})$ is $r^2$, which is exactly the full column rank.
\end{IEEEproof}

\section{Proof of Proposition \ref{prop:sufficient}}\label{sec:proof_sufficient}
\begin{IEEEproof}
\subsubsection{(Generic)}
According to Proposition \ref{prop:uniqueness}, to prove the sufficiency of condition 1), it suffices to show that
\begin{equation}\label{subspace_contain}
{\rm Span}\{\M{U}_{\rm opt}\}\subseteq {\rm Span}\{\M{U}_{\rm R}\}.
\end{equation}
Let us first rewrite problem \eqref{opt_problem} as follows
\begin{align}\label{opt_problem_transform}
\min_{\M{A},\widetilde{\M{R}}_{\RM{X}}} &~~\tr{\M{A}^{-1}}, \nonumber \\
{\rm s.t.}&~~\M{A}=\M{\Phi}_{P_{\rm T}M}(\widetilde{\M{R}}_{\RM{X}}),\nonumber \\
&~~\tr{\widetilde{\M{R}}_{\RM{X}}}=P_{\rm T}M,~\widetilde{\M{R}}_{\RM{X}}\succeq \M{0},~\widetilde{\M{R}}_{\RM{X}}=\widetilde{\M{R}}_{\RM{X}}^{\rm H}.
\end{align}
The \ac{kkt} conditions of problem \eqref{opt_problem_transform} can be expressed as
\begin{subequations}
\begin{align}
&\M{\Phi}_{P_{\rm T}M}^{\rm a}(\M{A}^{-2})=\lambda\M{I}-\M{Z}_{\rm R}, \label{lagrangian_gradient}\\
&\M{\Phi}_{P_{\rm T}M}(\widetilde{\M{R}}_{\RM{X}})=\M{A},~\tr{\widetilde{\M{R}}_{\RM{X}}}=P_{\rm T}M,~\widetilde{\M{R}}_{\RM{X}}\succeq \M{0},\\
&\tr{\M{Z}_{\rm R}\widetilde{\M{R}}_{\RM{X}}}=0, \label{complementary_slackness}\\
&\M{Z}_{\rm R}\succeq \M{0} \label{dual_feasibility}.
\end{align}
\end{subequations}
From \eqref{lagrangian_gradient}, \eqref{complementary_slackness} and \eqref{dual_feasibility}, we see that
\begin{equation}\label{maximum_eigenspace}
\M{\Phi}_{P_{\rm T}M}^{\rm a}(\M{A}^{-2})\widetilde{\M{R}}_{\RM{X}} = \lambda \widetilde{\M{R}}_{\RM{X}},
\end{equation}
and that $\lambda$ is the maximum eigenvalue of $\M{\Phi}_{P_{\rm T}M}^{\rm a}(\M{A}^{-2})$. Applying \eqref{maximum_eigenspace} to the maximal-rank solution $\M{R}_{\rm opt}$, we have
\begin{equation}\label{maximum_eigenspace2}
\M{U}_{\rm opt}^{\rm H}\M{\Phi}_{P_{\rm T}M}^{\rm a}(\M{A}^{-2})\M{U}_{\rm opt} = \lambda\M{I}.
\end{equation}
From \eqref{maximum_eigenspace2}, we see that it now suffices to show that
\begin{equation}\label{condition_a}
\M{U}_{\rm R}^{\rm H} \M{\Phi}_{P_{\rm T}M}^{\rm a}(\M{A}^{-2})\M{U}_{\rm R} = \lambda \M{I}_{r_{\rm A}},
\end{equation}
where $r_{\rm A}$ is the multiplicity of the maximum eigenvalue of $\M{\Phi}_{P_{\rm T}M}^{\rm a}(\M{A}^{-2})$. To this end, we write the Lagrange dual function of problem \eqref{opt_problem_transform} as follows
$$
\begin{aligned}
&g(\tilde{\lambda},\M{Z}_{\rm A},\M{Z}_{\rm R}) \\
&\hspace{3mm}= \inf_{\M{A},\widetilde{\M{R}}_{\RM{X}}} \Big\{\tr{\M{A}^{-1}} +\tr{\M{Z}_{\rm A}\M{A}} -\tr{\M{Z}_{\rm R}\widetilde{\M{R}}_{\RM{X}}} \\
&\hspace{10mm}-\tr{\M{Z}_{\rm A}\M{\Phi}_{P_{\rm T}M}(\widetilde{\M{R}}_{\RM{X}})}+\tilde{\lambda}\Big(\tr{\widetilde{\M{R}}_{\RM{X}}}-P_{\rm T}M\Big)\Big\}\\
&\hspace{3mm}= \inf_{\M{A},\widetilde{\M{R}}_{\RM{X}}} \Big\{\tr{\M{A}^{-1}} +\tr{\M{Z}_{\rm A}\M{A}}-\tilde{\lambda} P_{\rm T}M \\
&\hspace{10mm}+\tr{\widetilde{\M{R}}_{\RM{X}}\Big(-\M{Z}_{\rm R}-\M{\Phi}_{P_{\rm T}M}^{\rm a}(\M{Z}_{\rm A})+\tilde{\lambda}\M{I}\Big)}\Big\}\\
&\hspace{3mm}=\left\{
                \begin{array}{ll}
                  2\tr{\M{Z}_{\rm A}^{\frac{1}{2}}}-\tilde{\lambda} P_{\rm T}M, & \hbox{$\M{\Phi}_{P_{\rm T}M}^{\rm a}(\M{Z}_{\rm A})=\tilde{\lambda}\M{I}-\M{Z}_{\rm R}$;} \\
                  -\infty, & \hbox{otherwise.}
                \end{array}
              \right.
\end{aligned}
$$
Hence the dual problem of \eqref{opt_problem_transform} can be written as
\begin{align}
\max_{\tilde{\lambda},\M{Z}_{\rm A}}&~~2\tr{\M{Z}_{\rm A}^{\frac{1}{2}}}-\tilde{\lambda} P_{\rm T}M \nonumber \\
{\rm s.t.}&~~ \M{\Phi}_{P_{\rm T}M}^{\rm a}(\M{Z}_{\rm A})-\tilde{\lambda}\M{I}\preceq \M{0},~\M{Z}_{\rm A}\succeq \M{0},~\M{Z}_{\rm A}=\M{Z}_{\rm A}^{\rm H},
\end{align}
which is equivalent to \eqref{dual_problem} after a maximization over $\tilde{\lambda}$, while $\lambda$ is the optimal value of $\tilde{\lambda}$. Obviously, strong duality holds for the primal-dual pair \eqref{opt_problem_transform} and \eqref{dual_problem}, which implies that
\begin{equation}
\M{\Phi}_{P_{\rm T}M}^{\rm a}(\M{Z}_{\rm A})=\M{\Phi}_{P_{\rm T}M}^{\rm a}(\M{A}^{-2}),
\end{equation}
if $\M{A}$ is an optimal solution of \eqref{opt_problem_transform} and $\M{Z}_{\rm A}$ is an optimal solution of \eqref{dual_problem}. Since $\lambda$ is the maximum eigenvalue of $\M{\Phi}_{P_{\rm T}M}^{\rm a}(\M{Z}_{\rm A})$, we have
\begin{equation}
\M{U}_{\rm R}^{\rm H} \M{\Phi}_{P_{\rm T}M}^{\rm a}(\M{Z}_{\rm A})\M{U}_{\rm R} = \lambda \M{I}_{r_{\rm A}},
\end{equation}
which implies \eqref{condition_a}.

\subsubsection{($K\geq M$)}
Note that $\M{\Xi}\in\mathbb{C}^{K^2\times M^2}$, therefore, it is indeed possible that $\M{\Xi}$ has full column rank when $K\geq M$. When this is true, we also have that
\begin{equation}
{\rm rank}\{\M{\Xi}(\M{U}_{\rm R}^*\otimes \M{U}_{\rm R})\} =r^2,
\end{equation}
which implies that ${\rm rank}\{\M{\Xi}(\M{U}_{\rm R}^*\otimes \M{U}_{\rm R})\}$ has full column rank, since $r^2 \leq M^2 \leq K^2$. Thus we can conclude that problem \eqref{opt_problem} has a unique optimal solution according to Proposition \ref{prop:uniqueness}.

\subsubsection{($K=1$)}
When $K=1$, the optimal $\M{Z}_{\rm A}$ of the dual problem \eqref{dual_problem} is a scalar, denoted as $z$. Furthermore, $\M{\Xi}$ is now a row-vector, denoted as $\V{\xi}^{\rm T}$. Hence the generic sufficient condition for solution-uniqueness becomes
\begin{equation}\label{no_null_k1}
(\M{U}_{\rm R}^*\otimes \M{U}_{\rm R})^{\rm H} \V{\xi} \neq 0,
\end{equation}
where $\M{U}_{\rm R}$ contains the eigenvectors corresponding to the maximum eigenvalue of $\M{\Phi}_{P_{\rm T}M}^{\rm a}(z)$, which can now be expressed as (according to \eqref{adjoint_phi})
\begin{equation}
\M{\Phi}_{P_{\rm T}M}^{\rm a}(z) = z\sum_{i=1}^{r_3} \bar{\M{G}}_i^{\rm H}\bar{\M{F}}_i.
\end{equation}
Furthermore, since $\M{J}_{\rm P}$ is now a scalar (denoted as $J_{\rm P}$), the Choi representation of $\M{\Phi}_{\rm P}(\gamma,\M{A})$ is now given by
\begin{equation}
\widetilde{\M{\Psi}}_{\rm P}(\gamma) = \frac{J_{\rm P}}{\gamma}\M{I},
\end{equation}
which has no impact on the eigenspace structure of $\widetilde{\M{\Psi}}_{\gamma}$. As for the transpose map $\M{\Phi}_{\rm T}(\cdot)$, we now have
\begin{equation}
\M{\Phi}_1(\M{\Phi}_{\rm T}(\M{A}))=\sum_{i=1}^{r_1} \widetilde{\M{F}}_i^*\M{A}\widetilde{\M{F}}_i^{\rm T},
\end{equation}
according to \eqref{phi_1t}, since $\widetilde{\M{F}}_i$'s are now row-vectors. Hence we obtain
\begin{align}
\M{\Phi}_{P_{\rm T}M}^{\rm a}(z) &= k\Big(\sum_{i=1}^{r_1} (\widetilde{\M{F}}_i^{\rm H} \widetilde{\M{F}}_i)^{\rm T} + \sum_{j=1}^{r_2} \widetilde{\M{G}}_j^{\rm H} \widetilde{\M{G}}_j\Big) \nonumber \\
&=k\M{B}_1,
\end{align}
where $k$ is a positive constant with respect to $\M{B}_1$. Finally, let us consider the case when $\M{U}_{\rm R}$ has a single column, which implies that the eigenspace corresponding to the maximum eigenvalue of $\M{\Phi}_{P_{\rm T}M}^{\rm a}(z)$ has dimensionality $1$. In this case, \eqref{no_null_k1} always holds, otherwise the optimal solution of \eqref{opt_problem} would be unbounded, since $\M{\Phi}_{P_{\rm T}M}(\widetilde{\M{R}}_{\RM{X}})$ is not invertible. Thus the proof is completed.
\end{IEEEproof}

\section{Proof of Corollary \ref{coro:psc_rank}}\label{sec:proof_psc_rank}
\begin{IEEEproof}
It is straightforward that ${\rm rank}(\widetilde{\M{R}}_{\RM{X}}) \leq M$ when $K\geq M$. Hence it suffices to consider the case of $K<M$ only. Consider the eigendecomposition of the optimal $\widetilde{\M{R}}_{\RM{X}}$ as follows
\begin{equation}
\widetilde{\M{R}}_{\RM{X}} = \M{U}_{\rm opt}\M{\Lambda}_{\rm opt}\M{U}_{\rm opt}^{\rm H}.
\end{equation}
According to Appendix \ref{sec:proof_uniqueness}, when the optimal solution of \eqref{opt_problem} is unique, the column rank of $\M{\Xi}(\M{U}_{\rm opt}^*\otimes \M{U}_{\rm opt})$ is $({\rm rank}(\widetilde{\M{R}}_{\RM{X}}))^2$, which implies that
\begin{equation}
r^2\leq K^2,
\end{equation}
since $\M{\Xi}(\M{U}_{\rm opt}^*\otimes \M{U}_{\rm opt})\in\mathbb{C}^{K^2\times ({\rm rank}(\widetilde{\M{R}}_{\RM{X}}))^2}$.
\end{IEEEproof}

\section{Proof of Theorem \ref{thm:sensing_limited}}\label{sec:proof_sensing_limited}
\begin{IEEEproof}
Under the aforementioned assumptions, rate $R_{\rm SC}$ may be expressed as
\begin{equation}
R_{\rm SC} \!=\! \max_{p_{\RM{X}}(\M{X})} \frac{1}{T}I(\RM{Y}_{\rm c};\RM{X}|\RM{H}_{\rm c}),~{\rm s.t.}~\RM{X}\RM{X}^{\rm H} \!=\! T\widetilde{\M{R}}_{\RM{X}}^{\rm SC}.
\end{equation}
Consider the singular value decomposition of $\RM{H}_{\rm c}\RM{X}$:
\begin{equation}
\RM{H}_{\rm c}\RM{X} = \sqrt{T}\RM{U}\RM{\Sigma}\RM{V}_{\rm HX}^{\rm H}.
\end{equation}
Note that only $\RM{V}_{\rm HX}^{\rm H}$ is not yet determined, since we have
$$
\begin{aligned}
\RM{H}_{\rm c}\widetilde{\M{R}}_{\RM{X}}^{\rm SC}\RM{H}_{\rm c}^{\rm H}&=T^{-1}\RM{H}_{\rm c}\RM{X}\RM{X}^{\rm H} \RM{H}_{\rm c}^{\rm H} \\
&=\RM{U}\RM{\Sigma}\RM{\Sigma}^{\rm H}\RM{U}^{\rm H},
\end{aligned}
$$
which does not depend on $\RM{V}_{\rm HX}^{\rm H}$. Let us denote the first $\rv{M}_{\rm SC}$ columns of $\RM{U}$ corresponding to the non-zero singular values as $\widetilde{\RM{U}}$. Without loss of generality, we may apply a linear combiner $\widetilde{\RM{U}}^{\rm H}$ at the receiver side, and obtain
\begin{equation}\label{sufficient_statistic}
\widetilde{\RM{U}}^{\rm H}\RM{Y}_{\rm c} = \sqrt{T}\widetilde{\RM{\Sigma}}\widetilde{\RM{V}}_{\rm HX}^{\rm H} + \widetilde{\RM{U}}^{\rm H}\RM{Z}_{\rm c},
\end{equation}
where $\widetilde{\RM{U}}^{\rm H}\RM{Z}_{\rm c}\in\mathbb{C}^{\rv{M}_{\rm SC}\times T}$ has \ac{iid} entries which are circularly symmetric complex Gaussian distributed, $\widetilde{\RM{\Sigma}}\in\mathbb{C}^{\rv{M}_{\rm SC}\times \rv{M}_{\rm SC}}$  denotes the $\rv{M}_{\rm SC}\times \rv{M}_{\rm SC}$ submatrix  containing non-zero singular values, and $\widetilde{\RM{V}}_{\rm HX}$ contains the first $\rv{M}_{\rm SC}$ columns of $\RM{V}_{\rm HX}$. Note that $\widetilde{\RM{U}}^{\rm H}\RM{Y}_{\rm c}$ is a sufficient statistic of $\RM{Y}_{\rm c}$ for the estimation of $\RM{X}$, hence the mutual information $I(\RM{Y}_{\rm c}|\RM{H}_{\rm c};\RM{X})$ may be expressed as
\begin{align}
I(\RM{Y}_{\rm c};\RM{X}|\RM{H}_{\rm c}) &= I(\widetilde{\RM{U}}^{\rm H}\RM{Y}_{\rm c};\RM{X}|\RM{H}_{\rm c}) \nonumber \\
&=h(\widetilde{\RM{U}}^{\rm H}\RM{Y}_{\rm c}|\RM{H}_{\rm c}) - h(\widetilde{\RM{U}}^{\rm H}\RM{Y}_{\rm c}|\RM{H}_{\rm c},\RM{X}).
\end{align}
Upon denoting $\widetilde{\RM{Y}}_{\rm c}=\widetilde{\RM{U}}^{\rm H}\RM{Y}_{\rm c}$, $ \widetilde{\RM{X}}=\sqrt{T}\widetilde{\RM{\Sigma}}\widetilde{\RM{V}}_{\rm HX}^{\rm H}$, and $\widetilde{\RM{Z}}_{\rm c}=\widetilde{\RM{U}}^{\rm H}\RM{Z}_{\rm c}$, from \eqref{sufficient_statistic} we have
\begin{equation}\label{observation_algebra}
\widetilde{\RM{Y}}_{\rm c} = \widetilde{\RM{X}}+\widetilde{\RM{Z}}_{\rm c},
\end{equation}
and hence
\begin{subequations}
\begin{align}
I(\RM{Y}_{\rm c};\RM{X}|\RM{H}_{\rm c}) &= I(\widetilde{\RM{Y}}_{\rm c};\RM{X}|\RM{H}_{\rm c}) \\
&=h(\widetilde{\RM{Y}}_{\rm c}|\RM{H}_{\rm c})-h(\widetilde{\RM{Y}}_{\rm c}|\RM{H}_{\rm c},\RM{X}). \label{upper_direction}
\end{align}
\end{subequations}
From \eqref{upper_direction} we obtain
\begin{align}\label{upper_bound_mi}
&\max_{p_{\RM{X}}(\M{X})}I(\RM{Y}_{\rm c};\RM{X}|\RM{H}_{\rm c}) \nonumber \\
&\hspace{3mm}= \max_{p_{\RM{X}}(\M{X})}\Big(h(\widetilde{\RM{Y}}_{\rm c}|\RM{H}_{\rm c})-h(\widetilde{\RM{Y}}_{\rm c}|\RM{H}_{\rm c},\RM{X})\Big) \nonumber \\
&\hspace{3mm}=\max_{p_{\RM{X}}(\M{X})}h(\widetilde{\RM{Y}}_{\rm c}|\RM{H}_{\rm c}) - T\rv{M}_{\rm SC}\log(\pi e \sigma_{\rm c}^2).
\end{align}
Next, we provide estimates of $h(\widetilde{\RM{Y}}_{\rm c}|\RM{H}_{\rm c})$ based on the following lemma.
\begin{lemma}\label{lem:log_volume}
The logarithmic volume of the rescaled complex Stiefel manifold
\begin{align}\label{rescaled_stiefel}
\RS{S}&=\big\{\sqrt{T}\widetilde{\RM{\Sigma}}\M{Q}|\M{Q}\in\mathbb{C}^{\rv{M}_{\rm SC}\times T},~\M{Q}\M{Q}^{\rm H}=\M{I}_{\rv{M}_{\rm SC}}\big\} \nonumber \\
&=\big\{\widetilde{\M{Q}}\in\mathbb{C}^{\rv{M}_{\rm SC}\times T}|\widetilde{\M{Q}}\widetilde{\M{Q}}^{\rm H}=T\widetilde{\RM{\Sigma}}^2\big\}
\end{align}
is given by
\begin{equation}\label{rescaled_volume}
\log {\rm Vol}(\RS{S})=\Big(T-\frac{1}{2}\rv{M}_{\rm SC}\Big)\log \Big|T\widetilde{\RM{\Sigma}}^2\Big|+\log V_{T,\rv{M}_{\rm SC}},
\end{equation}
where $V_{T,\rv{M}_{\rm SC}}$ denotes the volume of the complex Stiefel manifold (without rescaling), given by
\begin{equation}\label{stiefel_classical}
V_{T,\rv{M}_{\rm SC}} = \prod_{k=T-\rv{M}_{\rm SC}+1}^T \frac{2\pi^k}{(k-1)!}.
\end{equation}
The result \eqref{stiefel_classical} is known in \cite{noncoherent_lizhong}.
\begin{IEEEproof}
Please refer to Appendix \ref{sec:proof_log_volume}.
\end{IEEEproof}
\end{lemma}

Now, let us consider the following set
\begin{equation}
\RS{S}_{\epsilon} = \Big\{\M{A}+\epsilon\M{B}| \M{A}\in\RS{S},~\|\M{B}\|_{\rm F} \leq 1\Big\},
\end{equation}
which is known as the $\epsilon$-\emph{tube} about the manifold $\RS{S}$ \cite{tube}. According to Theorem 9.23 in \cite{tube}, for a small $\epsilon$, the volume of the $\epsilon$-tube can be approximated as follows
\begin{equation}
{\rm Vol}(\RS{S}_{\epsilon})=\frac{(\pi \epsilon^2)^{\frac{1}{2}\rv{M}_{\rm SC}^2}}{\Gamma(\frac{1}{2}\rv{M}_{\rm SC}^2+1)}{\rm Vol}(\RS{S})(1+O(\epsilon^2)).
\end{equation}
Hence we have
\begin{equation}
\begin{aligned}
\log \frac{\partial{\rm Vol}(\RS{S}_{\epsilon})}{\partial \epsilon} &= \log {\rm Vol}(\RS{S})-\log\Gamma\Big(\frac{1}{2}\rv{M}_{\rm SC}^2+1\Big)\\
&\hspace{3mm}+\frac{1}{2}\rv{M}_{\rm SC}^2\log\pi +\frac{1}{2}(\rv{M}_{\rm SC}^2-1)\log \epsilon^2 \\
&\hspace{3mm}+\log \rv{M}_{\rm SC}^2 + O(\epsilon^2).
\end{aligned}
\end{equation}
According to \eqref{observation_algebra}, the maximum differential entropy can be expressed as
$$
\max_{p_{\RM{X}}(\M{X})}~h(\RM{Y}_{\rm c}|\RM{H}_{\rm c})=\underbrace{\int \log \frac{\partial{\rm Vol}(\RS{S}_{\epsilon})}{\partial \epsilon} f(\epsilon){\rm d}\epsilon}_{h(\RM{Y}_{\rm c}|\RM{H}_{\rm c},\rv{\epsilon})} + h(\rv{\epsilon}),
$$
where $\rv{\epsilon}$ is a random variable such that $\rv{x} = 2\rv{\epsilon}^2\sigma_{\rm c}^{-2}$ is a chi-squared distribution having \ac{dof} $\rv{M}_{\rm SC}^2$,\footnote{The random variable $\rv{\epsilon}$ represents $\|\widetilde{\RM{Z}}_{\rm c}^{\perp}\|_{\rm F}$, where $\widetilde{\RM{Z}}_{\rm c}^{\perp}$ is the component of $\widetilde{\RM{Z}}_{\rm c}$ being orthogonal to the tangent space of $\RS{S}$ at $\RM{X}$. The dimensionality of $\widetilde{\RM{Z}}_{\rm c}^{\perp}$ is $\rv{M}_{\rm SC}^2$.} whose probability density function is given by
$$
f(x) = \frac{1}{2^{\rv{M}_{\rm SC}^2/2}\Gamma(\rv{M}_{\rm SC}^2/2)}x^{\rv{M}_{\rm SC}^2/2-1}e^{-x/2}.
$$
 We may now express the maximum differential entropy $h(\widetilde{\RM{Y}}_{\rm c}|\RM{H}_{\rm c})$ as follows
\begin{align}
&\max_{p_{\RM{X}}(\M{X})} ~h(\widetilde{\RM{Y}}_{\rm c}|\RM{H}_{\rm c})= \int \log \frac{\partial{\rm Vol}(\RS{S}_{\epsilon})}{\partial \epsilon} f(\epsilon){\rm d}\epsilon+h(\rv{\epsilon})\nonumber \\
&\hspace{2mm}=\log {\rm Vol}(\RS{S})\!-\!\log\Gamma\Big(\frac{1}{2}\rv{M}_{\rm SC}^2\!+\!1\Big)\!+\!\frac{1}{2}\rv{M}_{\rm SC}^2\log \pi\!+\!\log\rv{M}_{\rm SC}^2 \nonumber\\
&\hspace{6mm}+\frac{1}{2}(\rv{M}_{\rm SC}^2-1)\Big(\mathbb{E}\{\log \rv{x}\}+\log \frac{\sigma_{\rm c}^2}{2}\Big)+h\big(\sqrt{\rv{x}}\big) \nonumber \\
&\hspace{6mm}+\frac{1}{2}\log (\sigma_{\rm c}^2/2)+O(\sigma_{\rm c}^2)\nonumber \\
&\hspace{2mm}= \Big(T-\frac{1}{2}\rv{M}_{\rm SC}\Big)\log \Big|T\widetilde{\RM{\Sigma}}^2\Big|+\frac{1}{2}\rv{M}_{\rm SC}^2\log\pi + \log V_{T,\rv{M}_{\rm SC}} \nonumber\\
&\hspace{4mm} -\log \Gamma\Big(\frac{1}{2}\rv{M}_{\rm SC}^2\!+\!1\Big)\!+\!\frac{1}{2}(\rv{M}_{\rm SC}^2\!-\!1)\int f(x) \log (x\sigma_{\rm c}^2/2){\rm d}x  \nonumber\\
&\hspace{4mm} +\frac{1}{2}\Big([\rv{M}_{\rm SC}^2\!-\!(\rv{M}_{\rm SC}^2\!-\!1)\psi(\rv{M}_{\rm SC}^2/2)]\log e \!+\!\log (\sigma_{\rm c}^2/2)\Big)\nonumber\\
&\hspace{4mm}+ \log\Gamma(\rv{M}_{\rm SC}^2/2)+\log\rv{M}_{\rm SC}^2 -\frac{1}{2}\log 2+ O(\sigma_{\rm c}^2),
\end{align}
where
$$
\psi(x)=-\gamma-\frac{1}{x}+\sum_{n=1}^\infty \left(\frac{1}{n}-\frac{1}{n+x}\right)
$$
is the digamma function \cite{tables}, with $\gamma=0.572\dotsc$ being the Euler-Mascheroni constant. Note that
\begin{equation}
\int f(x) \log x{\rm d}x = \psi(\rv{M}_{\rm SC}^2/2)\log e+\log 2,
\end{equation}
and thus we obtain
\begin{align}\label{upper_entropy}
\max_{p_{\RM{X}}(\M{X})}~h(\widetilde{\RM{Y}}_{\rm c}|\RM{H}_{\rm c})&= \Big(T-\frac{1}{2}\rv{M}_{\rm SC}\Big)\log \Big|T\widetilde{\RM{\Sigma}}^2\Big|+\log V_{T,\rv{M}_{\rm SC}} \nonumber \\
&\hspace{3mm}+\frac{1}{2}\rv{M}_{\rm SC}^2\log(\pi e \sigma_{\rm c}^2)+O(\sigma_{\rm c}^2),
\end{align}
after some simplification, which implies that
\begin{align}
&\max_{p_{\RM{X}}(\M{X})}I(\RM{Y}_{\rm c};\RM{X}|\RM{H}_{\rm c})\\
&\hspace{3mm}= \mathbb{E}\Big\{\Big(T-\frac{1}{2}\rv{M}_{\rm SC}\Big)\log\Big|\sigma_{\rm c}^{-2}\widetilde{\RM{\Sigma}}^2\Big|\Big\}+\log V_{T,\rv{M}_{\rm SC}} \nonumber \\
&\hspace{5mm} + \mathbb{E}\Big\{\rv{M}_{\rm SC}\Big(T-\frac{1}{2}\rv{M}_{\rm SC}\Big)\log \Big(\frac{T}{\pi e}\Big) \Big\}+ O(\sigma_{\rm c}^2).
\end{align}
Next, we will show that the term
\begin{equation}
\rv{c}_1:=\log V_{T,\rv{M}_{\rm SC}}+\rv{M}_{\rm SC}\Big(T-\frac{1}{2}\rv{M}_{\rm SC}\Big)\log \Big(\frac{T}{\pi e}\Big)
\end{equation}
is on the order of $o(T)$ as $T\rightarrow \infty$. To see this, we first note that
\begin{equation}
\log V_{T,\rv{M}_{\rm SC}} = \sum_{i=1}^{\rv{M}_{\rm SC}} \log V_{T-i+1,1},
\end{equation}
which implies that
\begin{equation}
\begin{aligned}
\log V_{T,\rv{M}_{\rm SC}}&=\rv{M}_{\rm SC}\log 2+\rv{M}_{\rm SC}\Big(T-\frac{1}{2}\rv{M}_{\rm SC}\Big)\log \pi \\
&\hspace{3mm}+ \frac{1}{2}\rv{M}_{\rm SC}\log \pi -  \sum_{i=1}^{\rv{M}_{\rm SC}}\sum_{j=1}^{T-1} \log j,
\end{aligned}
\end{equation}
and hence
\begin{align}
\rv{c}_1&=\rv{M}_{\rm SC}\Big(T-\frac{1}{2}\rv{M}_{\rm SC}\Big)\log T/e -  \sum_{i=1}^{\rv{M}_{\rm SC}}\sum_{j=1}^{T-1} \log j\\
&\hspace{4mm}+\rv{M}_{\rm SC}\log (2\sqrt{\pi}) \nonumber\\
&=\rv{M}_{\rm SC}\Big[\Big(T-\frac{\rv{M}_{\rm SC}}{2}\Big)\log T/e - \sum_{j=1}^{T-1} \log j+\log(2\sqrt{\pi})\Big].
\end{align}
Now we have
\begin{align}\label{approx_error}
\frac{\rv{c}_1}{T} &= \rv{M}_{\rm SC} \Big[\Big(1-\frac{\rv{M}_{\rm SC}}{2T}\Big)\log T/e - \frac{1}{T}\sum_{j=1}^{T-1} \log j+\frac{\log(2\sqrt{\pi})}{T}\Big] \nonumber\\
&\rightarrow \rv{M}_{\rm SC} \Big[\log T/e - \frac{1}{T}\log \Gamma(T)\Big] \nonumber\\
&\rightarrow 0,
\end{align}
as $T\rightarrow \infty$, where the last line follows from the fact that
\begin{equation}
\begin{aligned}
\lim_{T\rightarrow \infty} \ln T - \frac{1}{T}\ln \Gamma(T)&=\lim_{T\rightarrow \infty}1+\ln T -\psi(T) \\
&=1,
\end{aligned}
\end{equation}
Thus we arrive at
\begin{equation}
\begin{aligned}
R_{\rm SC} &= \mathbb{E}\Big\{\Big(1-\frac{\rv{M}_{\rm SC}}{2T}\Big)\log \Big|\sigma_{\rm c}^{-2}\widetilde{\RM{\Sigma}}^2\Big| -\frac{\rv{M}_{\rm SC}}{T}\log \Gamma(T)\\
&\hspace{3mm}+\rv{M}_{\rm SC}\Big[\Big(1-\frac{\rv{M}_{\rm SC}}{2T}\Big)\log\frac{T}{e}+\frac{\log(2\sqrt{\pi})}{T}\Big]\Big\} + O(\sigma_{\rm c}^2).
\end{aligned}
\end{equation}
Hence the proof is completed.
\end{IEEEproof}

\section{Proof of Lemma \ref{lem:log_volume}}\label{sec:proof_log_volume}
\begin{IEEEproof}
Let us first consider the Riemannian metric tensor on $\RS{S}$, which can be written in a matrix form in the tangent space $\Set{T}_{\widetilde{\M{Q}}} \RS{S}$ at each point $\widetilde{\M{Q}}\in\RS{S}$, denoted as $\M{G}_{\widetilde{\M{Q}}}$. The volume of $\Set{V}$ can then be computed as \cite[Sec. 1.2]{riemannian_geometry}
\begin{equation}
{\rm Vol}(\RS{S}) = \int_{\RS{S}} \sqrt{|\M{G}_{\widetilde{\M{Q}}}|} ({\rm d}\widetilde{\M{Q}})^{\wedge},
\end{equation}
where ${\rm d}\widetilde{\M{Q}}$ can be viewed as a tangent vector in the tangent space $\Set{T}_{\widetilde{\M{Q}}} \RS{S}$, the term $\sqrt{|\M{G}_{\widetilde{\M{Q}}}|} ({\rm d}\widetilde{\M{Q}})^{\wedge}$ is a differential form known as the volume form \cite[Sec. 1.2]{riemannian_geometry} on the manifold $\RS{S}$ induced by its Riemannian metric $\M{G}_{\widetilde{\M{Q}}}$, and $({\rm d}\widetilde{\M{Q}})^{\wedge}$ is the exterior product over all components in ${\rm d}\widetilde{\M{Q}}$, which serves as a volume form on the tangent space (as a Euclidean space). According to \cite[Sec. 1.2]{riemannian_geometry}, \eqref{rescaled_stiefel} may be viewed as an alternative parametrization of the original Stiefel manifold
\begin{equation}
\Set{V} = \big\{\M{Q}\in\mathbb{C}^{\rv{M}_{\rm SC}\times T}|\M{Q}\M{Q}^{\rm H}=\M{I}_{\rv{M}_{\rm SC}}\big\}.
\end{equation}
Upon denoting the Riemannian metric of $\Set{V}$ at $\M{Q}$ as $\widetilde{\M{G}}_{\M{Q}}$, we have
\begin{equation}
{\rm Vol}(\Set{V}) = \int_{\Set{V}} \sqrt{|\widetilde{\M{G}}_{\M{Q}}|} ({\rm d}\M{Q})^{\wedge},
\end{equation}
and hence
\begin{align}
{\rm Vol}(\RS{S})&=\int_{\RS{S}} \sqrt{|\M{G}_{\widetilde{\M{Q}}}|} ({\rm d}\widetilde{\M{Q}})^{\wedge} \nonumber \\
&=\int_{\Set{V}} \big|\M{J}_{\M{Q},\widetilde{\M{Q}}}\big|\sqrt{|\widetilde{\M{G}}_{\M{Q}}|} ({\rm d}\M{Q})^{\wedge},
\end{align}
where $|\M{J}_{\M{Q},\widetilde{\M{Q}}}|$ is the Jacobian determinant of the transformation from $\M{Q}$ on $\Set{V}$ to $\widetilde{\M{Q}}$ on $\RS{S}$. Especially, when the Jacobian determinant does not depend on the specific points $\M{Q}$ and $\widetilde{\M{Q}}$, in the sense that $\big|\M{J}_{\M{Q},\widetilde{\M{Q}}}\big|=|\M{J}|$ for all $\M{Q}\in\Set{V}$ and the corresponding $\widetilde{\M{Q}}\in\RS{S}$, we have
\begin{equation}\label{volume_transformation}
{\rm Vol}(\RS{S}) = |\M{J}|{\rm Vol}(\Set{V}).
\end{equation}

Next, let us consider the structure of ${\rm d}\M{Q}$ and ${\rm d}\widetilde{\M{Q}}$. For the differential ${\rm d}\M{Q}$, we see that
\begin{align}\label{differential_stiefel}
{\rm d}(\M{Q}\M{Q}^{\rm H}) &=\M{0} \nonumber \\
&=({\rm d}\M{Q})\M{Q}^{\rm H}+\M{Q}{\rm d}\M{Q}^{\rm H}.
\end{align}
Without loss of generality, we consider the tangent space at $\M{Q}_0=[\M{I},~\M{0}_{\rv{M}_{\rm SC}\times (T-\rv{M}_{\rm SC})}]$. Thus from \eqref{differential_stiefel} we have
\begin{equation}
{\rm d}\M{Q}_0 =[\M{\Delta}_{\parallel},~\M{\Delta}_{\perp}],
\end{equation}
where $\M{\Delta}_{\parallel}\in\mathbb{C}^{\rv{M}_{\rm SC}\times \rv{M}_{\rm SC}}$ is a skew-Hermitian matrix, and $\M{\Delta}_{\perp}\in\in\mathbb{C}^{\rv{M}_{\rm SC}\times (T-\rv{M}_{\rm SC})}$ is an arbitrary matrix. As for the tangent space at an arbitrary $\M{Q}\in\Set{V}$, we have $\M{Q}=\M{Q}_0\M{U}$, where $\M{U}$ is a $T\times T$ unitary matrix, and hence
\begin{equation}
{\rm d}\M{Q} = ({\rm d}\M{Q}_0)\M{U}=[\M{\Delta}_{\parallel},~\M{\Delta}_{\perp}]\M{U}.
\end{equation}
Similarly, the matrix $\widetilde{\M{Q}}_0\in\RS{S}$ corresponding to $\M{Q}_0$ takes the form of
\begin{equation}
\widetilde{\M{Q}}_0=[\sqrt{T}\widetilde{\RM{\Sigma}},~\M{0}_{\rv{M}_{\rm SC}\times (T-\rv{M}_{\rm SC})}].
\end{equation}
Its differential
$$
{\rm d}\widetilde{\M{Q}}_0 = [\widetilde{\M{\Delta}}_{\parallel},~\widetilde{\M{\Delta}}_{\perp}]
$$
satisfies
\begin{subequations}
\begin{align}
\widetilde{\M{\Delta}}_{\parallel}&=\sqrt{T}\widetilde{\RM{\Sigma}}\M{\Delta}_{\parallel}, \\
\widetilde{\M{\Delta}}_{\perp}&=\sqrt{T}\widetilde{\RM{\Sigma}}\M{\Delta}_{\perp}.
\end{align}
\end{subequations}
For a generic $\widetilde{\M{Q}}\in\RS{S}$, we also have
\begin{equation}
{\rm d}\widetilde{\M{Q}} = ({\rm d}\widetilde{\M{Q}}_0)\M{U}=[\widetilde{\M{\Delta}}_{\parallel},~\widetilde{\M{\Delta}}_{\perp}]\M{U}.
\end{equation}
Note that the inner product between matrices is preserved by right-multiplying unitary matrices, and thus the Riemannian metric of $\Set{V}$ at $\M{Q}$ is equal to that at $\M{Q}_0$, namely we have $\M{G}_{\M{Q}}=\M{G}_{\M{Q}_0}$. Moreover, the exterior product is also preserved by right-multiplying unitary matrices. Consequently, the volume form $\sqrt{|\widetilde{\M{G}}_{\M{Q}}|} ({\rm d}\M{Q})^{\wedge}$ on $\Set{V}$ is preserved by right-multiplying unitary matrices. Similarly, the volume form $\sqrt{|\M{G}_{\widetilde{\M{Q}}}|} ({\rm d}\widetilde{\M{Q}})^{\wedge}$ on $\RS{S}$ at $\widetilde{\M{Q}}=\widetilde{\M{Q}}_0\M{U}$ should be equal to that at $\widetilde{\M{Q}}_0$ as well. Thus the result \eqref{volume_transformation} is now applicable, and we have
\begin{equation}\label{volume_transformation2}
{\rm Vol}(\RS{S}) = \big|\M{J}_{\widetilde{\M{Q}}_0,\widetilde{\M{Q}}_0}\big|{\rm Vol}(\Set{V}).
\end{equation}

In order to obtain the Jacobian determinant $\big|\M{J}_{\M{Q}_0,\widetilde{\M{Q}}_0}\big|$, we consider the real vector representations of ${\rm d}\M{Q}_0$ and ${\rm d}\widetilde{\M{Q}}_0$. Naturally, we may choose the vector representation of ${\rm d}\M{Q}_0$ as
\begin{equation}
{\rm d}\V{q}_0 = [\V{v}_{\parallel},~{\rm Re}\{{\rm vec}(\M{\Delta}_{\perp})\},~{\rm Im}\{{\rm vec}(\M{\Delta}_{\perp})\}]^{\rm T},
\end{equation}
where $\V{v}_{\parallel}\in\mathbb{R}^{\rv{M}_{\rm SC}^2}$ is characterized by
\begin{equation}
\M{B}\V{v}_{\parallel}={\rm vec}(\M{\Delta}_{\parallel}),
\end{equation}
$\M{B}$ is a matrix constituted by orthonormal columns forming a basis of the space of all  $\rv{M}_{\rm SC}\times \rv{M}_{\rm SC}$ skew-Hermitian matrices. Specifically, we may express $\M{B}$ as follows
\begin{equation}
\footnotesize
\M{B}_{:,i} = \left\{
                \begin{array}{ll}
                  \frac{1}{\sqrt{2}}{\rm vec}(\M{E}_{jk}-\M{E}_{kj}), & \hbox{$i=j\rv{M}_{\rm SC}+k$;} \\
                  \frac{\sqrt{-1}}{\sqrt{2}}{\rm vec}(\M{E}_{jk}+\M{E}_{kj}), & \hbox{$i=\frac{\rv{M}_{\rm SC}(\rv{M}_{\rm SC}-1)}{2}+j\rv{M}_{\rm SC}+k$;} \\
                   \frac{1}{\sqrt{2}}{\rm vec}(\M{E}_{kk}), & \hbox{$i=\rv{M}_{\rm SC}(\rv{M}_{\rm SC}-1)+k$},
                \end{array}
              \right.
\end{equation}
where $j<k$. Now, we may write the vector representation of ${\rm d}\widetilde{\M{Q}}_0$ as
\begin{equation}
{\rm d}\widetilde{\V{q}}_0 = [\widetilde{\V{v}}_{\parallel},~{\rm Re}\{\widetilde{\V{v}}_{\perp}\},~{\rm Im}\{\widetilde{\V{v}}_{\perp}\}]^{\rm T},
\end{equation}
where
\begin{subequations}
\begin{align}
\widetilde{\V{v}}_{\parallel}&=\M{B}^{\rm H}(\M{I}_{\rv{M}_{\rm SC}}\otimes \sqrt{T}\widetilde{\RM{\Sigma}})\M{B}\V{v}_{\parallel}\\
\widetilde{\V{v}}_{\perp}&=(\M{I}_{T-\rv{M}_{\rm SC}}\otimes \sqrt{T}\widetilde{\RM{\Sigma}}){\rm vec}(\M{\Delta}_{\perp}).
\end{align}
\end{subequations}
Thus we obtain the Jacobian determinant as follows
\begin{align}
\big|\M{J}_{\M{Q}_0,\widetilde{\M{Q}}_0}\big|&=|\M{B}^{\rm H}(\M{I}_{\rv{M}_{\rm SC}}\otimes \sqrt{T}\widetilde{\RM{\Sigma}})\M{B}|\cdot |\M{I}_{T-\rv{M}_{\rm SC}}\otimes \sqrt{T}\widetilde{\RM{\Sigma}}|^2 \nonumber \\
&=|\M{I}_{\rv{M}_{\rm SC}}\otimes \sqrt{T}\widetilde{\RM{\Sigma}}|\cdot |\M{I}_{T-\rv{M}_{\rm SC}}\otimes \sqrt{T}\widetilde{\RM{\Sigma}}|^2 \nonumber \\
&=\Big|T\widetilde{\RM{\Sigma}}^2\Big|^{T-\frac{1}{2}\rv{M}_{\rm SC}},
\end{align}
where the second line follows from the fact that $\M{B}\in\mathbb{C}^{\rv{M}_{\rm SC}^2\times \rv{M}_{\rm SC}^2}$ is constituted by orthonormal columns, and that it is a square matrix, hence $\M{B}$ is unitary.

Finally, using \eqref{volume_transformation2}, we arrive at
\begin{equation}
{\rm Vol}(\RS{S})=\Big|T\widetilde{\RM{\Sigma}}^2\Big|^{T-\frac{1}{2}\rv{M}_{\rm SC}}{\rm Vol}(\Set{V}),
\end{equation}
which implies \eqref{rescaled_volume}.
\end{IEEEproof}

\section{Proof of Corollary \ref{coro:shc} and Corollary \ref{coro:zero_knowledge}}\label{sec:proof_shc}
\subsection{Proof of Corollary \ref{coro:shc}}
\begin{IEEEproof}
Observe that $\widetilde{\RM{\Sigma}}\RM{Q}$ in \eqref{shc_waveform} has an identical distribution as that of the optimal $\widetilde{\RM{\Sigma}}\widetilde{\RM{V}}_{\rm HX}^{\rm H}$ in Appendix \ref{sec:proof_sensing_limited} that maximizes the mutual information $I(\widetilde{\RM{U}}^{\rm H}\RM{Y}_{\rm c};\RM{X})$ in the high-\ac{snr} regime. Hence the $P_{\rm SC}$--achieving $\RM{X}$ satisfies
\begin{equation}\label{equation_x}
\RM{H}_{\rm c}\RM{X} = \sqrt{T}\widetilde{\RM{U}}\widetilde{\RM{\Sigma}}\RM{Q}.
\end{equation}
One of the solutions to \eqref{equation_x} is
\begin{subequations}
\begin{align}
\RM{X}&=\RM{X}_{\rm inf}+\RM{X}_{\perp}, \\
\RM{X}_{\rm inf}&=\sqrt{T}\RM{H}_{\rm c}^\dagger \widetilde{\RM{U}}\widetilde{\RM{\Sigma}}\RM{Q},
\end{align}
\end{subequations}
where $\RM{X}_{\rm inf}$ denotes the part of the signal that actually carries the information, while $\RM{X}_{\perp}$ serves as a non-informative padding in a subspace of ${\rm Col}(\widetilde{\M{R}}_{\RM{X}}^{\rm SC})$ that is orthogonal to ${\rm Col}(\RM{H}_{\rm c}\RM{X})$, which can be any matrix that satisfies
\begin{subequations}
\begin{align}
\RM{X}_{\perp}\RM{X}_{\rm inf}^{\rm H} &= \M{0},\\
T^{-1}(\RM{X}_{\perp}\RM{X}_{\perp}^{\rm H} + \RM{X}_{\rm inf}\RM{X}_{\rm inf}^{\rm H})&=\widetilde{\M{R}}_{\RM{X}}^{\rm SC}.
\end{align}
\end{subequations}
This is exactly the signalling scheme given in \eqref{shc_waveform}.

As for the uniform sampling procedure detailed in Corollary \ref{coro:shc}, it is a well-known method for obtaining a random sample from uniform distribution with respect to the Haar measure over the complex Stiefel manifold $\Set{V}$ defined in \eqref{stiefel_complex_definition} \cite{qr_steifel}. Thus the proof is completed.
\end{IEEEproof}

\subsection{Proof of Corollary \ref{coro:zero_knowledge}}

\begin{IEEEproof}
First let us introduce the following lemma.
\begin{lemma}[Unitary Equivalence of Hermitian Square Roots]\label{lem:unitary_equivalence}
Given a positive semidefinite Hermitian matrix $\M{H}\in\mathbb{C}^{m\times m}$, any pair of its square roots $\M{A}\in\mathbb{C}^{m\times n}$ and $\M{B}\in\mathbb{C}^{m\times n}$ (where $m\le n$) satisfying $\M{A}\M{A}^{\rm H}=\M{B}\M{B}^{\rm H}=\M{H}$, are equivalent up to a unitary transformation, in the sense that there exists a unitary matrix $\M{U}\in\mathbb{C}^{n\times n}$ satisfying $\M{A}=\M{B}\M{U}$.
\begin{IEEEproof}
Consider the singular value decomposition of $\M{A}$ and $\M{B}$ as follows
\begin{equation}
\M{A}=\M{U}_{\rm A} \M{\Sigma}_{\rm A} \M{V}_{\rm A}^{\rm H},~\M{B}=\M{U}_{\rm B} \M{\Sigma}_{\rm B} \M{V}_{\rm B}^{\rm H},
\end{equation}
where $\M{U}_{\rm A}\in\mathbb{C}^{m\times r_{\rm H}}$, $\M{U}_{\rm B}\in\mathbb{C}^{m\times r_{\rm H}}$, $\M{V}_{\rm A}\in\mathbb{C}^{n\times r_{\rm H}}$, $\M{V}_{\rm B}\in\mathbb{C}^{n\times r_{\rm H}}$ are semi-unitary matrices, $r_{\rm H}$ is the rank of $\M{H}$, while $\M{\Sigma}_{\rm A}\in\mathbb{C}^{r_{\rm H}\times r_{\rm H}}$ and $\M{\Sigma}_{\rm B}\in\mathbb{C}^{r_{\rm H}\times r_{\rm H}}$ are positive definite diagonal matrices. Since $\M{A}\M{A}^{\rm H}=\M{B}\M{B}^{\rm H}$, we have
\begin{equation}
    \M{U}_{\rm A}\M{\Sigma}_{\rm A}^2\M{U}_{\rm A}^{\rm H} =\M{U}_{\rm B}\M{\Sigma}_{\rm B}^2\M{U}_{\rm B}^{\rm H},
\end{equation}
and hence
\begin{equation}
    \M{U}_{\rm A}\M{\Sigma}_{\rm A}\M{U}_{\rm A}^{\rm H} =\M{U}_{\rm B}\M{\Sigma}_{\rm B}\M{U}_{\rm B}^{\rm H}.
\end{equation}
We can now rewrite $\M{A}$ and $\M{B}$ as
\begin{equation}\label{polar_decomp}
    \M{A} = \M{P}\M{W}_{\rm A},~\M{B}=\M{P}\M{W}_{\rm B},
\end{equation}
where $\M{P}=\M{U}_{\rm A}\M{\Sigma}_{\rm A}\M{U}_{\rm A}^{\rm H} =\M{U}_{\rm B}\M{\Sigma}_{\rm B}\M{U}_{\rm B}^{\rm H}$, while $\M{W}_{\rm A}=\M{U}_{\rm A}\M{V}_{\rm A}^{\rm H}\in\mathbb{C}^{m\times n}$ and $\M{W}_{\rm B}=\M{U}_{\rm B}\M{V}_{\rm B}^{\rm H}\in\mathbb{C}^{m\times n}$ are semi-unitary matrices. We may then construct the following augmented matrices
\begin{equation}
    \widetilde{\M{W}}_{\rm A}=[\M{W}_{\rm A}^{\rm T}~\M{\Delta}_{\rm A}^{\rm T}]^{\rm T},~\widetilde{\M{W}}_{\rm B}=[\M{W}_{\rm A}^{\rm T}~\M{\Delta}_{\rm B}^{\rm T}]^{\rm T},
\end{equation}
such that $\widetilde{\M{W}}_{\rm A}$ and $\widetilde{\M{W}_{\rm B}}$ are unitary. Let $\M{U}=\widetilde{\M{W}}_{\rm B}^{\rm H}\widetilde{\M{W}}_{\rm A}$, we see that $\widetilde{\M{W}}_{\rm A}=\widetilde{\M{W}}_{\rm B}\M{U}$, and $\M{U}$ is a unitary matrix. This also implies that $\M{W}_{\rm A}=\M{W}_{\rm B}\M{U}$, and hence we obtain $\M{A}=\M{B}\M{U}$ according to \eqref{polar_decomp}. This completes the proof.
\end{IEEEproof}
\end{lemma}

Now, observe that
$$
\begin{aligned}
&(\RM{H}_{\rm c}^{\dagger}\RM{H}_{\rm c}\RM{X}_{\rm SC,1})(\RM{H}_{\rm c}^{\dagger}\RM{H}_{\rm c}\RM{X}_{\rm SC,1})^{\rm H} \\
&\hspace{3mm}= (\sqrt{T}\RM{H}_{\rm c}^\dagger\widetilde{\RM{U}}\widetilde{\RM{\Sigma}}\RM{Q})(\sqrt{T}\RM{H}_{\rm c}^\dagger\widetilde{\RM{U}}\widetilde{\RM{\Sigma}}\RM{Q})^{\rm H}\\
&\hspace{3mm}=T\RM{H}_{\rm c}^{\dagger}\RM{H}_{\rm c}\widetilde{\M{R}}_{\RM{X}}^{\rm SC}(\RM{H}_{\rm c}^{\dagger}\RM{H}_{\rm c})^{\rm H}.
\end{aligned}
$$
According to Lemma \ref{lem:unitary_equivalence}, we have
\begin{equation}
    \RM{H}_{\rm c}^{\dagger}\RM{H}_{\rm c}\RM{X}_{\rm SC,1}\M{U}\!=\!\sqrt{T}\RM{H}_{\rm c}^{\dagger}\RM{H}_{\rm c}\M{U}_{\rm s}\M{\Lambda}_{\rm s}^{\frac{1}{2}}\RM{Q}\M{U}\!=\!\sqrt{T}\RM{H}_{\rm c}^\dagger\widetilde{\RM{U}}\widetilde{\RM{\Sigma}}\RM{Q},
\end{equation}
for some unitary matrix $\M{U}$. We see that $\sqrt{T}\RM{H}_{\rm c}^{\dagger}\RM{H}_{\rm c}\M{U}_{\rm s}\M{\Lambda}_{\rm s}^{\frac{1}{2}}\RM{Q}\M{U}$ has the same distribution as $\sqrt{T}\RM{H}_{\rm c}^{\dagger}\RM{H}_{\rm c}\M{U}_{\rm s}\M{\Lambda}_{\rm s}^{\frac{1}{2}}\RM{Q}$ does, since $\RM{Q}$ is sampled from the Haar measure on the complex Stiefel manifold $\Set{V}$, which is invariant under multiplication of unitary matrices. Finally, we may choose $\RM{X}_{\perp}$ as follows
$$
\RM{X}_{\perp} = (\M{I}-\RM{H}_{\rm c}^{\dagger}\RM{H}_{\rm c})\RM{X}_{\rm SC,1},
$$
which ensures that $\RM{X}_{\rm SC,1}$ is a valid candidate of $\RM{X}_{\rm SC}$.
\end{IEEEproof}

\section{Proof of Theorem \ref{thm:sensing_dofloss}}\label{sec:proof_sensing_dofloss}
\begin{IEEEproof}
First, let us denote
\begin{equation}
\widetilde{\M{\Phi}}(\M{A}) = \M{\Phi}(\M{A}) - \widetilde{\M{J}}_{\rm P}.
\end{equation}
According to \eqref{phi_def}, $\widetilde{\M{\Phi}}(\cdot)$ is a positive linear map characterized by
\begin{equation}
\widetilde{\M{\Phi}}(\M{A})=\sum_{i=1}^{r_1}\widetilde{\M{F}}_i\M{A}^{\rm T} \widetilde{\M{F}}_i^{\rm H}+\sum_{j=1}^{r_2}\widetilde{\M{G}}_j\M{A}\widetilde{\M{G}}_j^{\rm H}.
\end{equation}
Consider the eigendecomposition of $\widetilde{\M{R}}_{\RM{X}}$
\begin{equation}
\widetilde{\M{R}}_{\RM{X}} = \M{U}\M{\Lambda}\M{U}^{\rm H},
\end{equation}
where $\M{\Lambda}\in\mathbb{R}^{\mathrm{rank}(\widetilde{\RM{R}}_{\RM{X}})\times \mathrm{rank}(\widetilde{\RM{R}}_{\RM{X}})}$ is a diagonal matrix containing all non-zero eigenvalues of $\widetilde{\M{R}}_{\RM{X}}$, and $\M{U}\in\mathbb{C}^{M\times \mathrm{rank}(\widetilde{\M{R}}_{\RM{X}})}$ is a semi-unitary matrix constituted by the corresponding eigenvectors. Now, observe that the sample covariance matrix $\RM{R}_{\RM{X}}$ may be expressed as
\begin{equation}
\RM{R}_{\RM{X}} = \M{U}\M{\Lambda}^{\frac{1}{2}}\RM{W}\M{\Lambda}^{\frac{1}{2}}\M{U}^{\rm H},
\end{equation}
where $\RM{W}$ satisfies $T\RM{W}\sim\mathcal{CW}_{\mathrm{rank}(\widetilde{\M{R}}_{\RM{X}})}(\M{I},T)$, namely $T\RM{W}$ is complex Wishart distributed. Next, we construct the linear map
\begin{equation}\label{phi_unital}
\M{\Phi}_{\rm unital}(\RM{W}) = \M{L}^{-1}\widetilde{\M{\Phi}}(\M{U}\M{\Lambda}^{\frac{1}{2}}\RM{W}\M{\Lambda}^{\frac{1}{2}}\M{U}^{\rm H})(\M{L}^{\rm H})^{-1},
\end{equation}
which $\M{L}$ is obtained by the Cholesky decomposition
\begin{equation}\label{def_l}
\M{L}\M{L}^{\rm H} = \widetilde{\M{\Phi}}(\widetilde{\M{R}}_{\RM{X}}).
\end{equation}
The inverse $\M{L}^{-1}$ exists since $\M{\Phi}(\widetilde{\M{R}}_{\RM{X}})$ is invertible. Note that $\M{\Phi}_{\rm unital}$ is a unital map in the sense that
\begin{equation}
\M{\Phi}_{\rm unital}(\M{I}) = \M{L}^{-1}\widetilde{\M{\Phi}}(\widetilde{\M{R}}_{\RM{X}})(\M{L}^{\rm H})^{-1}=\M{I}.
\end{equation}
Moreover, $\M{\Phi}_{\rm unital}$ maps one positive semidefinite matrix to another, and hence it is a positive unital linear map. According to Choi \cite{choi_positive}, such a map satisfies
\begin{equation}\label{choi_inequality}
[\M{\Phi}_{\rm unital}(\RM{W})]^{-1}\preceq \M{\Phi}_{\rm unital}(\RM{W}^{-1}),
\end{equation}
which implies that
\begin{equation}
\M{L}^{\rm H}[\widetilde{\M{\Phi}}(\RM{R}_{\RM{X}})]^{-1}\M{L}\preceq \M{\Phi}_{\rm unital}(\RM{W}^{-1}),
\end{equation}
and hence
\begin{equation}
[\widetilde{\M{\Phi}}(\RM{R}_{\RM{X}})]^{-1} \preceq [\widetilde{\M{\Phi}}(\widetilde{\M{R}}_{\RM{X}})]^{-1}\widetilde{\M{\Phi}}(\M{U}\M{\Lambda}^{\frac{1}{2}}\RM{W}^{-1}\M{\Lambda}^{\frac{1}{2}}\M{U}^{\rm H})[\widetilde{\M{\Phi}}(\widetilde{\M{R}}_{\RM{X}})]^{-1}.
\end{equation}
Since $T\RM{W}\sim \mathcal{CW}_{\mathrm{rank}(\widetilde{\M{R}}_{\RM{X}})}(\M{I},T)$, we obtain
\begin{equation}
\mathbb{E}\{\RM{W}^{-1}\} = T\mathbb{E}\{(T\RM{W})^{-1}\} =\frac{T}{T-\mathrm{rank}(\widetilde{\M{R}}_{\RM{X}})}\M{I}.
\end{equation}
Thus we have
\begin{align}\label{r_bound}
&\mathbb{E}\{[\widetilde{\M{\Phi}}(\RM{R}_{\RM{X}})]^{-1}\} \nonumber \\
&\hspace{3mm}\preceq [\widetilde{\M{\Phi}}(\widetilde{\M{R}}_{\RM{X}})]^{-1}\widetilde{\M{\Phi}}(\M{U}\M{\Lambda}^{\frac{1}{2}}\mathbb{E}(\RM{W}^{-1})\M{\Lambda}^{\frac{1}{2}}\M{U}^{\rm H})[\widetilde{\M{\Phi}}(\widetilde{\M{R}}_{\RM{X}})]^{-1} \nonumber \\
&\hspace{3mm}= \frac{T}{T-\mathrm{rank}(\widetilde{\M{R}}_{\RM{X}})}[\widetilde{\M{\Phi}}(\widetilde{\M{R}}_{\RM{X}})]^{-1}\widetilde{\M{\Phi}}(\widetilde{\M{R}}_{\RM{X}})[\widetilde{\M{\Phi}}(\widetilde{\M{R}}_{\RM{X}})]^{-1} \nonumber \\
&\hspace{3mm}= \frac{T}{T-\mathrm{rank}(\widetilde{\M{R}}_{\RM{X}})}[\widetilde{\M{\Phi}}(\widetilde{\M{R}}_{\RM{X}})]^{-1}.
\end{align}

Next, we will show that when $K\leq {\rm rank}(\widetilde{\M{R}}_{\RM{X}})$, we have
\begin{equation}\label{k_bound}
\mathbb{E}\{[\widetilde{\M{\Phi}}(\RM{R}_{\RM{X}})]^{-1}\}\preceq \frac{T}{T-K}[\widetilde{\M{\Phi}}(\widetilde{\M{R}}_{\RM{X}})]^{-1}.
\end{equation}
Observe that
\begin{equation}
\M{\Phi}_{\rm unital}(\M{A})=\sum_{i=1}^{r_1}\underline{\M{F}}_i\M{A}^{\rm T} \underline{\M{F}}_i^{\rm H}+\sum_{j=1}^{r_2}\underline{\M{G}}_j\M{A}\underline{\M{G}}_j^{\rm H},
\end{equation}
where $\underline{\M{F}}_i = \M{L}^{-1}\widetilde{\M{F}}_i\M{U}\M{\Lambda}^{\frac{1}{2}},~\underline{\M{G}}_i = \M{L}^{-1}\widetilde{\M{G}}_i\M{U}\M{\Lambda}^{\frac{1}{2}}$. Using the singular value decompositions
\begin{equation}
\underline{\M{F}}_i = \M{U}_i \M{\Sigma}_i\M{V}_i^{\rm H},~\underline{\M{G}}_i = \M{U}_{i+r_1} \M{\Sigma}_{i+r_1}\M{V}_{i+r_1}^{\rm H},
\end{equation}
where for all $i$, $\M{U}_i\in \mathbb{C}^{K\times K}$, $\M{\Sigma}_i\in\mathbb{C}^{K\times K}$, and $\M{V}_i\in\mathbb{C}^{{\rm rank}(\widetilde{\M{R}}_{\RM{X}})\times K}$, we obtain an alternative representation of $\M{\Phi}_{\rm unital}(\M{A})$ as follows
\begin{equation}\label{dual_stinespring}
\M{\Phi}_{\rm unital}(\M{A}) = \underline{\M{F}}\M{V}(\M{A})\underline{\M{F}}^{\rm H},
\end{equation}
where
\begin{align}
\underline{\M{F}} &= [\M{U}_1 \M{\Sigma}_1,~\dotsc,~\M{U}_{r_1+r_2} \M{\Sigma}_{r_1+r_2}],\nonumber\\
\M{V}(\M{A}) &= {\rm blkdiag}(\M{V}_1^{\rm H}\M{A}^{\rm T}\M{V}_1,~\dotsc,~\M{V}_{r_1+r_2}^{\rm H}\M{A}\M{V}_{r_1+r_2}).
\end{align}
Since $\M{A}=\M{I}$ implies $\M{V}(\M{A})=\M{I}$, $\M{\Phi}_{\rm unital}(\M{A})$ may also be viewed as a unital positive linear map of $\M{V}(\M{A})$, based on \eqref{phi_unital}. Thus we have
\begin{align}\label{choi_inequality2}
{[\M{\Phi}_{\rm unital}(\RM{W})]}^{-1}&=[\underline{\M{F}}\M{V}(\RM{W})\underline{\M{F}}^{\rm H}]^{-1} \nonumber \\
&\preceq \underline{\M{F}}[\M{V}(\RM{W})]^{-1}\underline{\M{F}}^{\rm H}.
\end{align}
Note that
\begin{equation}
\mathbb{E}\{[\M{V}(\RM{W})]^{-1}\} = \frac{T}{T-K}\M{I}_{r_1+r_2}\otimes \M{I}_K,
\end{equation}
since each diagonal block in $\M{V}(\RM{W})$ satisfies
\begin{subequations}
\begin{align}
T\M{V}_i^{\rm H}\RM{W}^{\rm T}\M{V}_i&\sim \mathcal{CW}_K(\M{I},T),\\
T\M{V}_i^{\rm H}\RM{W}\M{V}_i&\sim \mathcal{CW}_K(\M{I},T).
\end{align}
\end{subequations}
Hence
\begin{align}
\mathbb{E}\{[\widetilde{\M{\Phi}}(\RM{R}_{\RM{X}})]^{-1}\}&\preceq \M{L}^{-1} \underline{\M{F}}\mathbb{E}\{[\M{V}(\RM{W})]^{-1}\}\underline{\M{F}}^{\rm H}(\M{L}^{\rm H})^{-1} \nonumber \\
&=\frac{T}{T-K}\M{L}^{-1}\M{\Phi}_{\rm unital}(\M{I})(\M{L}^{\rm H})^{-1} \nonumber \\
&=\frac{T}{T-K}[\widetilde{\M{\Phi}}(\widetilde{\M{R}}_{\RM{X}})]^{-1},
\end{align}
which implies \eqref{k_bound}. Combining \eqref{r_bound} and \eqref{k_bound}, we obtain
\begin{equation}\label{intermediate_result}
\mathbb{E}\{[\widetilde{\M{\Phi}}(\RM{R}_{\RM{X}})]^{-1}\}\preceq\frac{T[\widetilde{\M{\Phi}}(\widetilde{\M{R}}_{\RM{X}})]^{-1}}{T-\min\{K,{\rm rank}(\widetilde{\M{R}}_{\RM{X}})\}}.
\end{equation}

Next, let us consider the spectral decomposition of $\widetilde{\M{J}}_{\rm P}$
\begin{equation}\label{eigen_apriori}
\widetilde{\M{J}}_{\rm P} = \M{U}_{\rm P} \M{\Lambda}_{\rm P}\M{U}_{\rm P}^{\rm H}.
\end{equation}
Using \eqref{eigen_apriori}, we obtain
\begin{align}
&\mathbb{E}\Big\{{\rm tr}\Big[\Big(\widetilde{\M{\Phi}}(\RM{R}_{\RM{X}})+\widetilde{\M{J}}_{\rm P}\Big)^{-1}\Big]\Big\}\nonumber\\
&\hspace{3mm}=\mathbb{E}\Big\{{\rm tr}\Big[\Big(\M{U}_{\rm P}^{\rm H}\widetilde{\M{\Phi}}(\RM{R}_{\RM{X}})\M{U}_{\rm P}+\M{\Lambda}_{\rm P}\Big)^{-1}\Big]\Big\}\nonumber\\
&\hspace{3mm}=\mathbb{E}\Big\{{\rm tr}\Big[\M{\Lambda}_{\rm P}^{-\frac{1}{2}}\Big(\widetilde{\M{\Phi}}_0(\RM{R}_{\RM{X}})+\M{I}\Big)^{-1}\M{\Lambda}_{\rm P}^{-\frac{1}{2}}\Big]\Big\}\nonumber\\
&\hspace{3mm}={\rm tr}\Big\{\M{\Lambda}_{\rm P}^{-\frac{1}{2}}\mathbb{E}\Big[\Big(\widetilde{\M{\Phi}}_0(\RM{R}_{\RM{X}})+\M{I}\Big)^{-1}\Big]\M{\Lambda}_{\rm P}^{-\frac{1}{2}}\Big\},
\end{align}
where $\widetilde{\M{\Phi}}_0(\RM{R}_{\RM{X}})=\M{\Lambda}_{\rm P}^{-\frac{1}{2}}\M{U}_{\rm P}^{\rm H}\widetilde{\M{\Phi}}(\RM{R}_{\RM{X}})\M{U}_{\rm P}\M{\Lambda}_{\rm P}^{-\frac{1}{2}}$. Note that now it suffices to show that
$$
\mathbb{E}\Big[\Big(\widetilde{\M{\Phi}}_0(\RM{R}_{\RM{X}})+\M{I}\Big)^{-1}\Big]\preceq \frac{T\Big(\widetilde{\M{\Phi}}_0(\widetilde{\M{R}}_{\RM{X}})+\M{I}\Big)^{-1}}{T-\min\{K,{\rm rank}(\widetilde{\M{R}}_{\RM{X}})\}}.
$$
Let us define
\begin{equation}
\widetilde{\M{\Phi}}_1(\RM{R}_{\RM{X}}) = \widetilde{\M{\Phi}}_0(\RM{R}_{\RM{X}}) + \rv{\alpha}\M{I},
\end{equation}
which is a positive linear map, with the notation $\rv{\alpha}={\rm tr}\big[\RM{R}_{\RM{X}}\big]/\mathbb{E}\big\{{\rm tr}\big[\RM{R}_{\RM{X}}\big]\big\}$. Using again the previous arguments, we obtain
\begin{align}
\mathbb{E}\{[\widetilde{\M{\Phi}}_1(\RM{R}_{\RM{X}})]^{-1}\}&\preceq\frac{T[\widetilde{\M{\Phi}}_1(\widetilde{\M{R}}_{\RM{X}})]^{-1}}{T-\min\{K,{\rm rank}(\widetilde{\M{R}}_{\RM{X}})\}}\nonumber\\
&=\frac{T[\widetilde{\M{\Phi}}_0(\widetilde{\M{R}}_{\RM{X}})+\M{I}]^{-1}}{T-\min\{K,{\rm rank}(\widetilde{\M{R}}_{\RM{X}})\}}.
\end{align}
Next we will show that
$$
\mathbb{E}\Big[\Big(\widetilde{\M{\Phi}}_0(\RM{R}_{\RM{X}})+\M{I}\Big)^{-1}\Big]\preceq \mathbb{E}\{[\widetilde{\M{\Phi}}_1(\RM{R}_{\RM{X}})]^{-1}\},
$$
which can be rewritten as
\begin{equation}
\mathbb{E}\Big[\Big(\widetilde{\M{\Phi}}_0(\RM{R}_{\RM{X}})+\mathbb{E}(\rv{\alpha})\M{I}\Big)^{-1}\Big]\preceq\mathbb{E}\Big[\Big(\widetilde{\M{\Phi}}_0(\RM{R}_{\RM{X}})+\rv{\alpha}\M{I}\Big)^{-1}\Big].
\end{equation}
To this end, we consider the quantity
\begin{align}
&\mathbb{E}\Big[\Big(\widetilde{\M{\Phi}}_0(\RM{R}_{\RM{X}})+\mathbb{E}(\rv{\alpha})\M{I}\Big)^{-1}-\Big(\widetilde{\M{\Phi}}_0(\RM{R}_{\RM{X}})+\rv{\alpha}\M{I}\Big)^{-1}\Big]\nonumber\\
&\hspace{1mm}=\!\mathbb{E}\Big[(\rv{\alpha}\!-\!\mathbb{E}(\rv{\alpha}))\big[\big(\widetilde{\M{\Phi}}_0(\RM{R}_{\RM{X}})\!+\!\M{I}\big)\big(\widetilde{\M{\Phi}}_0(\RM{R}_{\RM{X}})\!+\!\rv{\alpha}\M{I}\big)\big]^{-1}\Big]\nonumber\\
&\hspace{1mm}=\!{\rm cov}\left\{\rv{\alpha}\M{I},\big[\big(\widetilde{\M{\Phi}}_0(\RM{R}_{\RM{X}})\!+\!\M{I}\big)\big(\widetilde{\M{\Phi}}_0(\RM{R}_{\RM{X}})\!+\!\rv{\alpha}\M{I}\big)\big]^{-1}\right\}\nonumber\\
&\hspace{1mm}=\!{\rm cov}\left\{\rv{\alpha}\M{I},\mathbb{E}\Big\{\big[\big(\widetilde{\M{\Phi}}_0(\RM{R}_{\RM{X}})\!+\!\M{I}\big)\big(\widetilde{\M{\Phi}}_0(\RM{R}_{\RM{X}})\!+\!\rv{\alpha}\M{I}\big)\big]^{-1}|\rv{\alpha}\Big\}\right\},
\end{align}
where ${\rm cov}(\cdot,\cdot)$ is defined as ${\rm cov}(\RM{A},\RM{B})=\mathbb{E}(\RM{A}\RM{B})-\mathbb{E}(\RM{A})\mathbb{E}(\RM{B})$, and the quantity
$$
\mathbb{E}\Big\{\big[\big(\widetilde{\M{\Phi}}_0(\RM{R}_{\RM{X}})\!+\!\M{I}\big)\big(\widetilde{\M{\Phi}}_0(\RM{R}_{\RM{X}})\!+\!\rv{\alpha}\M{I}\big)\big]^{-1}|\rv{\alpha}\Big\}
$$
can be seen to be a matrix-valued monotonically decreasing function of $\rv{\alpha}$. Next, note that ${\rm cov}\left\{\RM{A}(\rv{\alpha}),\RM{B}(\rv{\alpha})\right\}\succeq \M{0}$ when $\RM{A}(\rv{\alpha})$ and $\RM{B}(\rv{\alpha})$ are mutually commuting matrix-valued monotonically increasing functions of $\rv{\alpha}$ \cite{increasing_preserving}. Correspondingly, when $\M{A}(\rv{\alpha})$ is increasing and $\M{B}(\rv{\alpha})$ is decreasing, we would have ${\rm cov}\left\{\RM{A}(\rv{\alpha}),\RM{B}(\rv{\alpha})\right\}\preceq \M{0}$. This implies that
$$
\mathbb{E}\Big[\Big(\widetilde{\M{\Phi}}_0(\RM{R}_{\RM{X}})+\mathbb{E}(\rv{\alpha})\M{I}\Big)^{-1}-\Big(\widetilde{\M{\Phi}}_0(\RM{R}_{\RM{X}})+\rv{\alpha}\M{I}\Big)^{-1}\Big]\preceq \M{0},
$$
and hence the proof is completed.
\end{IEEEproof}

\section{Proof of Corollary \ref{coro:equality_dof}}\label{sec:proof_equality_dof}
\begin{IEEEproof}
From \eqref{dual_stinespring} we see that, when $K\leq {\rm rank}(\widetilde{\M{R}}_{\RM{X}})$, the equality in \eqref{sensing_dofloss} is achieved if
\begin{equation}\label{condition_equality1}
\mathbb{E}\{[\underline{\M{F}}\M{V}(\RM{W})\underline{\M{F}}^{\rm H}]^{-1}\} = \mathbb{E}\{\underline{\M{F}}[\M{V}(\RM{W})]^{-1}\underline{\M{F}}^{\rm H}\}.
\end{equation}
Note that $\underline{\M{F}}\underline{\M{F}}^{\rm H}=\M{I}$ since $\M{\Phi}_{\rm unital}$ is unital with respect to $\M{V}(\RM{W})$. Thus when $r_1+r_2=1$, we have that $\underline{\M{F}}\in\mathbb{C}^{K\times K}$, which implies that $\underline{\M{F}}$ is unitary, and hence \eqref{condition_equality1} holds.

Next, since $\widetilde{\M{J}}_{\rm P}$ can be neglected in the limit of $\sigma_{\rm s}\rightarrow 0$, let us assume that $\widetilde{\M{J}}_{\rm P}=\M{0}$ and obtain
\begin{equation}
\M{\Phi}(\M{A}) = \M{F}^{\rm st}\left[
                                  \begin{array}{cc}
                                    \M{I}\otimes \M{A}^{\rm T} & \M{0} \\
                                     \M{0} &  \M{I}\otimes \M{A} \\
                                  \end{array}
                                \right]
(\M{F}^{\rm st})^{\rm H}.
\end{equation}
Therefore, if $\M{F}^{\rm st}$ is unitary, we may conclude that
\begin{align}
&\mathbb{E}\{[\M{\Phi}(\RM{R}_{\RM{X}})]^{-1}\} \nonumber \\
&\hspace{3mm}=\M{F}^{\rm st}\left[
                                  \begin{array}{cc}
                                    \M{I}\otimes (\mathbb{E}\{\RM{R}_{\RM{X}}^{-1}\})^{\rm T} & \M{0} \\
                                     \M{0} &  \M{I}\otimes \mathbb{E}\{\RM{R}_{\RM{X}}^{-1}\} \\
                                  \end{array}
                                \right]
(\M{F}^{\rm st})^{\rm H} \nonumber \\
&\hspace{3mm}=\frac{T}{T-M}[\M{\Phi}(\widetilde{\M{R}}_{\RM{X}})]^{-1},
\end{align}
where the last line follows from the fact that
$$
T\RM{R}_{\RM{X}} \sim \mathcal{CW}_M(\widetilde{\M{R}}_{\RM{X}},T).
$$
This implies the equality in \eqref{sensing_dofloss}, since ${\rm rank}(\widetilde{\M{R}}_{\RM{X}})=M$ when $\widetilde{\M{R}}_{\RM{X}}$ is invertible.
\end{IEEEproof}

\section{Proof of Proposition \ref{prop:ts_sug}}\label{sec:proof_ts_sug}
\begin{IEEEproof}
To prove this proposition, it suffices to show that there exists at least one point $(\epsilon,R)$ on the Gaussian inner bound or the semi-unitary inner bound that lies above the line segment connecting $P_{\rm SC}$ and $P_{\rm CS}$. Once we found such a point $P=(\epsilon,R)$, we see that the time-sharing strategies between $P$, $P_{\rm SC}$ and $P_{\rm CS}$ readily constitute a tighter inner bound than the pentagon inner bound.

Next we show that there do exist such points, under mild assumptions. To this end, let us consider the slope of the semi-unitary--Gaussian inner bound at a specific value of $\alpha$. For our purpose, it suffices to show that at $P_{\rm SC}$ (resp. $P_{\rm CS}$), the slope of the semi-unitary--Gaussian inner bound is larger (resp. smaller) than the slope of the line segment connecting $P_{\rm SC}$ and $P_{\rm CS}$. First note that, as long as the optimal objective function value is not identical for all $\alpha\in[0,1]$, the slope of the outer bound \eqref{the_outer_bound} at a specific value of $\alpha$ is given by
\begin{equation}
\frac{\partial R_{\rm out}(\alpha)}{\partial \epsilon_{\rm out}(\alpha)} = \frac{T}{\sigma_{\rm s}^2} \cdot \lambda_{\alpha},
\end{equation}
where $\lambda_{\alpha}$ is the optimal Lagrange dual variable corresponding to the constraint \eqref{constraint_crb}. According to the correspondence between the formulations \eqref{opt_problem_pareto} and \eqref{opt_problem_pareto_alter} (c.f. \cite[Sec.~5.3.3]{convex_opt}), $\lambda_{\alpha}$ can be explicitly obtained as
\begin{equation}
\lambda_{\alpha} = \frac{1-\alpha}{\alpha},
\end{equation}
which tends to infinity as $\alpha\rightarrow 0$. Now, the slope of the semi-unitary inner bound can be expressed as
\begin{equation}
\frac{\partial R_{\rm in,U}(\alpha)}{\partial \epsilon_{\rm in,U}(\alpha)} = \Big(1-\frac{\rv{M}_{\rm U}}{2T}\Big)\frac{(1-\alpha) T}{\alpha\sigma_{\rm s}^2}+O(\sigma_{\rm c}^2),
\end{equation}
which also tends to infinity as $\alpha\rightarrow 0$. Since the semi-unitary--Gaussian inner bound equals to the semi-unitary inner bound around $P_{\rm SC}$, we may now conclude that the slope of the semi-unitary--Gaussian inner bound at $P_{\rm SC}$ is positive infinity. Following a similar line of reasoning, we can also show that the slope at $P_{\rm CS}$ is zero. Therefore, no matter what the slope of the line segment connecting $P_{\rm SC}$ and $P_{\rm CS}$ is, it should be upper and lower bounded by the slopes of the semi-unitary--Gaussian inner bound at $P_{\rm SC}$ and $P_{\rm CS}$, respectively. This completes the proof.
\end{IEEEproof}

\bibliographystyle{IEEEtran}
\bibliography{isac}
\end{document}